\documentclass[aps,physrev,reprint,superscriptaddress]{revtex4-2}

\usepackage{import}
\usepackage{tabularx}
\usepackage{graphicx}
\usepackage{hyperref}
\usepackage{color}
\usepackage{booktabs}

\begin{document}


\title{Shortest-path percolation on scale-free networks}


\author{Minsuk Kim}
\email{mk139@iu.edu}
\affiliation{Center for Complex Networks and Systems Research, Luddy School of Informatics, Computing, and Engineering, Indiana University, Bloomington, Indiana 47408, USA}
\author{Lorenzo Cirigliano}
\affiliation{Dipartimento di Fisica, Universit\`a \href{https://ror.org/02be6w209}{La Sapienza}, P.le
  A. Moro, 2, I-00185 Rome, Italy}
\author{Claudio Castellano}
\affiliation{\href{https://ror.org/05rcgef49}{Istituto dei Sistemi Complessi (ISC-CNR)}, Via dei Taurini
  19, I-00185 Rome, Italy}
\author{Hanlin Sun}
\affiliation{Nordita, KTH Royal Institute of Technology and Stockholm University, Hannes Alfv\'ens v\"ag 12, SE-106 91 Stockholm, Sweden}
\author{Robert Jankowski}
\affiliation{Departament de F\'isica de la Mat\`eria Condensada,
Universitat de Barcelona, Mart\'i i Franqu\`es 1, E-08028 Barcelona, Spain}
\affiliation{Universitat de Barcelona Institute of Complex Systems (UBICS), Universitat de Barcelona, Barcelona, Spain}
\author{Anna Poggialini}
\affiliation{Dipartimento di Fisica, Universit\`a \href{https://ror.org/02be6w209}{La Sapienza}, P.le  A. Moro, 2, I-00185 Rome, Italy}
\author{Filippo Radicchi}
\email{f.radicchi@gmail.com}
\affiliation{Center for Complex Networks and Systems Research, Luddy School of Informatics, Computing, and Engineering, Indiana University, Bloomington, Indiana 47408, USA}


\date{\today}

\begin{abstract}
The shortest-path-percolation (SPP) model aims at describing the consumption and eventual exhaustion of a network's resources. Starting from a network containing a macroscopic connected component, random pairs of nodes are sequentially selected, and if the length of the shortest path connecting the node pairs is smaller than a tunable budget parameter, then all edges along such a path are removed from the network. As edges are progressively removed, the network eventually breaks into multiple microscopic components, undergoing a percolation-like transition. It is known that the SPP transition on Erd\H{o}s-R\'enyi networks (ERNs) belongs to the same universality class as of the ordinary bond percolation if the budget parameter is finite; for an unbounded budget, instead, the SPP transition becomes more abrupt than the ordinary percolation transition. By means of large-scale numerical simulations and finite-size scaling analysis, here we study the SPP transition on random scale-free networks (SFNs) characterized by power-law degree distributions. We find, in contrast with ordinary percolation, that the transition is identical to the one observed on ERNs, denoting independence from the degree exponent. Still, we distinguish finite- and infinite-budget SPP universality classes.  Our findings follow from the fact that the SPP process drastically homogenizes the heterogeneous structure of SFNs before the SPP transition takes place.

\end{abstract}


\maketitle


\section{Introduction\label{intro}}
Percolation theory offers comprehensive insights into networked complex systems by revealing the relationship between a system's macroscopic connectedness and its microscopic structure. Traditionally, percolation theory has been studied to understand various physical phenomena such as diffusion processes in porous media, the expansion of forest fires, and the gelation of molecules, providing a profound understanding on the fundamental mechanisms underlying these systems in terms of phase transition and critical phenomena~\cite{stauffer2018introduction}.  
Later, percolation theory has been successfully adopted to characterize the robustness of various complex systems, represented as complex networks, against deletion or failure of microscopic elements, either nodes or edges~\cite{li2021percolation, artime2024robustness}. 
The key assumption linking percolation theory to the robustness of complex systems is that global connectivity enables proper function.

The rule used to remove or add the microscopic elements of the network defines the specific percolation model. The ordinary percolation model, which removes 
microscopic elements uniformly at random, has been extensively studied~\cite{stauffer2018introduction}. While simple, this model has provided a theoretical foundation for understanding the organization of complex systems~\cite{callaway2000network, cohen2002percolation}. Other percolation models and processes have been introduced and studied to consider more complex rules 
mimicking relevant scenarios in various complex systems. These models include targeted attacks~\cite{albert2000error}, cascading failures on interdependent networks~\cite{buldyrev2010catastrophic}, explosive percolation~\cite{achlioptas2009explosive}, traffic percolation~\cite{li2015percolation}, triadic percolation~\cite{sun2023dynamic}, and extended-range percolation~\cite{cirigliano2023extended,cirigliano2024general}.

Given a specific percolation model at hand, the outcome can significantly differ depending on 
the initial structure of the underlying network.
The most well-known example is the one of ordinary percolation on scale-free networks (SFNs).
On SFNs, in which the degree distribution follows a power law, i.e., $P(k)\sim k^{-\lambda}$, various anomalous behaviors have been observed, which are not reported in the case of Erd\H{o}s-R\'{e}nyi networks (ERNs)~\cite{erdos1959random, erdos1960evolution}. 

For example, the so-called ``vanishing critical threshold problem''—where the nonpercolating phase is reached only after removing every edge—has been observed
when $\lambda<3$. 
This is related to the divergence of the second moment of the degree distribution $P(k)$~\cite{cohen2002percolation}. Another anomalous behavior is the breaking of the site-bond percolation universality~\cite{radicchi2015breaking}. 
Moreover, other studies claim contradictory results in the analytical expression of some critical exponents; notably, the correlation exponent governing finite-size scaling (FSS)~\cite{cohen2002percolation, dorogovtsev2008critical, radicchi2015breaking, li2021percolation}, while recently reconciled in~\cite{cirigliano2024scaling}.
These anomalies are not only present in ordinary percolation models. For instance, the explosive percolation transition under the Achlioptas process displays a wide range of different critical exponents, depending on whether the network's degree distribution $P(k)$ follows a Poisson distribution or a power-law distribution with different values of $\lambda$~\cite{achlioptas2009explosive, radicchi2009explosive, cho2009percolation, radicchi2010explosive, dacosta2010explosive, riordan2011explosive}. These findings highlight the crucial role of network heterogeneity in the nature of the percolation transition.

Recently, the shortest-path-percolation (SPP) model, which is intended to describe the consumption and eventual depletion of network resources, was introduced and studied on ERNs~\cite{kim2024shortest}. In the model, one agent at a time demands a path from randomly selected origin and destination nodes; if the geodesic distance between the demanded pair of nodes does not exceed the budget $C$ available to the agent, then all edges along one of the shortest paths connecting origin and destination nodes are removed from the network; eventually, the network undergoes a percolation transition characterized by the disintegration of its macroscopic connected component into multiple microscopic clusters. $C$ is a tunable parameter of the SPP model; for $C=1$, the SPP reduces to the ordinary percolation model, however for $C>1$ the SPP model differs from the ordinary model as removed edges are topologically correlated. On ERNs, it was shown that the critical exponents of the SPP transition are different from the mean-field percolation exponents only if the parameter $C$ is unbounded~\cite{kim2024shortest, meng2025path}; mean-field exponents of the ordinary percolation transition are found instead for the finite-$C$ case~\cite{kim2024shortest}.

In this paper, we study the SPP transition on SFNs by means of extensive Monte Carlo simulations and FSS analysis.
For $C=1$, we clearly recover known results about the ordinary percolation model, where the transition on ERNs and SFNs is characterized by radically different behaviors as long as $\lambda < 4$. For $C>1$ however, we find that the SPP critical exponents on SFNs are identical to those for ERNs; this happens regardless of the specific value of the degree exponent $\lambda$. We conjecture and confirm that it is because the SPP process homogenizes the heterogeneous network structure before the SPP transition happens. Finally, we find that the correlation exponents vary from those for ERNs, depending on $\lambda$ and $C$, displaying a rich FSS behavior. We validate our numerical findings by performing data collapse across different system sizes.

The structure of the paper is as follows. In Sec.~\ref{sec:spp-model}, we introduce and define the SPP model. In Sec.~\ref{sec:methods}, we describe the methods of our work, including the network model, observables, and FSS analysis. In Sec.~\ref{sec:results}, we present results of the numerical analysis on SFNs with various values of $\lambda$. Finally, a summary of our work and discussion is included in Sec.~\ref{sec:discussion}.

\section{Shortest-Path-Percolation Model\label{sec:spp-model}}

The SPP model was introduced in Ref.~\cite{kim2024shortest}. The model is defined as follows:
For $t>0$, let $\mathcal{G}_{t} = (\mathcal{V}, \mathcal{E}_t)$ represent the undirected and unweighted network available at time $t$, where $N=|\mathcal{V}|$ is the number of nodes and $E_t=|\mathcal{E}_t|$ is the number of edges. At each time $t$, an agent demands a path connecting the origin-destination pair $o_t \to d_t$. If there exists a path between $o_t$ and $d_t$ in $\mathcal{G}_t$, we denote with $Q_t$ the shortest-path length. The shortest path connecting $o_t$ and $d_t$ can be supplied to the agent only if $Q_t \leq C$, where $C>0$ is a tunable parameter of the model. Here, $C$ can be interpreted as a maximum budget to traverse the network. If this condition is satisfied, we identify the shortest path, namely $(o_t = i_1 , i_2 , \ldots , i_{Q_{t}+1} = d_t)$, and remove all edges in the path from $\mathcal{G}_{t}$, i.e., $\mathcal{E}_{t}  \mapsto \mathcal{E}_{t} \setminus \bigcup_{q=1}^{Q_{t}} (i_{q},  i_{q+1})$. Note that one path is selected at random if more than one exists. If there is no path between $o_t$ and $d_t$ or if $Q_t > C$, no edge is removed from the network. In either case, we copy the network $\mathcal{G}_{t+1} \mapsto \mathcal{G}_{t}$ and then increase $t \mapsto t + 1$. The above-mentioned process is repeated until there is no more edge left in the network, i.e., $E_t=0$.
Assuming that the pair of nodes $o_t \to d_t$ is selected uniformly at random, the SPP model with $C=1$ corresponds to the ordinary bond-percolation model. 
In this work, we focus on the SPP model with finite budget $C=1,~2,~ 3,$ and unbounded budget $C\geq N-1$. For simplicity, we denote the unbounded budget case as $C=N$.

\section{Methods\label{sec:methods}}

\subsection{Uncorrelated scale-free networks}
We apply the SPP model to random SFNs characterized by the power-law degree distribution, $P(k)\sim{k^{-\lambda}}$ if $k \in [k_{\min}, k_{max}]$ and $P(k) = 0$ otherwise, generated according to the so-called uncorrelated configuration model (UCM)~\cite{catanzaro2005generation}. We set the minimum degree $k_{\text{min}}=4$ to ensure that the resulting network is composed of a single connected component. To ensure negligible degree-degree correlations, we impose the structural cutoff $k_{\text{max}}= \sqrt{N}$ when $2<\lambda < 3$ and the natural cut-off $k_{\text{max}}\sim N^{1/(\lambda-1)}$ for $\lambda \geq 3$.

\subsection{Observables}
By following the convention in previous percolation studies, we use the fraction of removed edges, namely $r$, as the control parameter~\cite{stauffer2018introduction}. This control parameter allows for immediate comparisons between the SPP model and other standard percolation models. We consider two standard observables in network percolation: the percolation strength $P$ and the average cluster size $S$~\cite{stauffer2018introduction}. $P$ serves as the order parameter of the SPP model and $S$ serves as a response function. By observing these quantities, we are able to characterize the cluster structure of a network under the SPP model. 
$P$ is defined as the size of the largest connected component of the network divided by the size of the network, thus 
\begin{equation}
P = \frac{s_Z}{\sum_{z=1}^Z s_z} \; ,
    \label{eq:perc_str}
\end{equation}
$S$ is instead defined as
\begin{equation}
S = \frac{\sum_{z=1}^{Z-1} s_z^2}{\sum_{z=1}^{Z-1} s_z} \; ,
    \label{eq:av_cluster}
\end{equation}
where $s_1 \leq s_2 \leq \cdots \leq s_{Z-1} \leq s_Z$ is the size of the $Z$ connected components observed in the network at a certain stage of the SPP model, i.e., $\sum_{z=1}^{Z} s_z=N$.

\subsection{Finite-size scaling in the event-based ensemble\label{subsec:method-fss}}
In our study, we rely on the so-called 
event-based ensemble~\cite{fan2020universal, li2023explosive, li2024explosive}.
Compared to the conventional ensemble, which samples the critical observables at the same value of the control parameter $r=r_c$, the event-based ensemble samples critical observables at the single-instance pseudocritical point $r=r_{c}^{*}(N)$ from each individual realization of the percolation process on a network of size $N$. The event-based ensemble is known to provide good estimates of critical exponents, especially when the percolation process entails substantial non-self-averaging effects~\cite{fan2020universal, li2023explosive, li2024explosive}.

Denote with $P(r, N)$ and $S(r, N)$ the values of the observables of Eqs.~(\ref{eq:perc_str}) and~(\ref{eq:av_cluster}), respectively, when exactly a fraction $r$ of edges is removed from a network of size $N$ during the SPP process. The single-instance pseudocritical point $r_c^*(N)$ is defined as
\begin{equation}
    r_c^* (N) =  \arg \max_r \, \left[ P(r,N) - P(r + 1/E, N) \right]  \;,
    \label{eq:ev_pseudo_single_threshold}
\end{equation}
where $E =E_0$ is the number of edges of the initial configuration of the network.
In other words, we locate the point when $P(r, N)$ undergoes the largest drop caused by the removal of a single edge from the network. By definition, Eq.~(\ref{eq:ev_pseudo_single_threshold}) implies that the observables are sampled just before the largest drop in $P$ happens, but observables can be sampled immediately after the largest drop as well~\cite{kim2024shortest}.

The single-instance pseudocritical point $r_c^*(N)$ is a random variable. Taking its ensemble average, we get an estimate of the pseudocritical point under the event-based ensemble
\begin{equation}
    r_c(N)=\langle r_c^*(N)\rangle \;.
    \label{eq:ev_pseudo_threshold}
\end{equation}
The quantity $r_c(N)$ follows the standard FSS ansatz 
\begin{equation}
    r_{c} (N) = r_{c} + b N^{-1/\bar\nu_{1}} \;,
    \label{eq:pseudo_scaling}
\end{equation}
where $r_c=r_c(\infty)$ is the critical point, $1/\bar\nu_{1}$ is the exponent related to the correlation length, and $b$ is a constant. The exponent $1/\bar\nu_{1}$ governs how fast the average $r_c(N)$ converges to its infinite-size limit $r_c$. However, in some cases, the random variable $r_c^*(N)$ cannot be described in terms of the sole average: we also need information on the fluctuations
\begin{equation}
    \sigma_{r_c^*}(N)=\sqrt{\langle {r_{c}^{*}}(N)^2\rangle - \langle r_{c}^{*}(N)\rangle^2} \sim N^{-1/\bar\nu_{2}},
    \label{eq:pseudo_variance}
\end{equation}
where $\bar\nu_{2}$ is another critical exponent determining how fast the random variable $r_c^{*}(N)$ converges to its average. In general, $1/\bar\nu_{1}=1/\bar\nu_{2}$ for ordinary percolation on various lattice and network structures~\cite{fan2020universal}. However, it was recently reported that $1/\bar\nu_{1}>1/\bar\nu_{2}$ for explosive percolation, while $1/\bar\nu_{1}<1/\bar\nu_{2}$ for ordinary bond percolation on lattices above the upper-critical dimension with free-boundary conditions~\cite{li2024crossover} and on SFNs~\cite{zhao2025finite}.

The critical observables under the event-based ensemble can be obtained by taking the ensemble average as follows:
\begin{equation}
    P_c(N) = \langle P(r_c^*(N), N)\rangle \;,
    \label{eq:ev_pseudo_gcc}
\end{equation}
and
\begin{equation}
    S_c(N) = \langle S(r_c^*(N), N)\rangle \;.
    \label{eq:ev_pseudo_avgs}
\end{equation}
They follow the FSS scaling
\begin{equation}
    P_c(N)   \sim N^{-\beta/\bar{\nu}} \;,
    \label{eq:ev_pseudo_crit_scaling1}
\end{equation}
and
\begin{equation}
    S_c(N)   \sim N^{\gamma/\bar{\nu}} \;,
    \label{eq:ev_pseudo_crit_scaling2}
\end{equation}
where $\beta/\bar\nu$ and $\gamma/\bar\nu$ are ratios of the critical exponents.
Note that we have not specified whether the exponent appearing here is either $\bar\nu_{1}$ or $\bar\nu_{2}$.
The scaling of the quantities $P_c(N)$ and $S_c(N)$ with the system size $N$
depends on which exponent between $\bar\nu_{1}$ and $\bar\nu_{2}$ is the largest.
If $1/\bar\nu_{1}>1/\bar\nu_{2}$, the fluctuations of the average around its infinite-size limit vanish much faster than the fluctuations of the random variable around the average, hence $\bar\nu=\bar\nu_{2}$.
If instead $1/\bar\nu_{1}<1/\bar\nu_{2}$, the random variable $r_c^{*}(N)$ quickly concentrates around its average $r_c(N)$,
while the average converges more slowly to the infinite-size value $r_c$; in this case $\bar\nu=\bar\nu_{1}$.

To estimate numerically $\beta/\bar\nu$, we perform a log-log transformation of the scaling relation
in Eq.~(\ref{eq:ev_pseudo_crit_scaling1}).
Specifically, by plotting $\ln P_c(N)$ versus $\ln N$, we expect a linear relationship of the form
\begin{equation}
    \ln P_c(N) = -\frac{\beta}{\bar\nu} \ln N + \text{constant} \;,
\end{equation}
from which the slope yields an estimate of $-\beta/\bar{\nu}$.
Similarly, $\gamma/\bar{\nu}$ can be obtained by plotting $\ln S_c(N)$ against $\ln N$;
the slope corresponds to $\gamma/\bar{\nu}$.
In both cases, the ratios of the critical exponents are estimated via least-squares fitting of the log-transformed data.

While the numerical estimate of $1/\bar\nu_2$ is readily obtained from Eq.~(\ref{eq:pseudo_variance}),
the estimation of the critical point $r_c$ and of the exponent $1/\bar\nu_{1}$ is a bit more elaborate.
From the FSS ansatz in Eq.~(\ref{eq:pseudo_scaling}), the deviation of the pseudocritical point $r_c(N)$ from the true critical point scales as $|r_c(N) - r_c| \sim N^{-1/\bar\nu_{1}}$.
Taking the logarithm, this yields a linear relationship:
\begin{equation}
    \ln |r_c(N) - r_c| = -\frac{1}{\bar\nu_{1}} \ln N + \text{constant} \;.
    \label{eq:critical_deviation}
\end{equation}
Based on this relation, we proceed as follows: given a tentative value $\hat{r}_c$, we compute the Pearson correlation coefficient $\rho$ between $\ln |\hat{r}_c - r_c(N)|$ and $\ln N$ over all available system sizes. We perform a grid search over a range of $\hat{r}_c$ values and, for each, record the corresponding $\rho$. Since the slope is expected to be negative, we identify the optimal estimate of $r_c$ as the value of $\hat{r}_c$ that yields $\rho$ closest to $-1$. Once $r_c$ is determined, we estimate $-1/\bar\nu_{1}$ as the slope of the best-fit line through the point $(\ln N, \ln |r_c(N) - r_c|)$.

\section{results\label{sec:results}}
\subsection{Estimation of $\beta/\bar\nu$ and $\gamma/\bar\nu$} \label{subsec:estimate_beta_gamma}
\begin{figure*}[ht!]
    \centering
    \includegraphics[width=\linewidth]{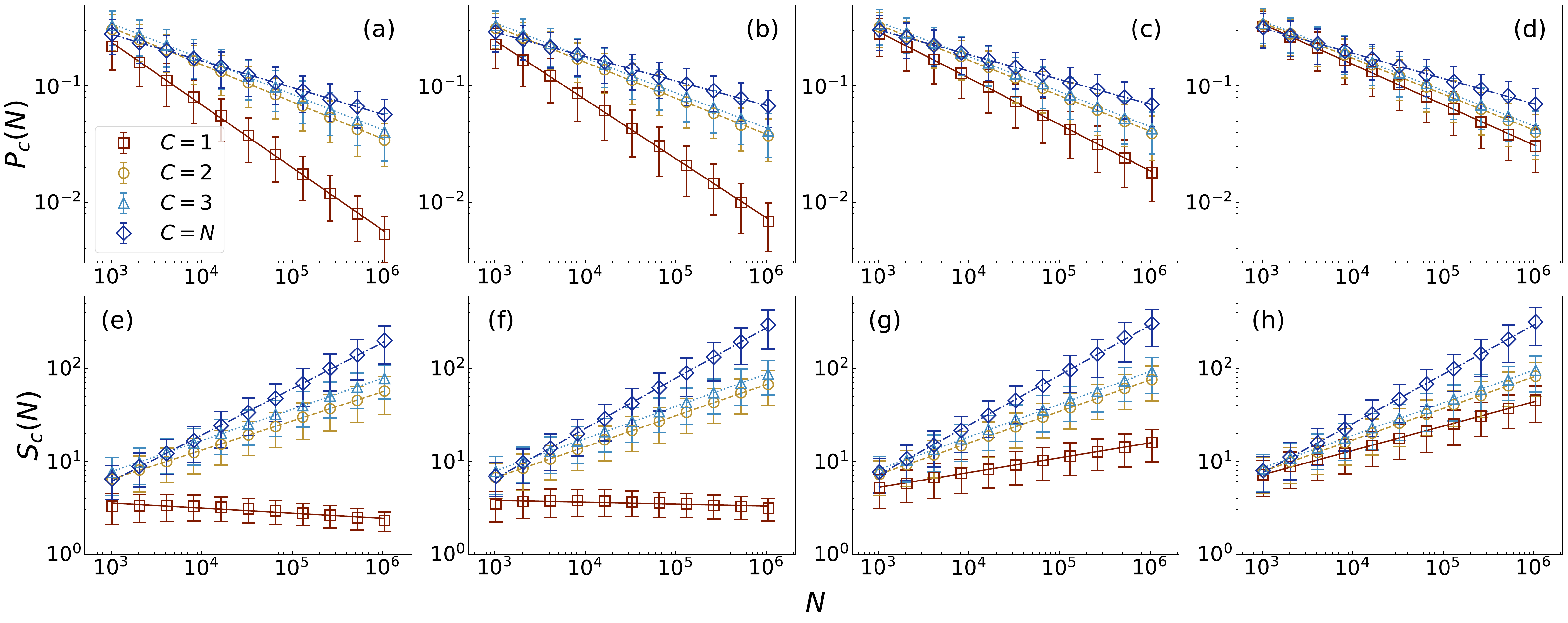}
    \caption{
    \textbf{Estimation of $\beta/\bar\nu$ and $\gamma/\bar\nu$.} We plot $P_c(N)$ as a function of $N$ for SFNs with 
    (a) $\lambda=2.1$, (b) $\lambda=2.7$, (c) $\lambda=3.5$, and (d) $\lambda=4.5$. 
    The slope of the straight lines on the log-log plot corresponds to the best estimate of the ratio $-\beta/\bar\nu$.
    Similarly, we plot $S_c(N)$ as a function of $N$ for SFNs with (e) $\lambda=2.1$, (f) $\lambda=2.7$, (g) $\lambda=3.5$, and (h) $\lambda=4.5$. 
    The slope of the straight lines on the log-log plot corresponds to the best estimate of the ratio $\gamma/\bar\nu$.
    Different symbols display results of the SPP model with $C=1$ (square), $C=2$ (circle), $C=3$ (triangle), and $C=N$ (diamond). Different line styles are used to display the best power-law fitting of the SPP model with $C=1$ (solid), $C=2$ (dashed), $C=3$ (dotted), and $C=N$ (dashed-dotted). 
    Symbols are the average values and error bars are the standard deviations.
    We report the number of realizations in Table~\ref{tab:simulation} of Ref.~\cite{supplementary}.
    }
    \label{fig1}
\end{figure*}

First, we estimate the critical exponent ratios $\beta/\bar\nu$ and $\gamma/\bar\nu$, which are respectively related to the relative size of the largest cluster $P$ and the average cluster size $S$ at criticality. In Figs.~\ref{fig1} (a)-(d), we plot $P_c(N)$ as a function of $N$ for different values of $\lambda$ and $C$. When $C=1$, the SPP model is identical to the ordinary bond percolation and we get $\beta/\bar\nu=0.54(1)$ and $\beta/\bar\nu=0.50(1)$ for $\lambda=2.1$ and $\lambda=2.7$, respectively. We can also obtain $\beta/\bar\nu = 0.40(1)$ and $\beta/\bar\nu=0.35(1)$ for $\lambda=3.5$ and $\lambda=4.5$, respectively. These values nicely align with the theoretical prediction in ~\cite{cohen2002percolation}, which are $\lambda$-dependent. Surprisingly, regardless of the value of $\lambda$, we find that the estimated values are $\beta/\bar\nu \approx 0.33$ when $C=2$ and $C=3$, which is compatible with the mean-field value $\beta/\bar\nu=1/3$. Moreover, we find that for the infinite-$C$ SPP transition $\beta/\bar\nu \approx 0.22$, regardless of the value of $\lambda$. 
This value was reported in previous studies on ERNs~\cite{kim2024shortest}. It is also compatible with the analytical prediction $\beta/\bar\nu=0.2$ by Meng \textit{et al.}~\cite{meng2025path}.

Similar results are found for $\gamma/\bar\nu$ as well [see Figs.~\ref{fig1} (e)-(h)]. In other words, our estimation reads $\gamma/\bar\nu\approx 0.33$ for $C=2$ and $C=3$, and $\gamma/\bar\nu\approx 0.55$ for infinite $C$, regardless of $\lambda$. 
When $C=1$ and $\lambda<3$, $S_c(N)$ does not diverge as $N \to \infty$ due to the lack of the subcritical phase,
thus displaying a decreasing behavior [see Figs.~\ref{fig1} (e) and (f)]. When $\lambda=3.5$, we could obtain $\gamma/\bar\nu=0.16(1)$ and when $\lambda=4.5$, we could get $\gamma/\bar\nu=0.26(1)$. FSS theory predicts $\gamma/\bar\nu=1/5$ when $\lambda=3.5$ and $\gamma/\bar\nu=1/3$ when $\lambda=4.5$, respectively, but these discrepancies might originate from finite-size effects and sample-to-sample fluctuations from different network realizations. 
Note that the hyperscaling relation $2\beta/\bar\nu + \gamma/\bar\nu=1$ roughly holds for the estimated exponents across different $\lambda$ and $C$. We report the details in Table~\ref{tab:summary}. 

\begin{figure}[!t]
    \centering
    \includegraphics[width=\linewidth]{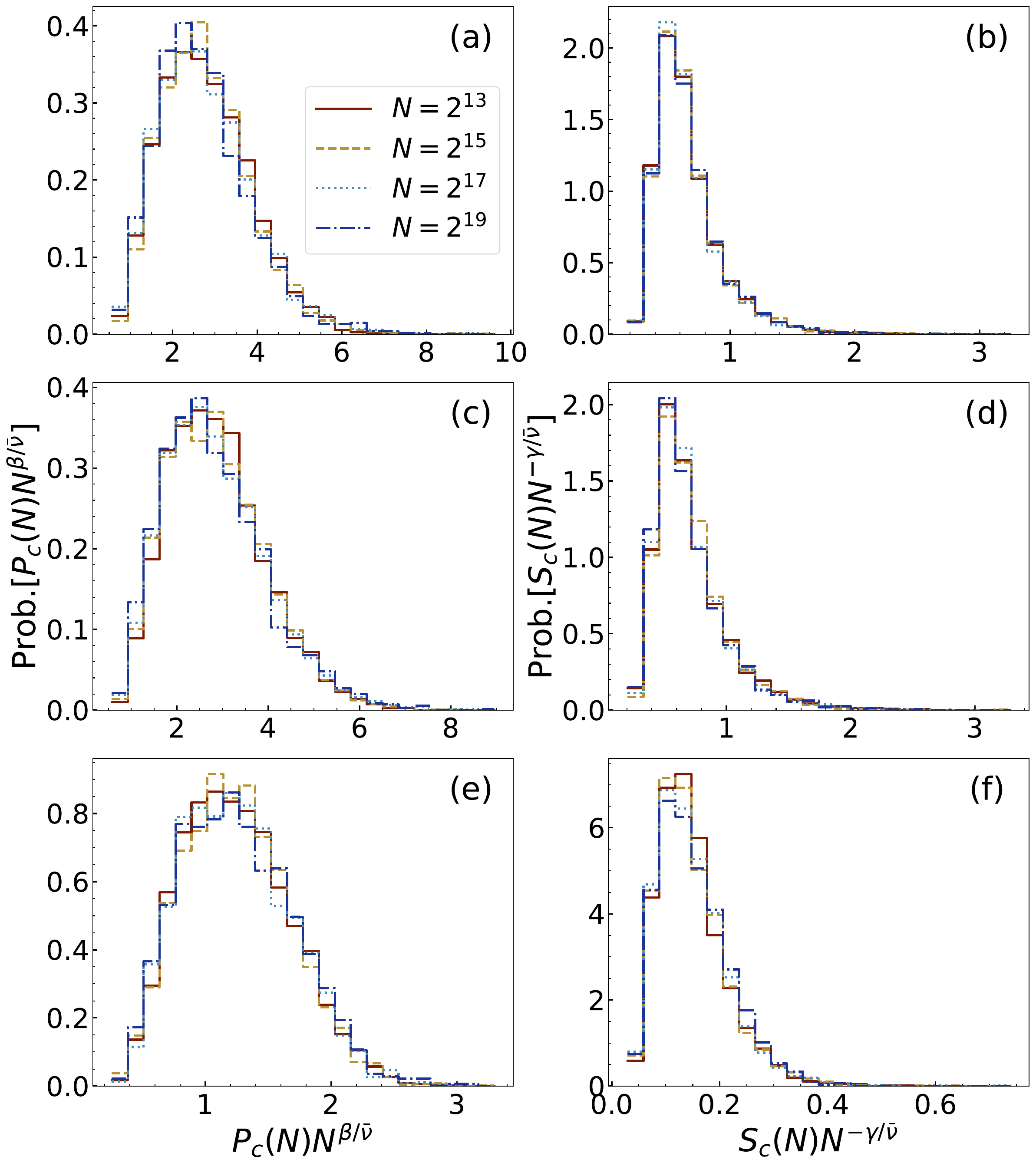}
    \caption{
    \textbf{Data collapse of probability density functions.} We plot the probability density function of the rescaled critical observables (a) $P_c(N)N^{\beta/\bar\nu}$ and (b) $S_c(N)N^{-\gamma/\bar\nu}$ for the SPP model with $C=2$ on SFNs with $\lambda=2.7$. We also plot similar results for (c-d) $C=3$, and (e-f) $C=N$. We use the best estimates of the critical exponent ratios $\beta/\bar\nu$ and $\gamma/\bar\nu$ reported in Table~\ref{tab:summary}. Note that we use 30 bins to plot the data.
    }
    \label{fig2}
\end{figure}

To verify our estimation, in Fig.~\ref{fig2}, we show the data-collapse plots of the probability density function of rescaled critical observables using the estimated critical exponents $\beta/\bar\nu$ and $\gamma/\bar\nu$ for SFNs with $\lambda=2.7$ and various values of $C$. We could obtain a satisfactory data collapse across different system sizes, which implies that our estimations of $\beta/\bar\nu$ and $\gamma/\bar\nu$ are fair. We could obtain clear data-collapse plots for different values of $\lambda$ as well (see Figs.~\ref{sm-fig:pdf_collapse_lambda2.1}, ~\ref{sm-fig:pdf_collapse_lambda3.5}, and ~\ref{sm-fig:pdf_collapse_lambda4.5} of Ref.~\cite{supplementary}).

\begin{table*}[t]
\renewcommand{\arraystretch}{1.3} 
\begin{ruledtabular}
\begin{tabular}{lllllll}
$\lambda$ & $C$ & $\beta/\bar{\nu}$ & $\gamma/\bar{\nu}$ & $r_c$ & $1/\bar{\nu}_1$ & $1/\bar{\nu}_2$ \\
\hline
{$2.1$} & $1$        & $0.54(1)$ & $-0.05(1)$ & $1.000(1)$ & $0.39(1)$ & $0.54(1)$ \\
                     & 2        & 0.32(1) & 0.31(1)    & 0.953(1) & 0.26(1) & 0.44(1) \\
                     & 3        & 0.30(1) & 0.33(1)    & 0.948(1) & 0.25(1) & 0.44(1) \\
                     & $\infty$ & 0.23(1) & 0.50(1)    & 0.957(1) & 0.21(1) & 0.48(1) \\
\hline
{2.7} & 1        & 0.51(1) & $-0.02(1)$ & 0.990(1) & 0.27(1) & 0.33(1) \\
                     & 2        & 0.31(1) & 0.33(1)    & 0.862(1) & 0.40(1) & 0.37(1) \\
                     & 3        & 0.30(1) & 0.35(1)    & 0.853(1) & 0.37(1) & 0.37(1) \\
                     & $\infty$ & 0.21(1) & 0.54(1)    & 0.843(1) & 0.49(1) & 0.48(1) \\
\hline
{3.5} & 1        & 0.40(1) & 0.16(1)    & 0.890(1) & 0.29(1) & 0.28(1) \\
                     & 2        & 0.30(1) & 0.33(1)    & 0.791(1) & 0.48(1) & 0.36(1) \\
                     & 3        & 0.30(1) & 0.35(1)    & 0.780(1) & 0.48(1) & 0.35(1) \\
                     & $\infty$ & 0.21(1) & 0.54(1)    & 0.774(1) & 0.51(1) & 0.51(1) \\
\hline
{4.5} & 1        & 0.35(1) & 0.26(1)    & 0.797(1) & 0.37(1) & 0.31(1) \\
                     & 2        & 0.31(1) & 0.34(1)    & 0.747(1) & 0.48(1) & 0.35(1) \\
                     & 3        & 0.30(1) & 0.35(1)    & 0.737(1) & 0.46(1) & 0.36(1) \\
                     & $\infty$ & 0.22(1) & 0.53(1)    & 0.736(1) & 0.57(1) & 0.52(1) \\
\end{tabular}
\end{ruledtabular}
\caption{\textbf{Critical points and exponents.} Reported estimates are valid for shortest-path percolation processes with maximum budget $C$ on scale-free networks with degree exponent $\lambda$.}
\label{tab:summary}
\end{table*}

\subsection{Shortest-path-percolation process homogenizes the network drastically}\label{subsec:spp_homogenization}

From the findings in Sec.~\ref{subsec:estimate_beta_gamma}, we conjecture and argue that the SPP model with $C>1$ drastically homogenizes the heterogeneous network structure before the SPP transition takes place. 
Betweenness centrality (BC), which quantifies how often a node appears on the shortest paths between other nodes, offers a useful lens to understand this homogenization~\cite{freeman1977set, newman2001scientific}. 
For SFNs, it has been well established that there is a power-law relation between the degree $k$ and the BC $g$ of a node, i.e., 

\begin{equation}
    g\sim k^{\frac{\lambda-1}{\eta-1}},
    \label{eq:betweenness_scaling}
\end{equation}
where $\lambda$ is the degree exponent and $\eta$ is the exponent related to the distribution of $g$, $P(g)\sim g^{-\eta}$, and $(\lambda-1)/(\eta-1)>0$~\cite{goh2001universal, barthelemy2004betweenness}. 
This relation implies that hubs mainly contribute to the formation of shortest paths in the network. As a result, they have a high chance of decreasing the number of their edges in the SPP process, as removed edges are selected based on the fact that they belong to the shortest path connecting two randomly chosen nodes.

One of the strongest indicators of the network's heterogeneity or scale-freeness is the scaling of the maximum degree. Thus, to validate our argument, we compare the maximum degree of the initial network, $k_{\max}(N) = k_{\max}(N,\ r=0)$ and the maximum degree of the network at some point $r=r^\dagger$ before the percolation transition takes place, $k_{\max}^{\dagger}(N)=k_{\max}(N,\ r=r^\dagger)$. Similar to Eq.~(\ref{eq:pseudo_scaling}), we assume
\begin{equation}
    \left\langle \frac{k^{\dagger}_{\max} (N)}{k_{\max} (N)} \right\rangle \equiv \phi(N)= \phi_{o}+a N^{-\delta},
    \label{eq:k_ratio_scaling}
\end{equation}
where $\left\langle \cdot \right\rangle$ denotes the ensemble average across different realizations, $\phi_o$ is the asymptotic limit of the ratio, $\delta$ is the scaling exponent, and $a$ is a constant. Equation~(\ref{eq:k_ratio_scaling}) implies that if a network still holds its scale-freeness at $r=r^\dagger$, we expect $k_{\max}^{\dagger}$ will scale with the size of the network $N$ in the same way as $k_{\max}$ does. In this case, we expect that $\phi_{o}$ is non-zero. On the other hand, we expect to obtain $\phi_o=0$ if the SPP process drastically homogenizes the network.

We sample the observables at $r=r^\dagger$ using two methods: we collect the observables (i) at a fixed value of $r^\dagger=q$ for each realization or (ii) at $r^\dagger=qr_c^{*}(N)$ for each realization, where $q\in[0, 1]$. Here, $q$ is a tunable parameter to decide on how far from the transition point we would like to sample the observables. We choose $q=0.7,\ 0.8,$ and $0.9$ to ensure that we are sampling the observables at the supercritical regime. For convenience, we add a subscript to differentiate the quantities in Eq.~(\ref{eq:k_ratio_scaling}) using different ensembles of observables, e.g., $\phi_{1}(N)$ and $\phi_{2}(N)$.

\begin{figure}[!ht]
    \centering
    \includegraphics[width=\linewidth]{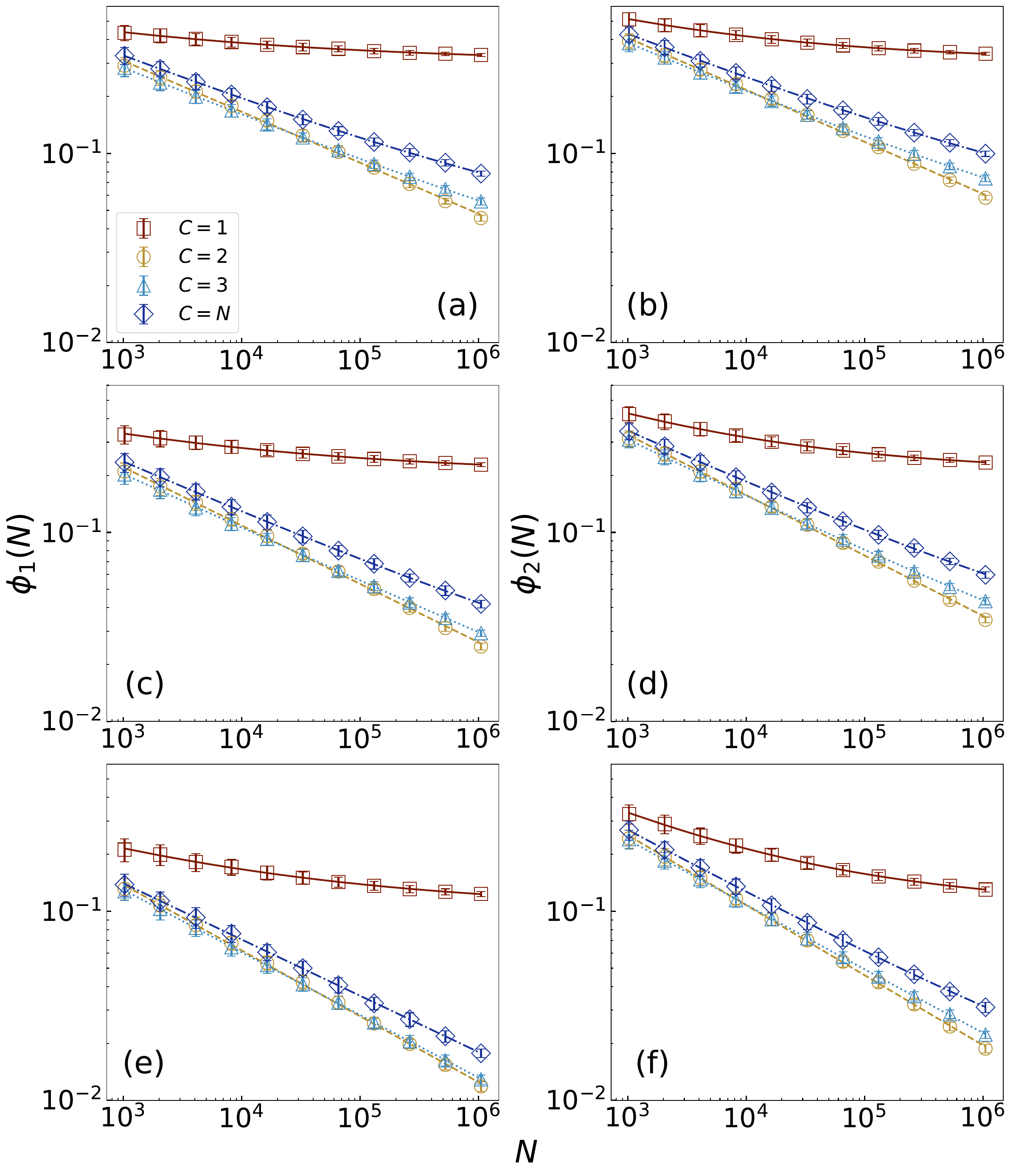}
    \caption{
    \textbf{Scaling of the $k_{\text{max}}$ ratios.}
    Plot of $\phi_{1}(N)$ as a function of $N$ for different values of $C$. Results are valid for SFNs with $\lambda=2.1$ using (a) $q=0.7$ (c) $q=0.8$, and (e) $q=0.9$. We also plot $\phi_{2}(N)$ as a function of $N$ for different values of $C$ using (b) $q=0.7$ (d) $q=0.8$, and (f) $q=0.9$. Plateau values and exponents are reported in Table~\ref{tab:k_max_slope}.
    Symbols are the average values and error bars are the standard deviations. 
    We report the number of realizations used for the scaling analysis in Table~\ref{tab:k_ratio_realization} of Ref.~\cite{supplementary}. 
    }
    \label{fig3}
\end{figure}

We test our hypothesis on SFNs with $\lambda=2.1$, which is the most heterogeneous network considered in our study.
In Fig.~\ref{fig3} (a) we use $q=0.7$ to plot $\phi_{1}(N)$ as a function of $N$ for different values of $C$. We could obtain a very clear power-law decay across different values of $q$ and $C$, while consistently observing nonzero values of $\phi_o$ when $C=1$.
The same explanation holds for results valid for $\phi_{2}$ as well. When we increase the value of $q$, for $C=1$, we find that $\phi_{o,1}$ and $\phi_{o,2}$ decrease since the observables are sampled closer to the transition point and the network becomes sparser. Also, the exponents $\delta_{1}$ and $\delta_{2}$ increase when we increase $q$, regardless of $C$. These findings support that the SPP process with $C>1$ can drastically homogenize the heterogeneous structures in SFNs before the percolation transition happens. Detailed estimations of $\phi_{o,1}$, $\phi_{o,2}$, $\delta_1$, and $\delta_2$ can be found in Table~\ref{tab:k_max_slope}. In addition, we perform a complementary analysis and present further evidence showing that the SPP process effectively homogenizes SFNs (see Appendix \ref{appendix:complementary_analysis} for details).

\renewcommand{\arraystretch}{1.3}
\begin{table*}[t]
\centering
\begin{tabular*}{\textwidth}{@{\extracolsep{\fill}}llllll}
\hline\hline
$q$ & $C$ & $\phi_{o,1}$ & $\delta_1$ & $\phi_{o,2}$ & $\delta_2$ \\
\hline
0.7 & 1 & 0.301(1) & 0.217(2) & 0.305(1) & 0.273(1) \\
    & 2 & 0.000(1) & 0.270(4) & 0.000(1) & 0.275(3) \\
    & 3 & 0.011(1) & 0.260(1) & 0.019(1) & 0.273(1) \\
    & $N$ & 0.026(1) & 0.252(1) & 0.032(1) & 0.253(1) \\
\hline
0.8 & 1 & 0.201(1) & 0.228(2) & 0.206(1) & 0.293(2) \\
    & 2 & 0.000(1) & 0.309(4) & 0.000(1) & 0.320(3) \\
    & 3 & 0.002(1) & 0.288(1) & 0.008(1) & 0.309(1) \\
    & $N$ & 0.009(1) & 0.278(1) & 0.015(1) & 0.286(1) \\
\hline
0.9 & 1 & 0.103(1) & 0.247(2) & 0.106(1) & 0.321(1) \\
    & 2 & 0.000(1) & 0.348(3) & 0.000(1) & 0.371(2) \\
    & 3 & 0.000(1) & 0.332(1) & 0.001(1) & 0.346(1) \\
    & $N$ & 0.000(1) & 0.298(1) & 0.006(1) & 0.338(1) \\
\hline\hline
\end{tabular*}
\caption{\textbf{Scaling of the $k_{\text{max}}$ ratios.} 
We report the estimated variables $\phi_{o,1}$, $\phi_{o,2}$, $\delta_1$, and $\delta_2$ using Eq.~(\ref{eq:k_ratio_scaling}).}
\label{tab:k_max_slope}
\end{table*}

\subsection{Estimation of $1/\bar\nu_1$ and $1/\bar\nu_{2}$}\label{subsection3}
So far, our findings reveal that the ratios of critical exponents $\beta/\bar\nu$ and $\gamma/\bar\nu$, related to the structure of the network at criticality, depend on whether $C$ is finite or infinite, regardless of the degree exponent $\lambda$. However, we find that the critical exponents $1/\bar\nu_{1}$ and $1/\bar\nu_{2}$, related to the convergence and fluctuation of pseudocritical points, vary depending on $\lambda$ and $C$ (see Sec.~\ref{subsec:method-fss} for details). 

\begin{figure}
    \centering
    \includegraphics[width=\linewidth]{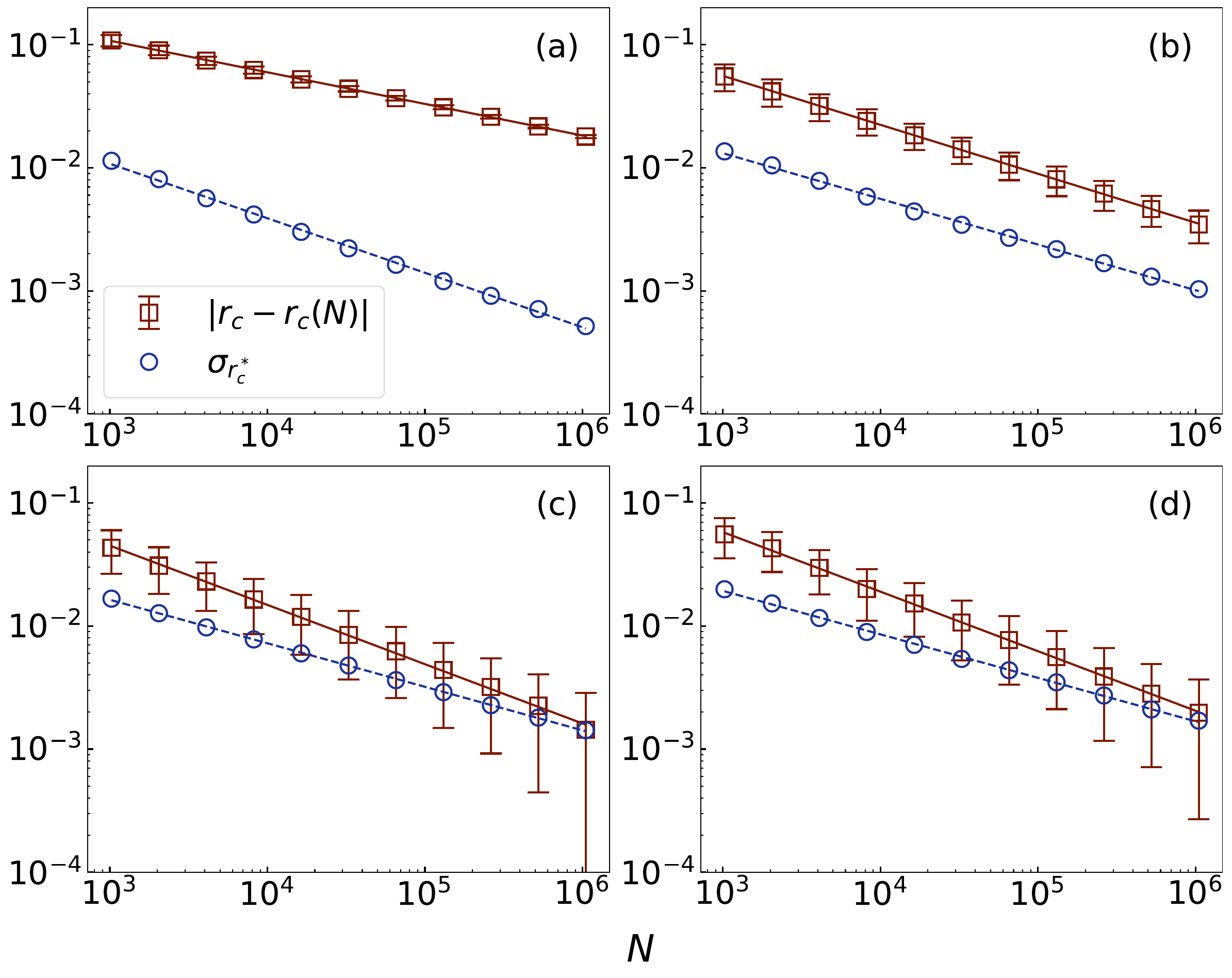}
    \caption{\textbf{Estimation of $1/\bar\nu$.} We plot $|r_c - r_c(N)|$ (square symbol) and $\sigma_{r^*_c}$ (circle symbol) as functions of $N$. The slope in the double-log scale corresponds to $1/\bar\nu_{1}$ (solid line) and $1/\bar\nu_{2}$ (dashed line), respectively. Results are valid for SPP with $C=2$ on SFNs with (a) $\lambda=2.1$, (b) $\lambda=2.7$, (c) $\lambda=3.5$, and (d) $\lambda=4.5$. 
    Symbols are the average values and error bars are the standard deviations.
    Note that error bars for $\sigma_{r^*_c}$ do not exist so they are omitted in the figure. Estimated critical exponents are reported in Table~\ref{tab:summary}.}
    \label{fig4}
\end{figure}

First, we report the case when $1/\bar\nu=1/\bar\nu_{1}= 1/\bar\nu_{2}$. This is the most common case that has been reported for ordinary percolation on ERNs, random-regular graphs, and low-dimensional lattices with periodic boundary conditions \cite{fan2020universal, li2024explosive}. 
On SFNs with $\lambda=2.7$, we get $1/\bar\nu\approx 0.37$ for finite $C$ and $1/\bar\nu\approx0.5$ for infinite $C$, respectively.
In Fig.~\ref{fig4} (b), we plot $|r_c-r_c(N)|$ and $\sigma_{r^*_c}$ as functions of $N$, respectively, where results are valid for SFNs with $\lambda=2.7$ and SPP with $C=2$ as an example. 

Second, when $\lambda>3$, we find that $1/\bar\nu_{1}>1/\bar\nu_{2}$ for finite $C >1$, where we get $1/\bar\nu_{1}\approx 0.5$ and $1/\bar\nu_{2}\approx 0.35$ [See Figs.~\ref{fig4} (c) and (d) for results for $C=2$]. Such a relation between $1/\bar\nu_{1}$ and $1/\bar\nu_{2}$ implies that the convergence of $r_c(N)$ towards $r_c$ is faster than the fluctuation of $r_c(N)$ decreasing as $N$ increases. The explosive percolation is a typical case where this type of discrepancy has been reported~\cite{fan2020universal, li2023explosive, li2024explosive}. For infinite $C$, $r_c(N)$ initially increases with $N$, but quickly saturates and forms a plateau, remaining nearly constant over a wide range of $N$ (see Fig.~\ref{sm-fig:pseudo_rc} of Ref.~\cite{supplementary}).
Due to this plateau, the quality of the least-squares fitting of Eqs.~(\ref{eq:pseudo_scaling}) and (\ref{eq:critical_deviation}) is compromised. Still, we find that the values of the plateau are within the error bounds of the estimated $r_c$.Two possible scenarios explain this: One possible explanation is that the size of the network considered is not large enough, and the plateau manifests finite-size effects. Another case is when the system size is already large enough; thus, the plateau value represents the true critical point at the infinite-$N$ limit. We confirm that our observation corresponds to the latter, which is supported by a clear data collapse using the obtained $r_c$.

Finally, on SFNs with $\lambda=2.1$, we find that $1/\bar\nu_{1}<1/\bar\nu_{2}$ both for infinite- and finite-$C$. For finite $C$, we get $1/\bar\nu_{1}\approx 0.25$ and $1/\bar\nu_{2}\approx 0.44$ while for infinite $C$, $1/\bar\nu_{1}\approx 0.21$ and $1/\bar\nu_{2}\approx 0.48$ [see Fig.~\ref{fig4} (a) for results for $C=2$]. This is the case reported for bond percolation on high-dimensional lattices above the upper-critical dimension with free-boundary condition \cite{li2024crossover}. Here, $1/\bar\nu_{1} < 1/\bar\nu_{2}$ implies that the single-instance pseudocritical points $r_{c}^{*}(N)$ fluctuate within a narrow window, while the pseudocritical point $r_c(N)$ is far away from $r_c$ at the infinite-$N$ limit. Results for other values of $C$ can be found in Figs.~\ref{sm-fig:nu_estimation_C1}, ~\ref{sm-fig:nu_estimation_C3}, and ~\ref{sm-fig:nu_estimation_CN} of Ref.~\cite{supplementary}.

\subsection{Data collapse of observables}
\begin{figure}
    \centering
    \includegraphics[width=\linewidth]{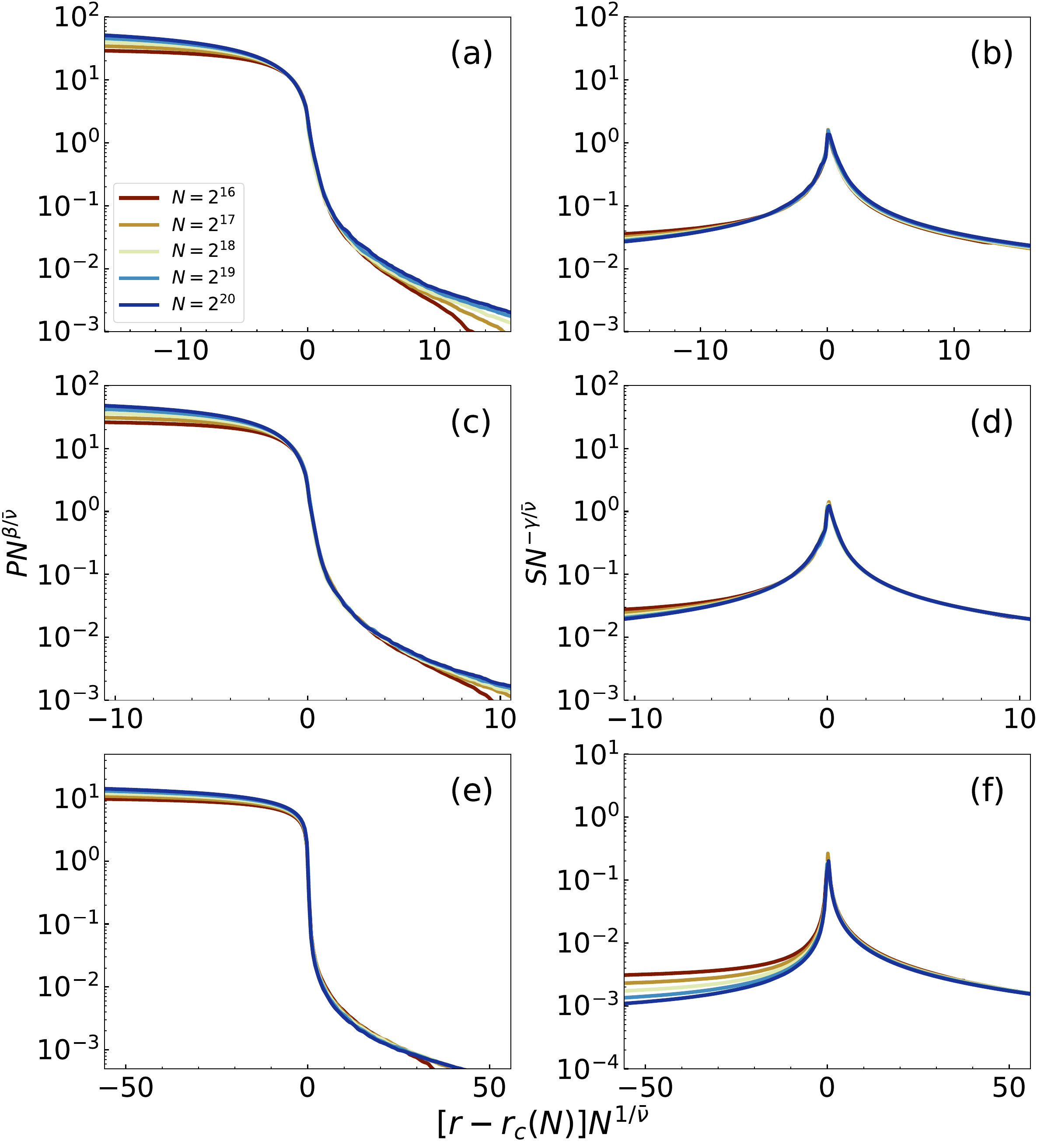}
    \caption{
    \textbf{Data collapse of $P$ and $S$.} We plot the rescaled observable $PN^{\beta/\bar\nu}$ as a function of $[r-r_c(N)]N^{1/\bar\nu}$ for different values of network size $N$. Results are valid for SFNs with $\lambda=2.7$ and $k_{\min}=4$ and SPP model with (a) $C=2$, (c) $C=3$, and (e) $C=N$. Similarly, we plot the rescaled observable $SN^{-\gamma/\bar\nu}$ as a function of $[r-r_c(N)]N^{1/\bar\nu}$ for (b) $C=2$, (d) $C=3$, and (f) $C=N$. Note that for SFNs with $\lambda=2.7$, we omit the subscript in $\bar\nu_{2}$ and used $1/\bar\nu$ since $1/\bar\nu_{1}\approx 1/\bar\nu_{2}$. The critical exponent ratios used in the figure are reported in Table~\ref{tab:summary}. The range of abscissa is adjusted to display 200 bins of the curve for $N=2^{20}$. We report the number of bins used for each network size in Table~\ref{tab:binning} of Ref.~\cite{supplementary} for clarity.
    }
    \label{fig5}
\end{figure}

According to the FSS theory, once we estimate the critical quantities in a fair manner, we can collapse the data across different system sizes by rescaling the data using these critical quantities. To verify our FSS analysis, we display the data-collapse plots in Fig.~\ref{fig5} for SFNs with $\lambda=2.7$ and $k_{\min}=4$, considering the event-based ensemble. By using the estimated critical exponents, we could fairly collapse the data across different orders of magnitude of network size $N$ for the SPP model with different values of $C$. We could also obtain a clear data collapse considering the conventional ensemble, which supports that our estimation of $r_c$ is fair, see Figs.~\ref{sm-fig:collapse_conventional_sfn2.7_C2}, ~\ref{sm-fig:collapse_conventional_sfn2.7_C3}, and ~\ref{sm-fig:collapse_conventional_sfn2.7_CN} of Ref. ~\cite{supplementary}. 
Moreover, the critical exponents imply the same universality class as of ordinary percolation on ERNs, i.e., $\beta/\bar\nu=\gamma/\bar\nu=1/\bar\nu=1/3$.

For SFNs with $\lambda=3.5$ and $\lambda=4.5$, for the SPP model with $C>1$, we find that using $1/\bar\nu_{2}$ instead of $1/\bar\nu_{1}$ gives better data collapse, regardless of considering the event-based ensemble (see Figs.~\ref{sm-fig:collapse_sfn3.5_C2}, ~\ref{sm-fig:collapse_sfn3.5_C3}, ~\ref{sm-fig:collapse_sfn3.5_CN}, ~\ref{sm-fig:collapse_sfn4.5_C2}, ~\ref{sm-fig:collapse_sfn4.5_C3}, and ~\ref{sm-fig:collapse_sfn4.5_CN} of Ref.~\cite{supplementary}) and the conventional ensemble (Figs.~\ref{sm-fig:collapse_conventional_sfn3.5_C2}, ~\ref{sm-fig:collapse_conventional_sfn3.5_C3}, ~\ref{sm-fig:collapse_conventional_sfn3.5_C-1} ~\ref{sm-fig:collapse_conventional_sfn4.5_C2}, ~\ref{sm-fig:collapse_conventional_sfn4.5_C3}, and ~\ref{sm-fig:collapse_conventional_sfn4.5_C-1} of Ref.~\cite{supplementary}). Especially for finite $C$, $1/\bar\nu_{2}\approx 1/3$ suggests the same universality class of ordinary percolation on ERNs. 
Note that for infinite $C$, we could achieve a fair data collapse by using $r_c$, where the plateau values are within the error bounds (Figs.~\ref{sm-fig:collapse_conventional_sfn3.5_C-1} and ~\ref{sm-fig:collapse_conventional_sfn4.5_C-1} of Ref.~\cite{supplementary}). 
This suggests that the plateau values have already converged close enough to the true critical point $r_c$.
For the case of $C=1$, which corresponds to the ordinary bond percolation, where $1/\bar\nu_{1} \approx 1/\bar\nu_{2}$, we could obtain a fair data collapse under the event-based ensemble as well as the conventional ensemble, see Figs.~\ref{sm-fig:collapse_sfn3.5_C1}, ~\ref{sm-fig:collapse_sfn4.5_C1}, ~\ref{sm-fig:collapse_conventional_sfn3.5_C1}, and ~\ref{sm-fig:collapse_conventional_sfn4.5_C1} of Ref.~\cite{supplementary}.

For SFNs with $\lambda=2.1$, we could obtain a fair data collapse under the event-based ensemble when using $1/\bar\nu_{2}$ for finite $C$ (see Figs.~\ref{sm-fig:collapse_sfn2.1_C2} and ~\ref{sm-fig:collapse_sfn2.1_C3} of Ref.~\cite{supplementary}). For infinite $C$, the quality of data collapse is compromised (see Fig.~\ref{sm-fig:collapse_sfn2.1_CN} of Ref.~\cite{supplementary}), but still the data collapse is decent in a narrow range. Under the conventional ensemble, however, we could not obtain a proper data collapse, using both $1/\bar\nu_{1}$ and $1/\bar\nu_{2}$. In addition, the location where the abscissa value corresponds to zero severely deviates from the peak position of $SN^{-\gamma/\bar\nu}$, which corresponds to the pseudocritical points (see Figs.~\ref{sm-fig:collapse_conventional_sfn2.1_C2}, ~\ref{sm-fig:collapse_conventional_sfn2.1_C3}, and ~\ref{sm-fig:collapse_conventional_sfn2.1_CN} of Ref.~\cite{supplementary}). As suggested from the relation $1/\bar\nu_{1} < 1/\bar\nu_{2}$, we suppose that the pseudocritical points are still far from the true critical point and a proper FSS behavior is not expected in this regime~\cite{li2024crossover}.

\section{Summary and Discussion\label{sec:discussion}}
In this work, we systematically investigated the phase transition of SPP on SFNs, considering various degree exponents $\lambda$ and maximum budget $C$. By means of extensive Monte Carlo simulations and FSS analysis, we observed rich behaviors in the SPP transition on SFNs.
For $C=1$, which corresponds to the ordinary bond percolation, we obtained results consistent with previous studies~\cite{cohen2002percolation, cirigliano2024scaling}. In particular, for $\lambda \leq 3$, the percolation phase transition occurs when all edges are removed from the network; also, for $\lambda \leq 4$, the critical exponents depend on $\lambda$. These anomalous, network-dependent properties are washed out by the SPP process. As long as $C>1$, the transition occurs when a critical fraction $r_c < 1$ of the network edges are removed. For finite $C$ values,  we find that the critical exponents describing the structure of the network at criticality are identical to the mean-field exponents on ERNs, i.e., $\beta/\bar\nu = \gamma/\bar\nu=1/3$, regardless of the value of $\lambda$. 
For the infinite-$C$ case, we obtain the critical exponents $\beta/\bar\nu\approx0.2$ and $\gamma/\bar\nu\approx0.6$, which have been observed on ERNs~\cite{kim2024shortest, meng2025path}, regardless of the value of $\lambda$. These findings lead to the hypothesis that the SPP with $C>1$ drastically homogenizes the heterogeneous structure of SFNs before the SPP transition takes place. We confirmed this hypothesis by analyzing the scaling of the ratio of the initial maximum degree and that before the percolating regime. 

The critical exponents $1/\bar\nu_{1}$ and $1/\bar\nu_{2}$, related to the convergence and fluctuation of the pseudocritical points, vary depending on $\lambda$ and $C$. To be specific, for finite $C>1$, we find that the relation between $1/\bar\nu_{1}$ and $1/\bar\nu_{2}$ has a dependency on $\lambda$. When $\lambda=2.1$, we observe $1/\bar\nu_{1} < 1/\bar\nu_{2}$, which has been reported in the case of bond percolation on lattice structure above the upper critical dimension with free boundary condition~\cite{li2024crossover}. However, for $\lambda=2.7$, we observe $1/\bar\nu_{1} \approx 1/\bar\nu_{2}$, in which most of the ordinary percolation belongs to. Finally, we observe $1/\bar\nu_{1} > 1/\bar\nu_{2}$ for SFNs with $\lambda=3.5$ and $\lambda=4.5$, which is the case of the explosive percolation. Unfortunately, there are only a few studies which report and explain these two exponents~\cite{fan2020universal, li2023explosive, kim2024shortest, li2024crossover}; thus, further investigations on various network structures are needed to better understand such a discrepancy. A recent study also reports rich and non-trivial estimation of these exponents on SFNs depending on the level of the degree cutoff and the degree heterogeneity~\cite{zhao2025finite}.

Our numerical findings reveal that the SPP process homogenizes the heterogeneous network structure before the percolation transition takes place.
However, an analytical description of our numerical findings is still missing. One of the biggest obstacles in formalizing an analytical framework is the topological correlation between edges forming the shortest paths. Unfortunately, typical frameworks such as the one based on generating functions~\cite{newman2001random, callaway2000network} and the one based on message passing~\cite{newman2023message} assume independence of edges. This limitation also prevents us from fully understanding the boundary of finite and infinite $C$. Possibly, some crossover behavior could occur as $C$ increases beyond some $N$-dependent threshold, which separates the finite-$C$ and infinite-$C$ regimes. In our previous work~\cite{kim2024shortest}, we numerically verified that the cases $C=N^{1/3}$ and $C=\ln N$ belong to the infinite-$C$ universality class. In the future, it would be a promising direction to develop an analytical framework considering ``path-based'' and ``long-range'' connectivity in percolation.

Finally, we would like to highlight the broad applicability of the SPP framework to various complex systems represented as supply-and-demand networks. These networks include, but are not limited to, transportation networks, communication networks, and supply chain networks, which incorporate resource consumption and depletion. 
For instance, the SPP has been extended to the minimum-cost percolation and applied to understand the consumption and exhaustion of resources in the U.S. air transportation system while serving the demands~\cite{kim2025modeling}. 
In addition, a similar model to the SPP model has been introduced to study quantum communication networks~\cite{meng2025path}.
Since various supply-and-demand network systems are characterized by ``scale-free'' structure with the existence of hubs, we believe the rich and complex behavior observed in this work can be found in these real-world systems as well.

\begin{acknowledgments}
We acknowledge the support of the AccelNet-MultiNet program, a project of the National Science
Foundation (Awards No. 1927425 and No. 1927418).
This work received partial support
from the Air Force Office of Scientific Research (Grant No.
FA9550-24-1-0039). 
CC acknowledges the PRIN Project No. 20223W2JKJ ``WECARE'', CUP B53D23003880006, financed by the Italian Ministry of University and Research (MUR), Piano Nazionale Di Ripresa e Resilienza (PNRR), Missione 4 ``Istruzione e Ricerca'' - Componente C2 Investimento 1.1, funded by the European Union - NextGenerationEU.
The funders had no role in study design, data collection and analysis, the decision to publish, or any opinions, findings, conclusions, or recommendations expressed in the manuscript.
\end{acknowledgments}

\section*{Data Availability}
The source code that supports the findings of this article is openly available~\cite{SPP_Source},

\appendix
\section{Complementary analysis for Sec.~\ref{subsec:spp_homogenization}}\label{appendix:complementary_analysis}
To provide additional evidence that the SPP process effectively homogenizes the heterogeneous network structure before the percolation transition takes place, we perform the following analysis. 
Consider the following quantity:
\begin{equation}
    \rho(N,\ r) = \frac{\langle k^2 \rangle}{\langle k \rangle^2},
\end{equation}
where $N$ is the size of the network, $\langle k \rangle$ is the average degree after removing a fraction $r$ of the edges, and $\langle k^{2} \rangle$ is the average of the squared degrees under the same removal. Note that we omitted the $r$-dependence of $k$ to avoid overcrowding the equation. $\rho(N,\ r)$ effectively captures the degree heterogeneity of the network.

Now, we compare $\rho(N,\ r)$ of the initial network, $\rho^o(N)=\rho(N,\ r=0)$, with the value obtained after removing a fraction $r=r^\dagger$ of edges before the percolation transition takes place, $\rho^\dagger(N)=\rho(N,\ r=r^\dagger)$. 
Then, we define the quantity $\Omega(N)$ and assume the functional form as follows:
\begin{equation}
    \bigg\langle \frac{\rho^\dagger(N)}{\rho^o(N)} \bigg\rangle\equiv\Omega(N)=\Omega_{o}+aN^{-\xi},
    \label{eq:variance_scaling}
\end{equation}
similar to Eq.~(\ref{eq:k_ratio_scaling}). 
To properly complement the analysis, we choose $r=r^\dagger$ (i) at a fixed value of $r^\dagger=q$ or (ii) $r^\dagger=qr_{c}^{*}(N)$ for each realization, using $q=0.7,\ 0.8,$ and $0.9$. 
We also add a subscript to distinguish the quantities in Eq.~(\ref{eq:variance_scaling}), i.e., $\Omega_1(N)$ and $\Omega_2(N)$. 
Note that nodes with zero degree are excluded from this computation.

\begin{figure}[!ht]
    \centering
    \includegraphics[width=\linewidth]{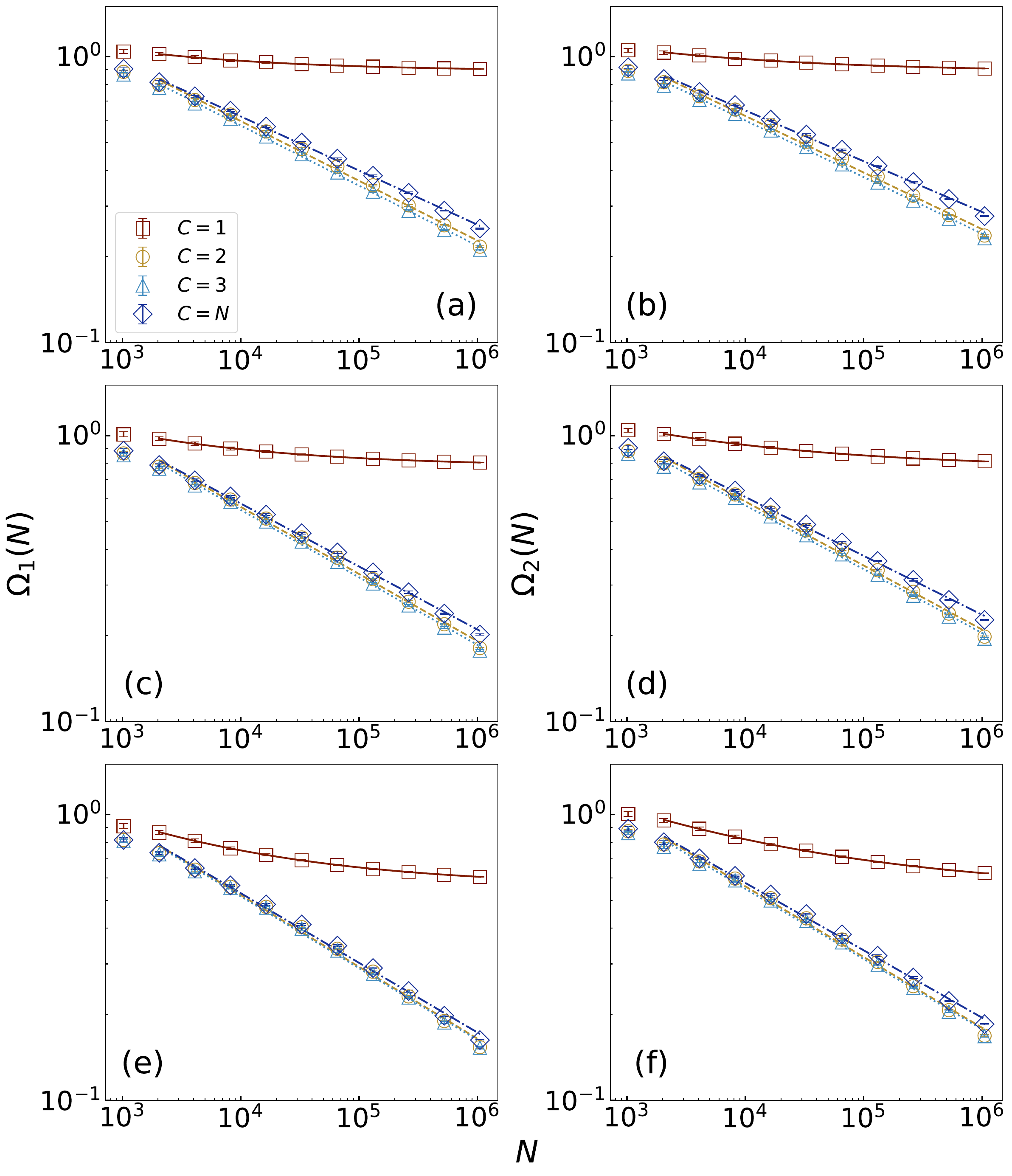}
    \caption{
    \textbf{Scaling of the $\rho$ ratios.}
    Plot of $\Omega_{1}(N)$ as a function of $N$ for different values of $C$. Results are valid for SFNs with $\lambda=2.1$ using (a) $q=0.7$ (c) $q=0.8$, and (e) $q=0.9$. We also plot $\Omega_{2}(N)$ as a function of $N$ for different values of $C$ using (b) $q=0.7$ (d) $q=0.8$, and (f) $q=0.9$.
     Plateau values and exponents are reported in Table~\ref{tab:k_variance_slope} of Ref.~\cite{supplementary}.
    Symbols are the average values and error bars are the standard deviations. 
    For the scaling analysis, we use 50 realizations for each data point.
    We also drop the data point corresponding to the smallest network size when performing the scaling analysis.
    }
    \label{fig:rho_scaling}
\end{figure}

We perform the analysis on SFNs with $\lambda=2.1$. In Fig.~\ref{fig:rho_scaling}, we plot $\Omega(N)$ as a function of $N$ for different values of $q$ and $C$. We also report the estimated variables $\Omega_{o,1}$, $\Omega_{o,2}$, $\xi_1$, and $\xi_2$ in Table~\ref{tab:k_variance_slope} of Ref.~\cite{supplementary}. These results clearly complement the analysis in Fig.~\ref{fig:rho_scaling}, indicating that the SPP process with $C>1$ effectively homogenizes the heterogeneous structure of SFNs.

\bibliographystyle{apsrev4-2} 

\pagebreak
\clearpage
\widetext
\begin{center}
\textbf{\Large Supplemental Materials: Shortest-path percolation on scale-free networks}

\large{Minsuk Kim, Lorenzo Cirigliano, Claudio Castellano, Hanlin Sun, Robert Jankowski, Anna Poggialini, and Filippo Radicchi}
\end{center}

\onecolumngrid


\setcounter{section}{0}
\setcounter{equation}{0}
\setcounter{figure}{0}
\setcounter{table}{0}
\setcounter{page}{1}
\makeatletter
\renewcommand{\theequation}{S\arabic{equation}}
\renewcommand{\thefigure}{S\arabic{figure}}
\renewcommand{\thetable}{S\arabic{table}}
\renewcommand{\bibnumfmt}[1]{[S#1]}
\renewcommand{\citenumfont}[1]{S#1}
\renewcommand{\thesection}{S\arabic{section}}

\section*{Supplementary Figures}
In this section, we provide detailed numerical results that are not displayed in the main paper.

\begin{figure}[h]
    \centering
    \includegraphics[width=0.6\linewidth]{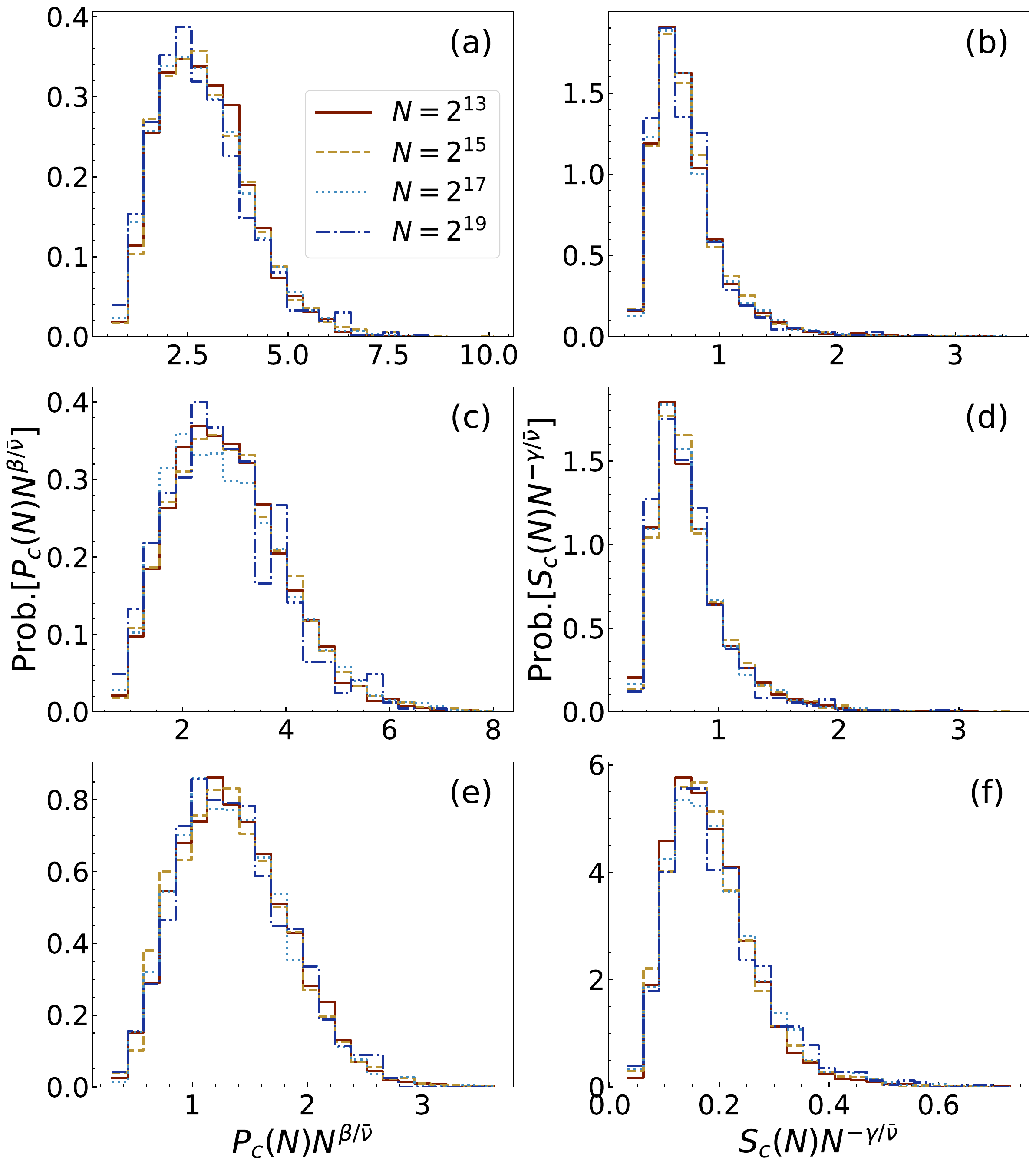}
    \caption{\textbf{Data Collapse of probability distributions of critical observables.} Similar to Fig.~\ref{fig2} in the main, but results are valid for SFNs with $\lambda=2.1$.}
    \label{sm-fig:pdf_collapse_lambda2.1}
\end{figure}

\begin{figure}[h]
    \centering
    \includegraphics[width=0.6\linewidth]{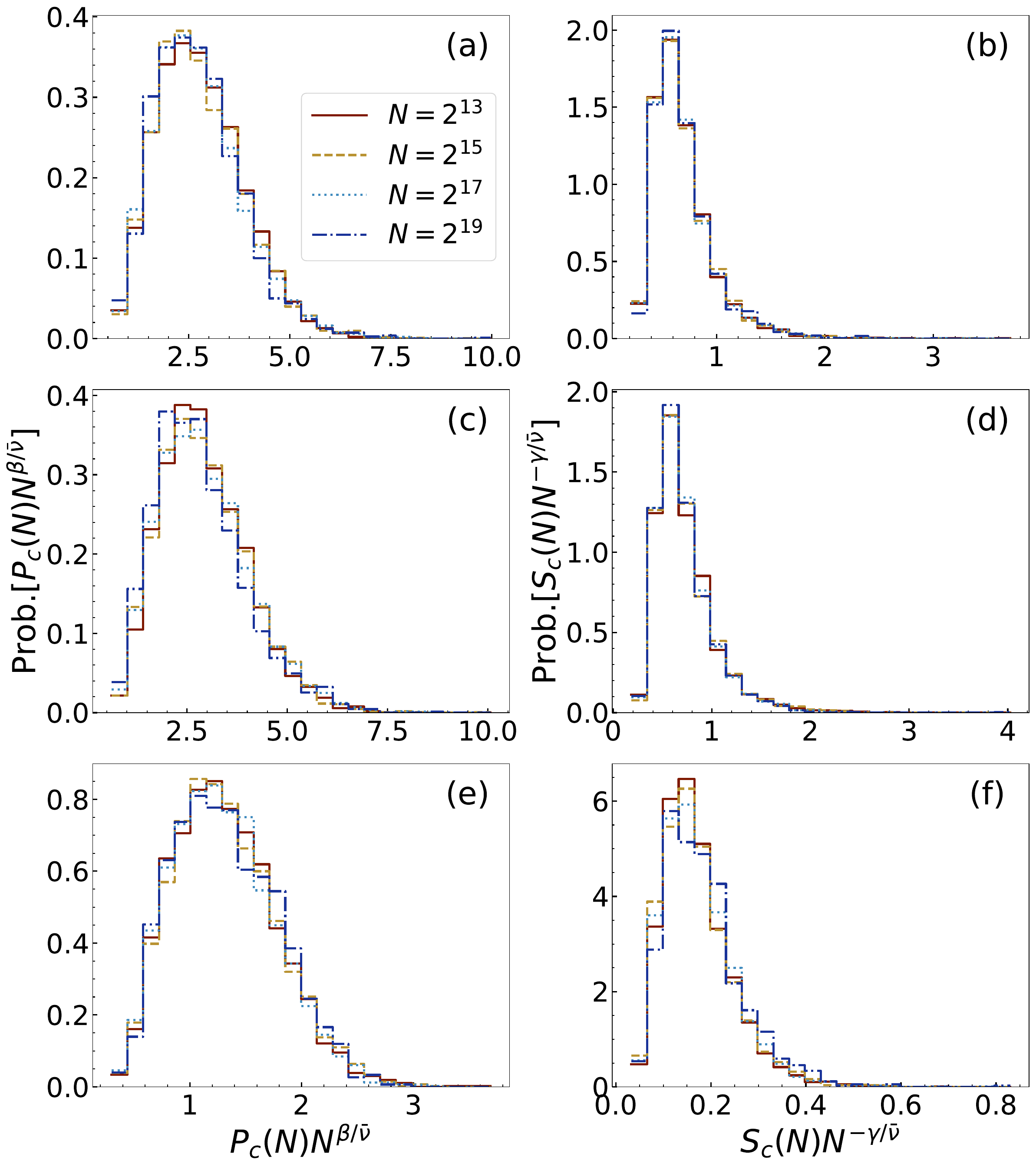}
    \caption{\textbf{Data Collapse of probability distributions of critical observables.} Similar to Fig.~\ref{fig2} in the main, but results are valid for SFNs with $\lambda=3.5$.}
    \label{sm-fig:pdf_collapse_lambda3.5}
\end{figure}

\begin{figure}[h]
    \centering
    \includegraphics[width=0.6\linewidth]{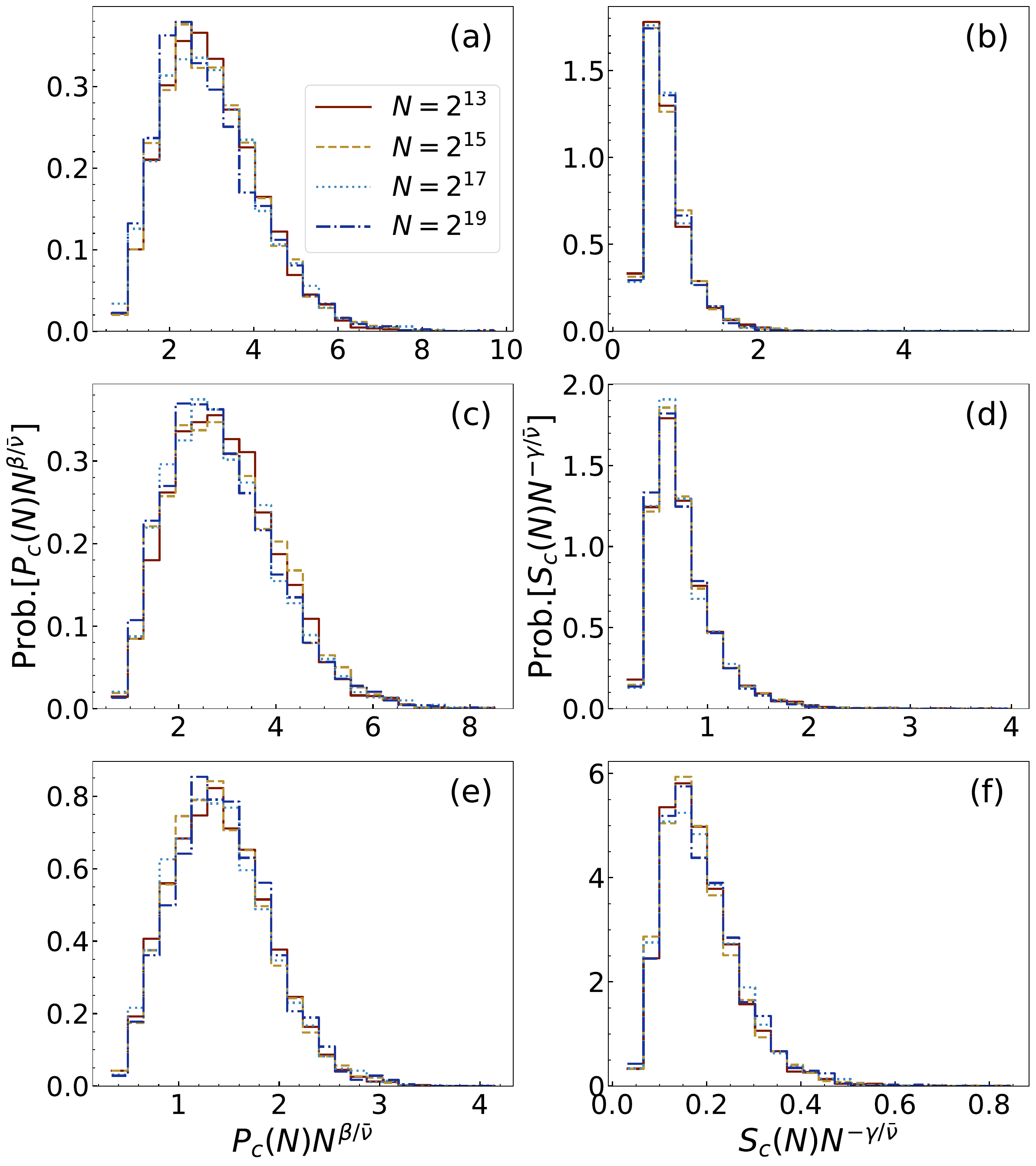}
    \caption{\textbf{Data Collapse of probability distributions of critical observables.} Similar to Fig.~\ref{fig2} in the main, but results are valid for SFNs with $\lambda=4.5$.}
    \label{sm-fig:pdf_collapse_lambda4.5}
\end{figure}

\begin{figure}
    \centering
    \includegraphics[width=0.7\linewidth]{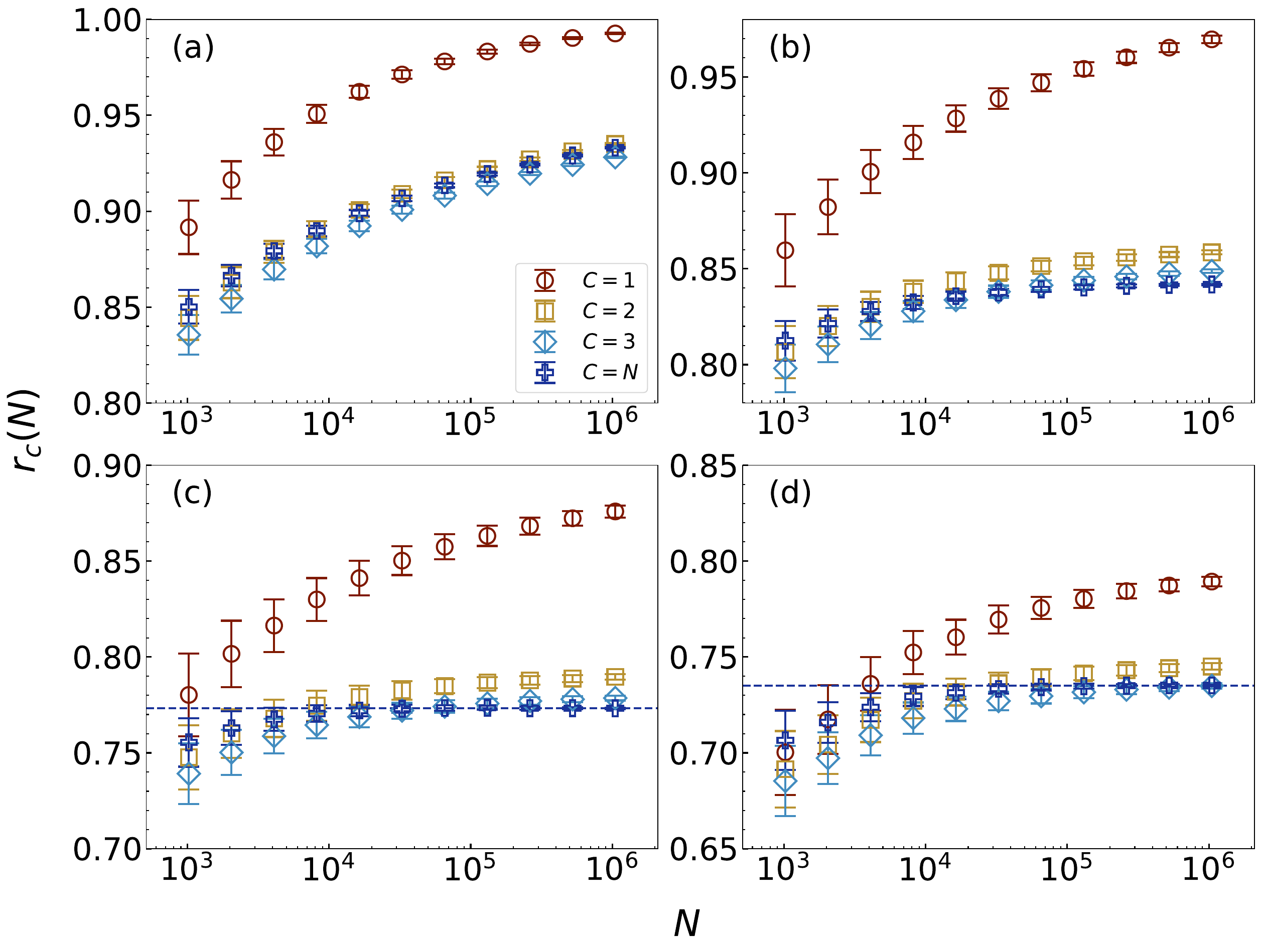}
    \caption{\textbf{Pseudocritical point $r_c(N)$}. We plot the pseudocritical point $r_c(N)$ as a function of network size $N$. Results are valid for SFNs with (a) $\lambda=2.1$, (b) $\lambda=2.7$, (c) $\lambda=3.5$, and (d) $\lambda=4.5$. 
    Note that the horizontal dashed lines in panel (c) and (d) are the plateau values of $r_c(N)$ for SPP with $C=N$. 
    These values are obtained by taking the average of $r_c(N)$ with $N>10^{5}$, which are within the error bounds reported in Table~\ref{tab:summary}.
    }
    \label{sm-fig:pseudo_rc}
\end{figure}

\pagebreak
\clearpage

\begin{figure}[h]
    \centering
    \includegraphics[width=0.7\linewidth]{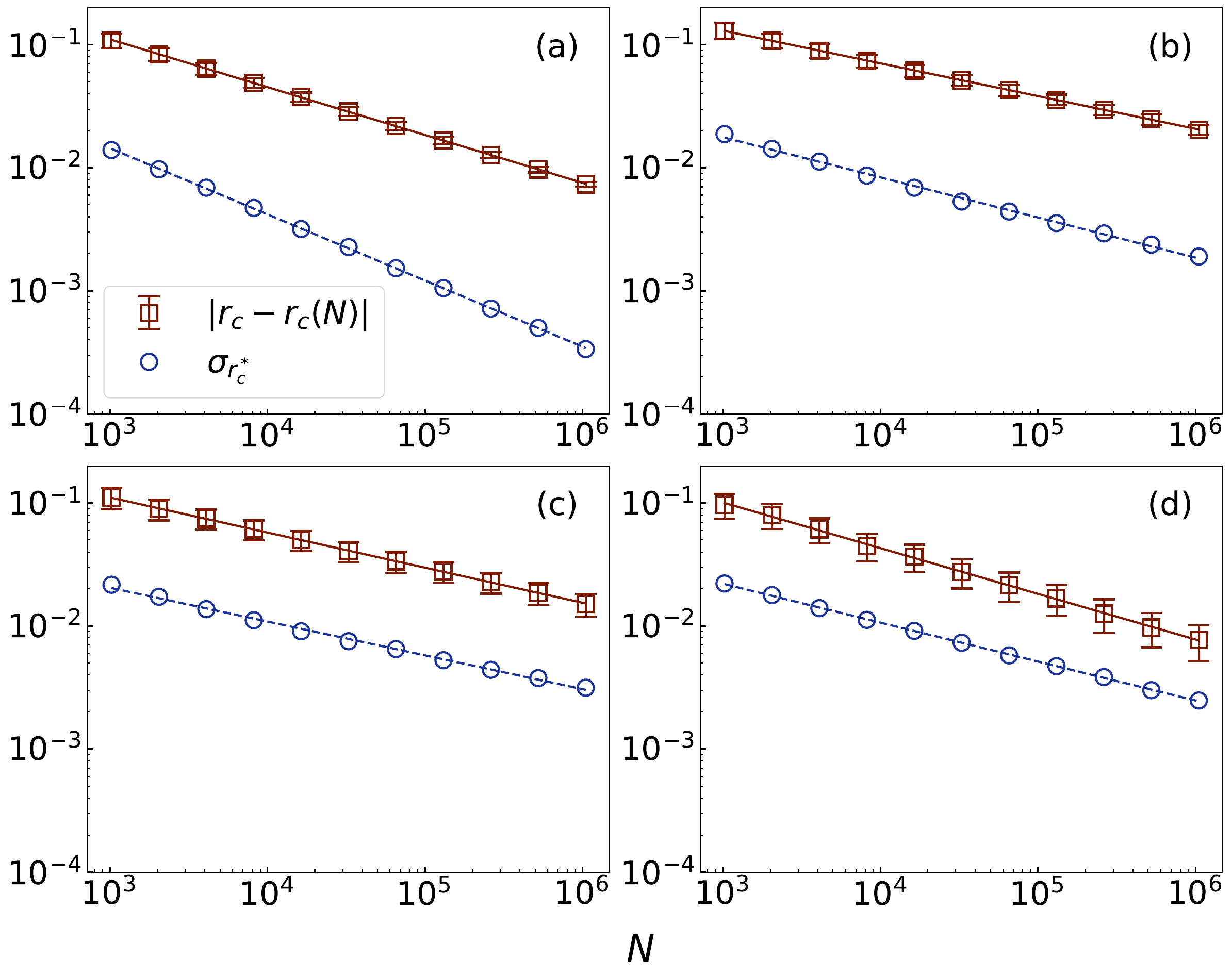}
    \caption{\textbf{Estimation of $1/\bar\nu$.} Similar to Fig.~\ref{fig4} in the main, but results are valid for SPP with $C=1$. Note that this corresponds to ordinary bond percolation.}
    \label{sm-fig:nu_estimation_C1}
\end{figure}

\begin{figure}[h]
    \centering
    \includegraphics[width=0.7\linewidth]{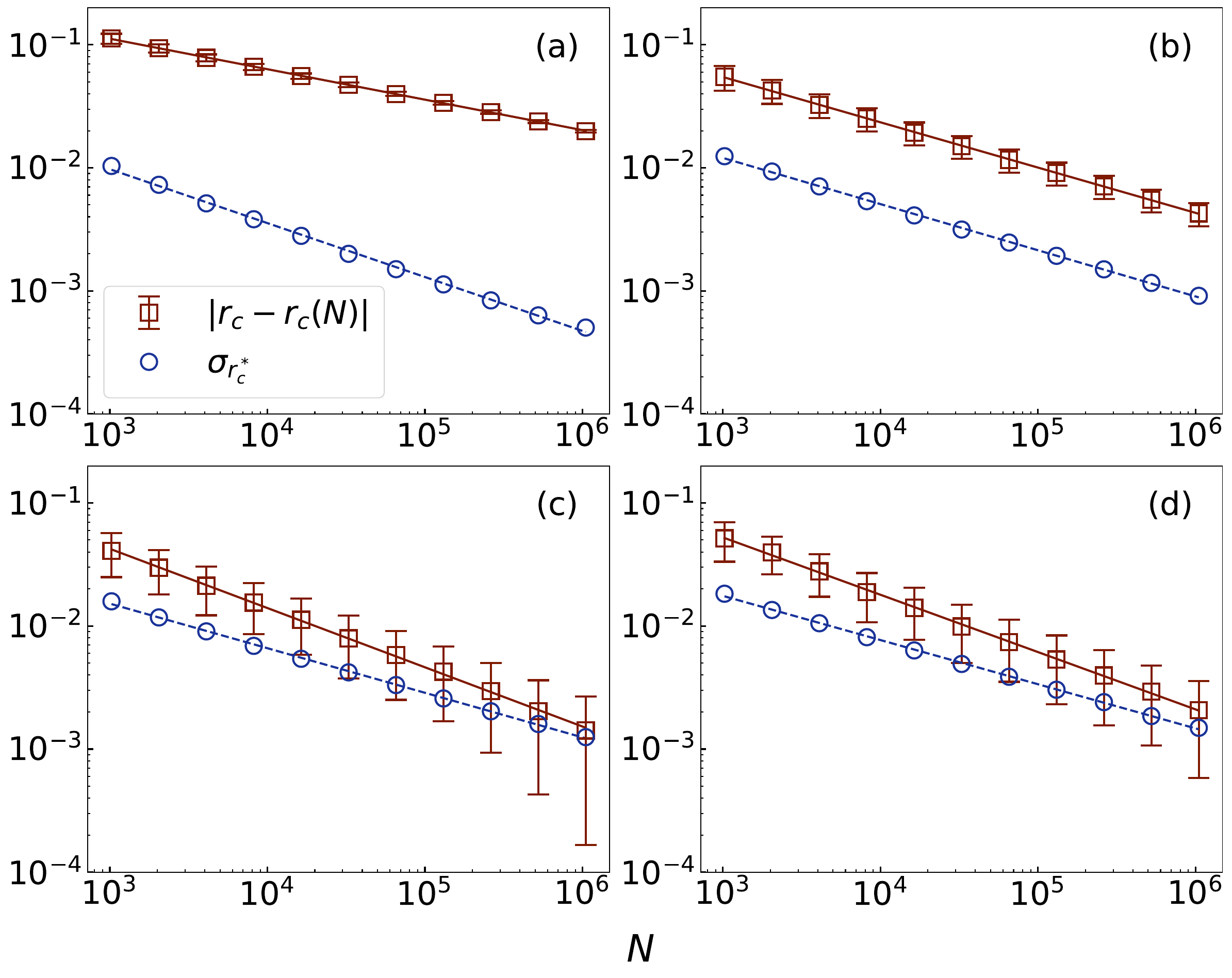}
    \caption{\textbf{Estimation of $1/\bar\nu$.} Similar to Fig.~\ref{fig4} in the main, but results are valid for SPP with $C=3$.}
    \label{sm-fig:nu_estimation_C3}
\end{figure}

\begin{figure}[h]
    \centering
    \includegraphics[width=0.7\linewidth]{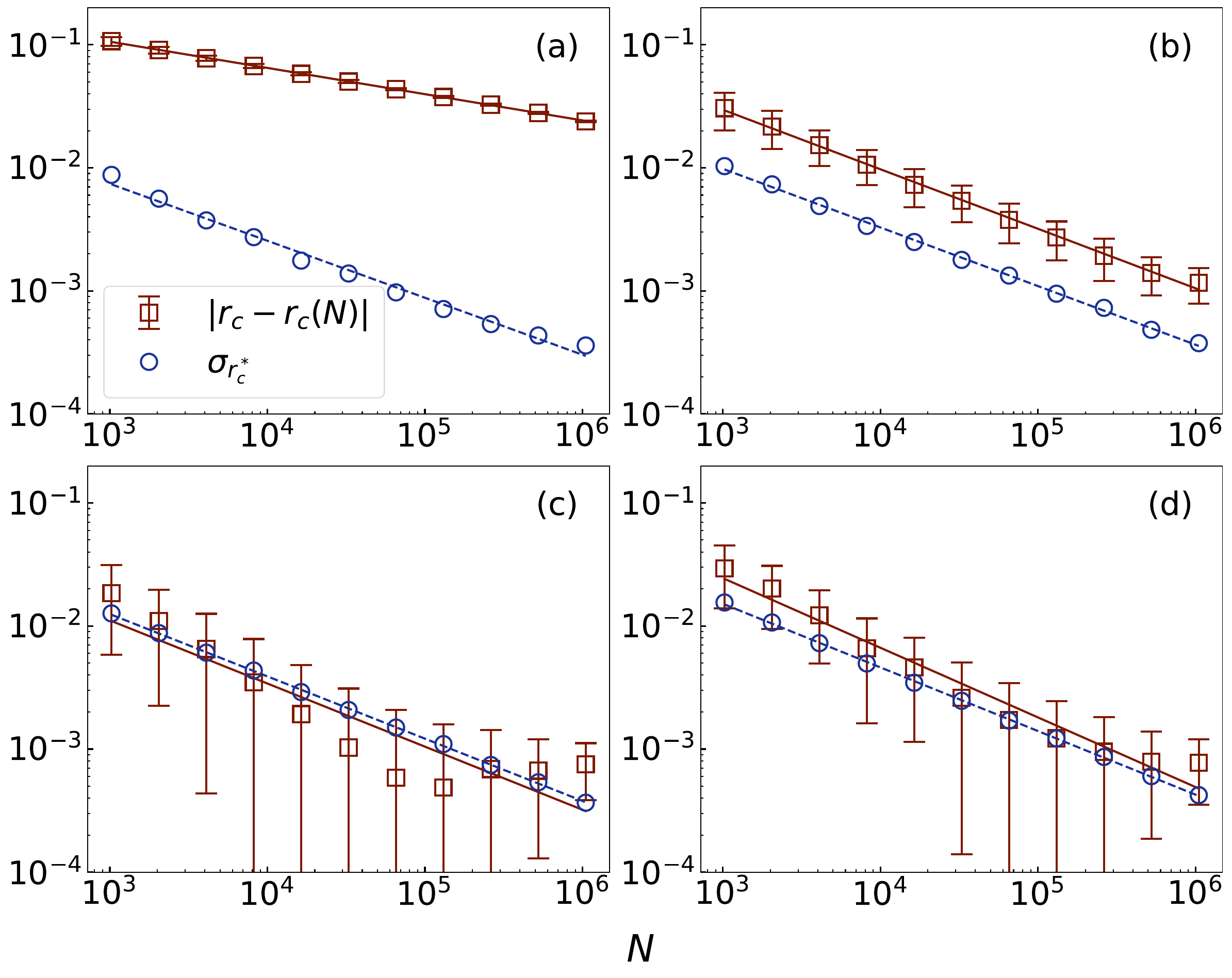}
    \caption{\textbf{Estimation of $1/\bar\nu$.} Similar to Fig.~\ref{fig4} in the main, but results are valid for SPP with $C=N$. 
    }
    \label{sm-fig:nu_estimation_CN}
\end{figure}

\pagebreak
\clearpage

\begin{figure}[h]
    \centering
    \includegraphics[width=0.7\linewidth]{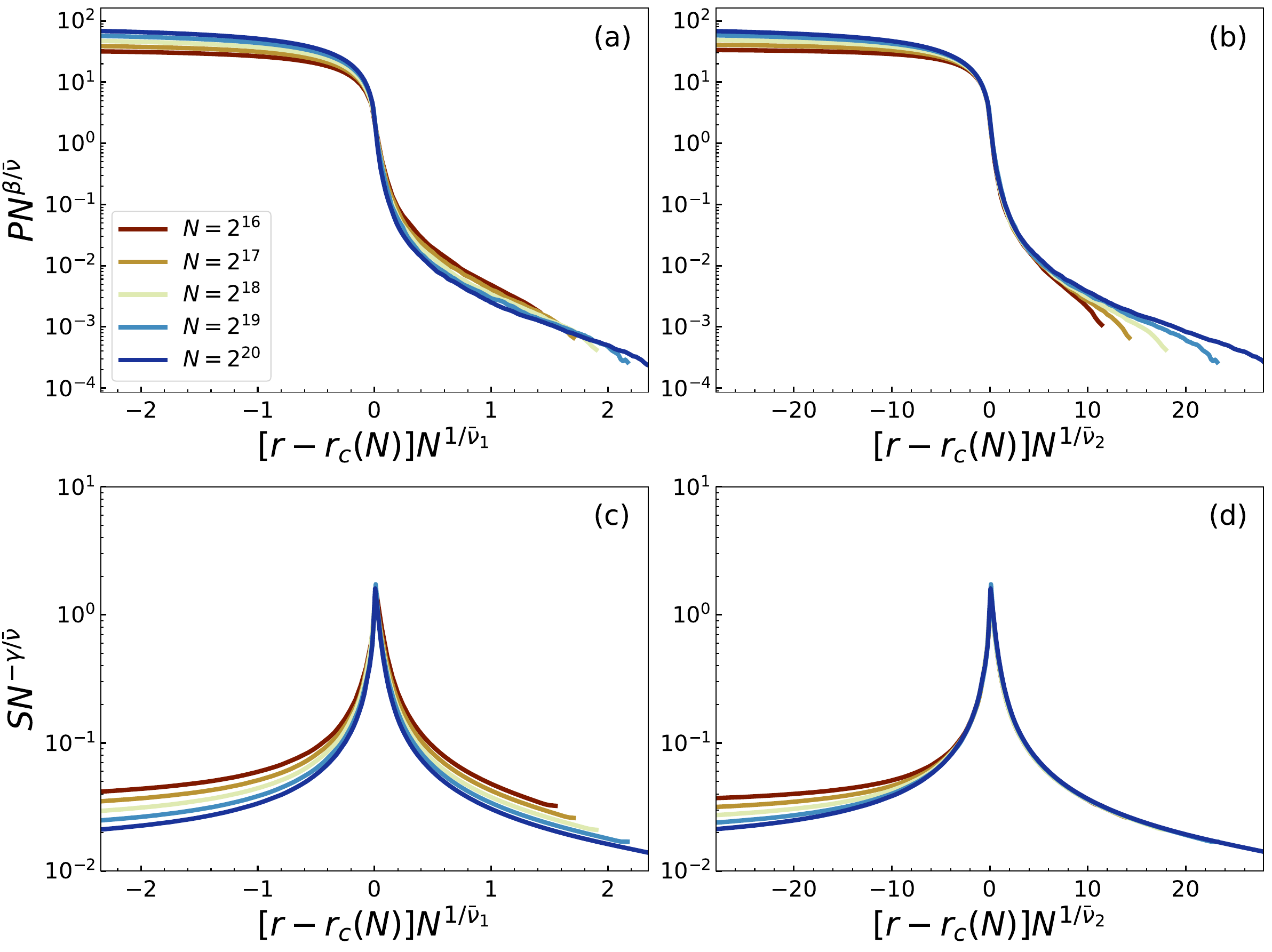}
    \caption{\textbf{Data Collapse of $P$ and $S$.} We plot the rescaled observable $PN^{\beta/\bar\nu}$ as a function of (a) $[r-r_c(N)]N^{1/\bar\nu_{1}}$ and (b) $[r-r_c(N)]N^{1/\bar\nu_{2}}$, respectively. Similarly, we plot the rescaled observable $SN^{-\gamma/\bar\nu}$ as a function of (c) $[r-r_c]N^{1/\bar\nu_{1}}$ and (d) $[r-r_c(N)]N^{1/\bar\nu_{2}}$, respectively. Results are valid for SFNs with $\lambda=2.1$ and $k_{min}=4$, and SPP model with $C=2$. 
    Note that in order to make a fair comparison between the data collapse using $1/\bar\nu_1$ and $1/\bar\nu_2$, the range of the abscissa is adjusted to display 200 bins of the curve for $N=2^{20}$.
    }
    \label{sm-fig:collapse_sfn2.1_C2}
\end{figure}

\begin{figure}[h]
    \centering
    \includegraphics[width=0.7\linewidth]{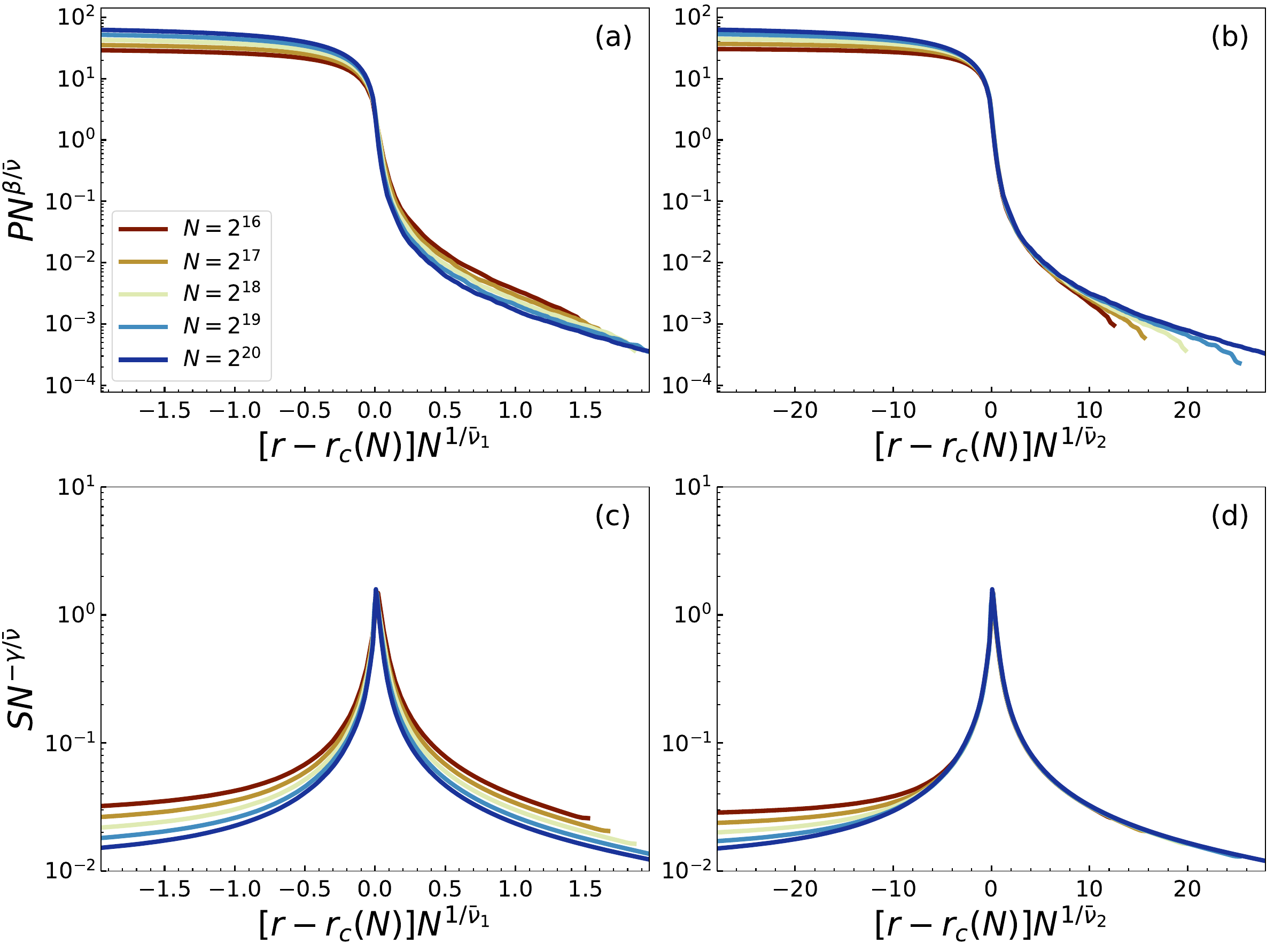}
    \caption{\textbf{Data Collapse of $P$ and $S$.} Similar to Fig.~\ref{sm-fig:collapse_sfn2.1_C2}, but results are valid for SPP model with $C=3$.}
    \label{sm-fig:collapse_sfn2.1_C3}
\end{figure}

\begin{figure}[h]
    \centering
    \includegraphics[width=0.7\linewidth]{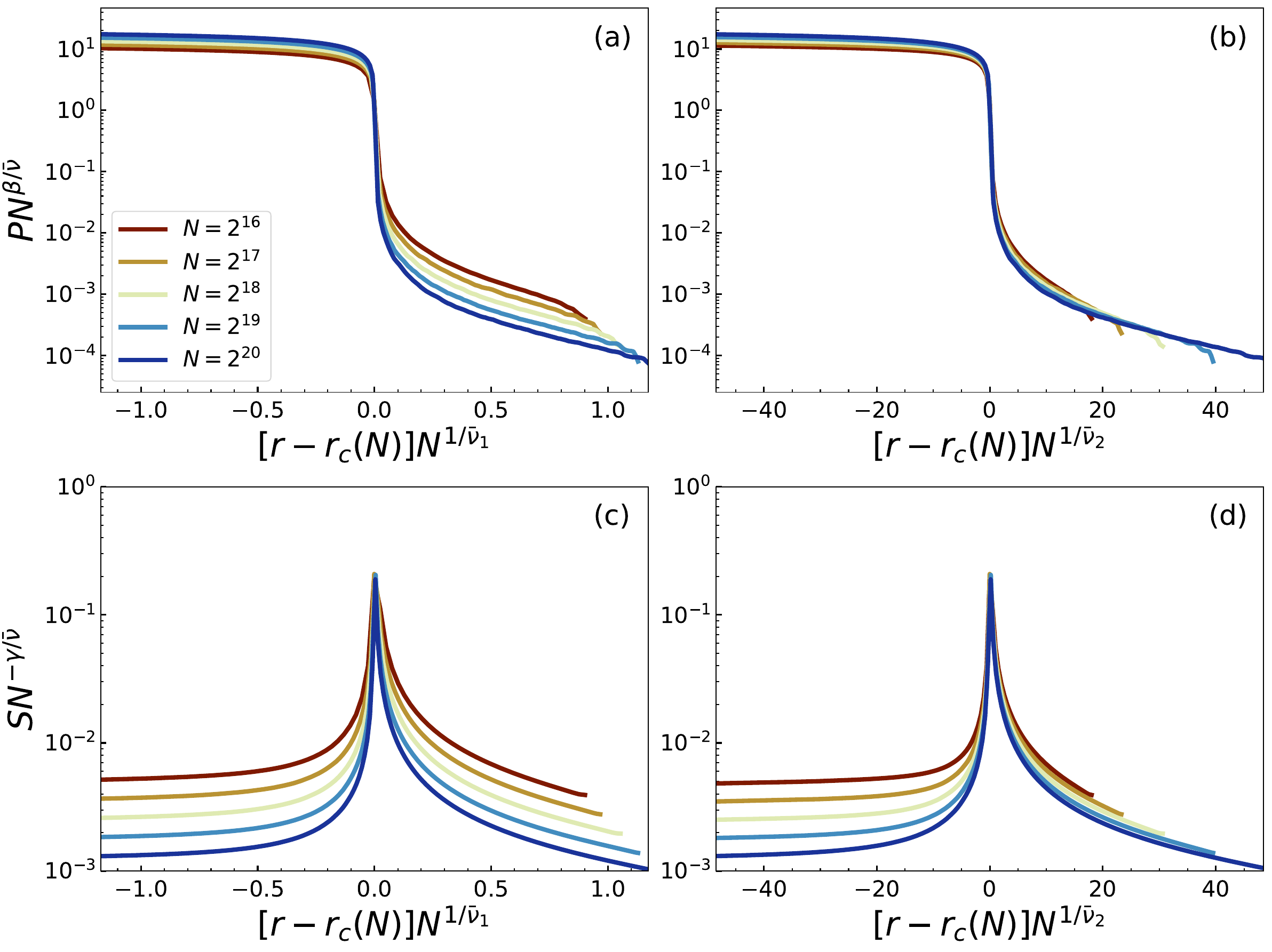}
    \caption{\textbf{Data Collapse of $P$ and $S$.} Similar to Fig.~\ref{sm-fig:collapse_sfn2.1_C2}, but results are valid for SPP model with $C=N$.}
    \label{sm-fig:collapse_sfn2.1_CN}
\end{figure}

\pagebreak
\clearpage

\begin{figure}[h]
    \centering
    \includegraphics[width=0.7\linewidth]{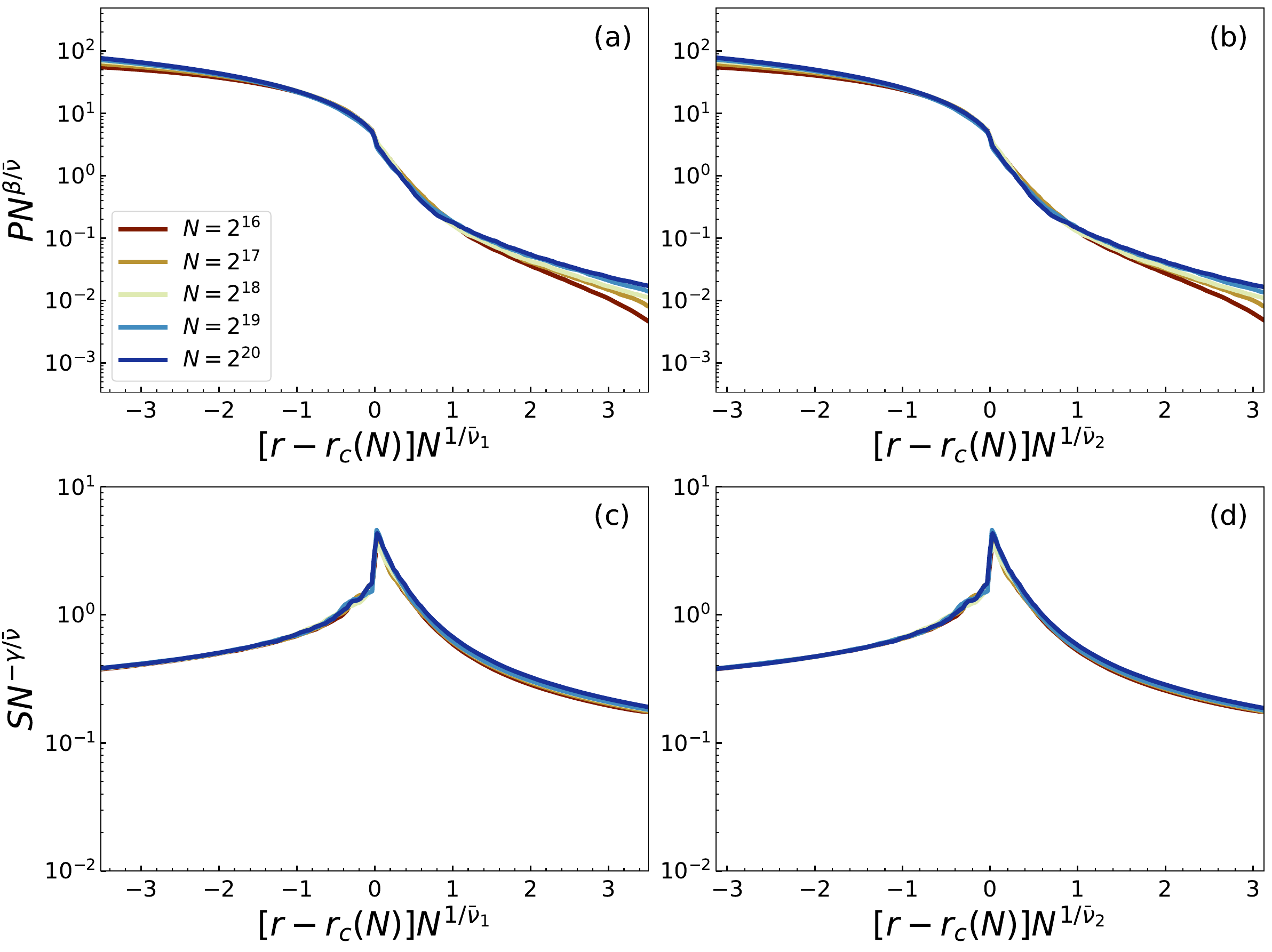}
    \caption{\textbf{Data Collapse of $P$ and $S$.} Similar to Fig.~\ref{sm-fig:collapse_sfn2.1_C2}, but results are valid for SFNs with $\lambda=3.5$ and $k_{min}=4$, and SPP model with $C=1$.}
    \label{sm-fig:collapse_sfn3.5_C1}
\end{figure}

\begin{figure}[h]
    \centering
    \includegraphics[width=0.7\linewidth]{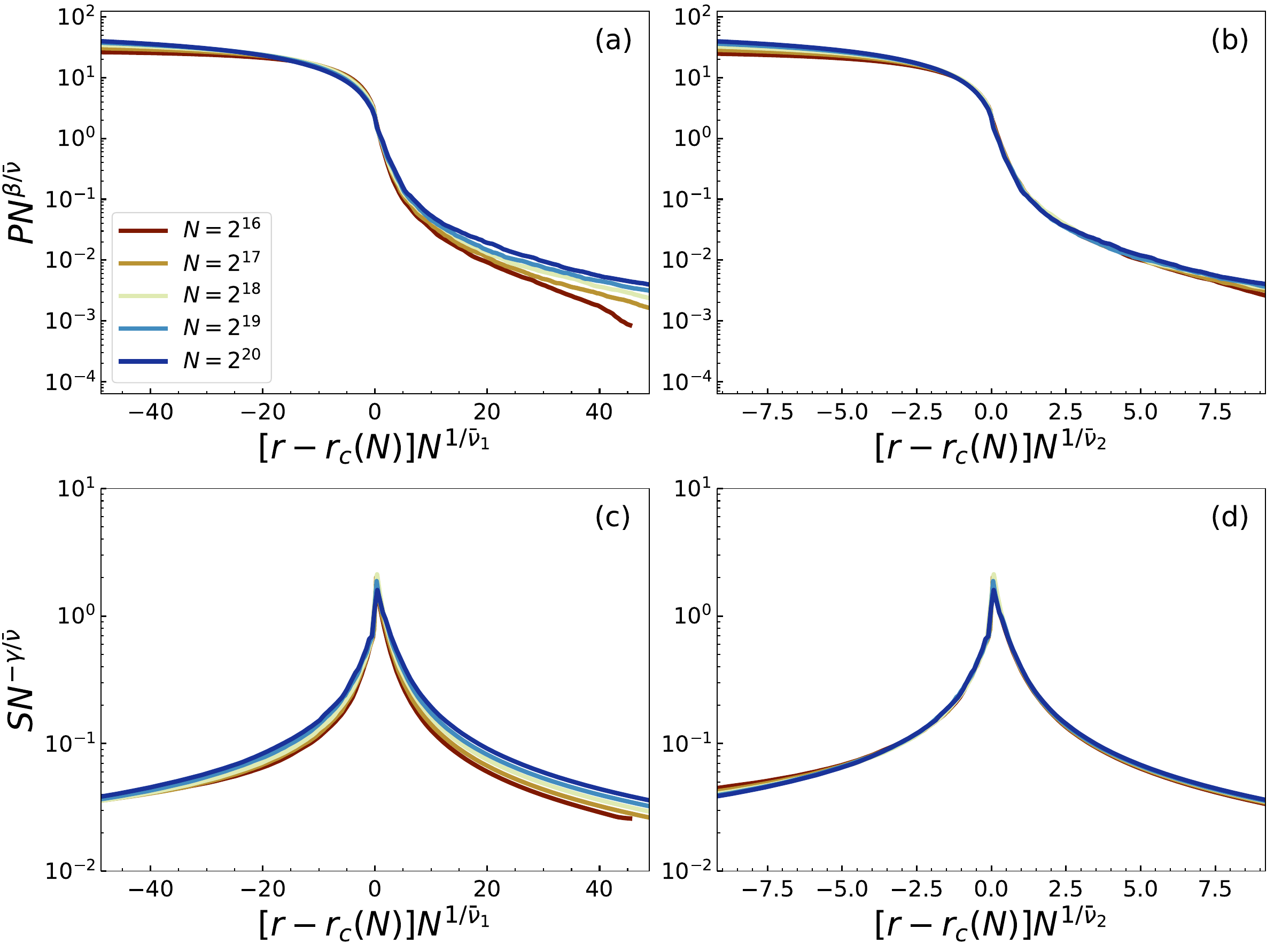}
    \caption{\textbf{Data Collapse of $P$ and $S$.} Similar to Fig.~\ref{sm-fig:collapse_sfn3.5_C1}, but results are valid for SPP model with $C=2$.}
    \label{sm-fig:collapse_sfn3.5_C2}
\end{figure}

\begin{figure}[h]
    \centering
    \includegraphics[width=0.7\linewidth]{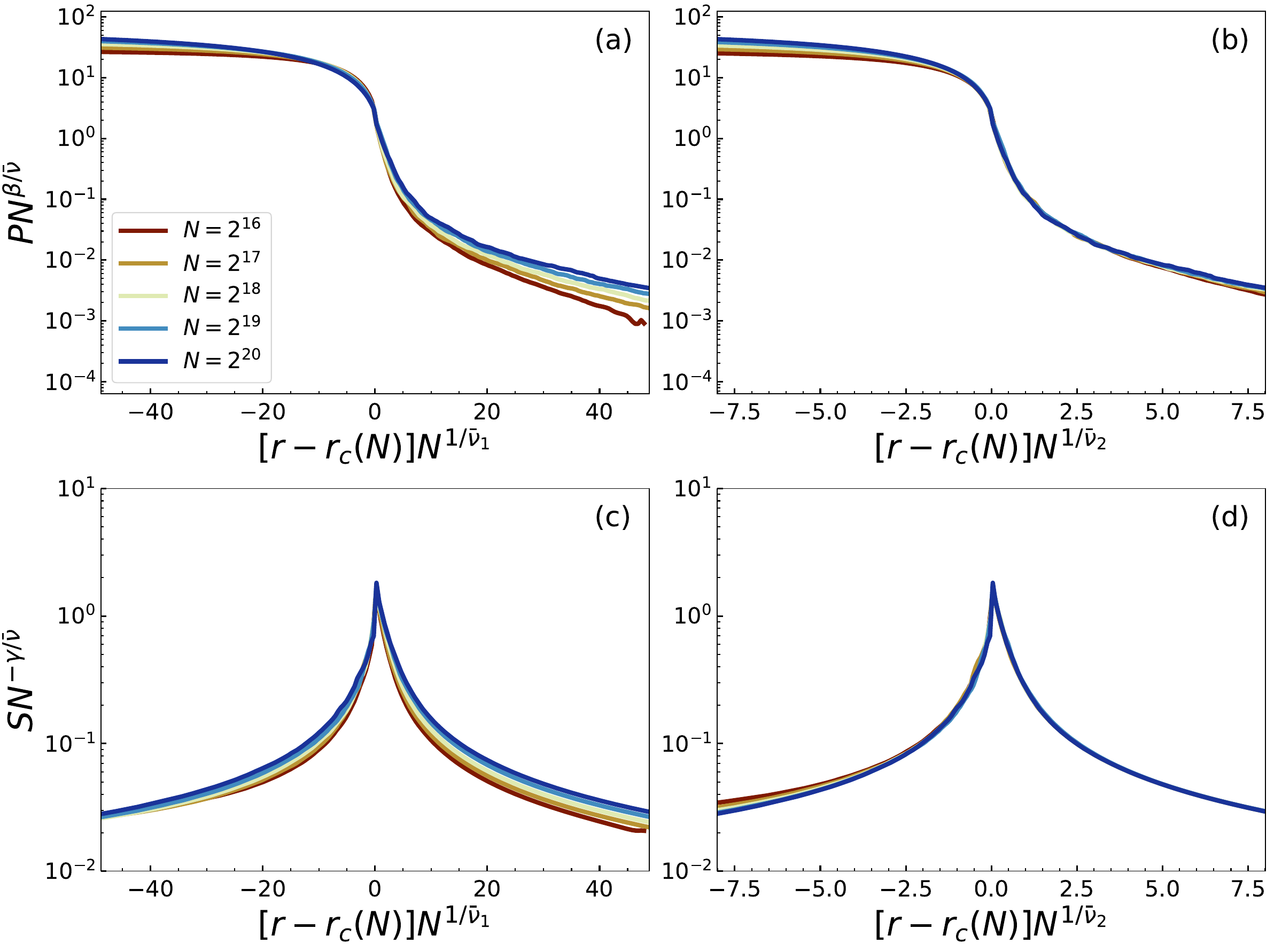}
    \caption{\textbf{Data Collapse of $P$ and $S$.} Similar to Fig.~\ref{sm-fig:collapse_sfn3.5_C1}, but results are valid for SPP model with $C=3$.}
    \label{sm-fig:collapse_sfn3.5_C3}
\end{figure}

\begin{figure}[h]
    \centering
    \includegraphics[width=0.7\linewidth]{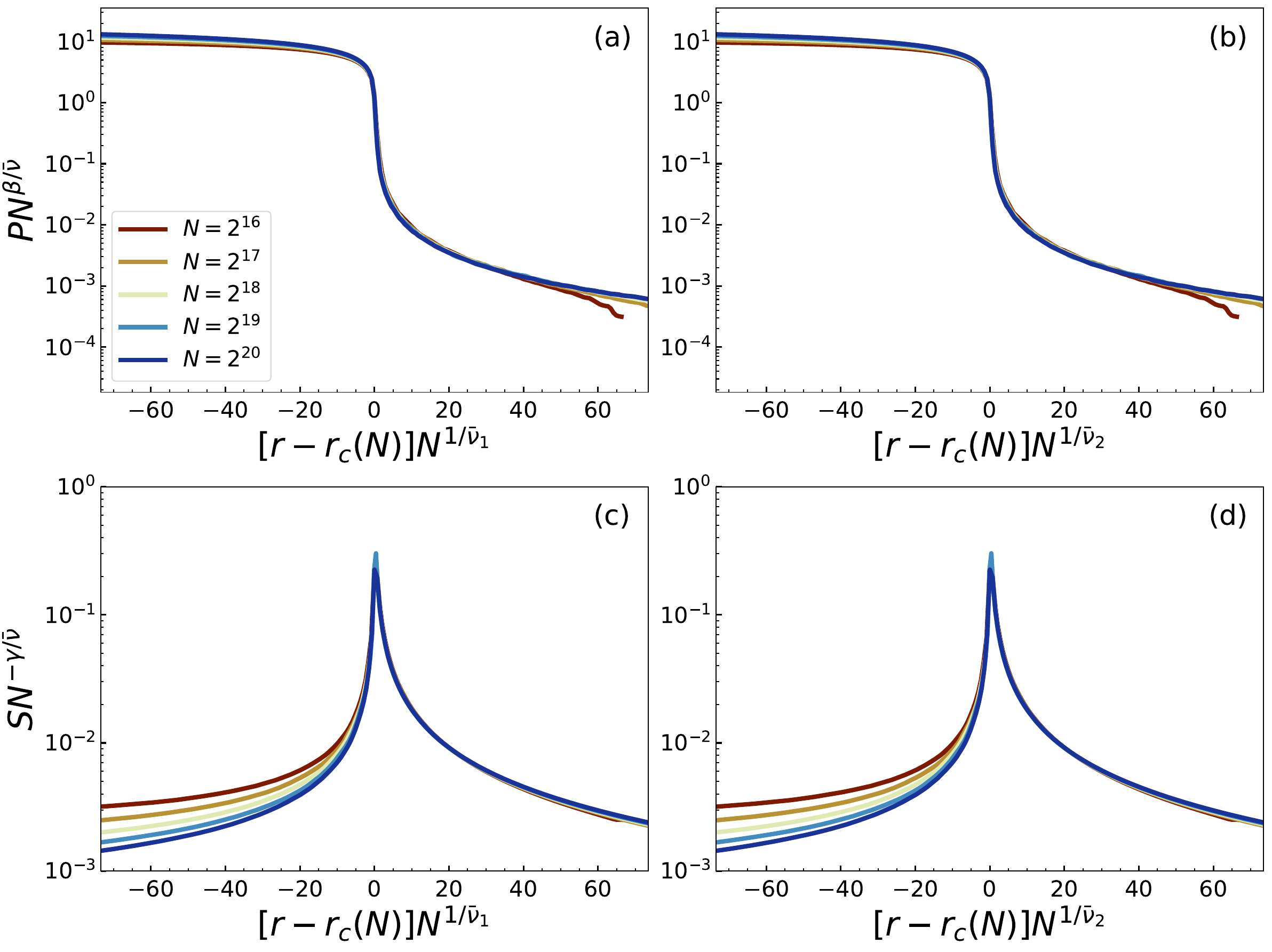}
    \caption{\textbf{Data Collapse of $P$ and $S$.} Similar to Fig.~\ref{sm-fig:collapse_sfn3.5_C1}, but results are valid for SPP model with $C=N$.
    }
    \label{sm-fig:collapse_sfn3.5_CN}
\end{figure}

\pagebreak
\clearpage

\begin{figure}[h]
    \centering
    \includegraphics[width=0.7\linewidth]{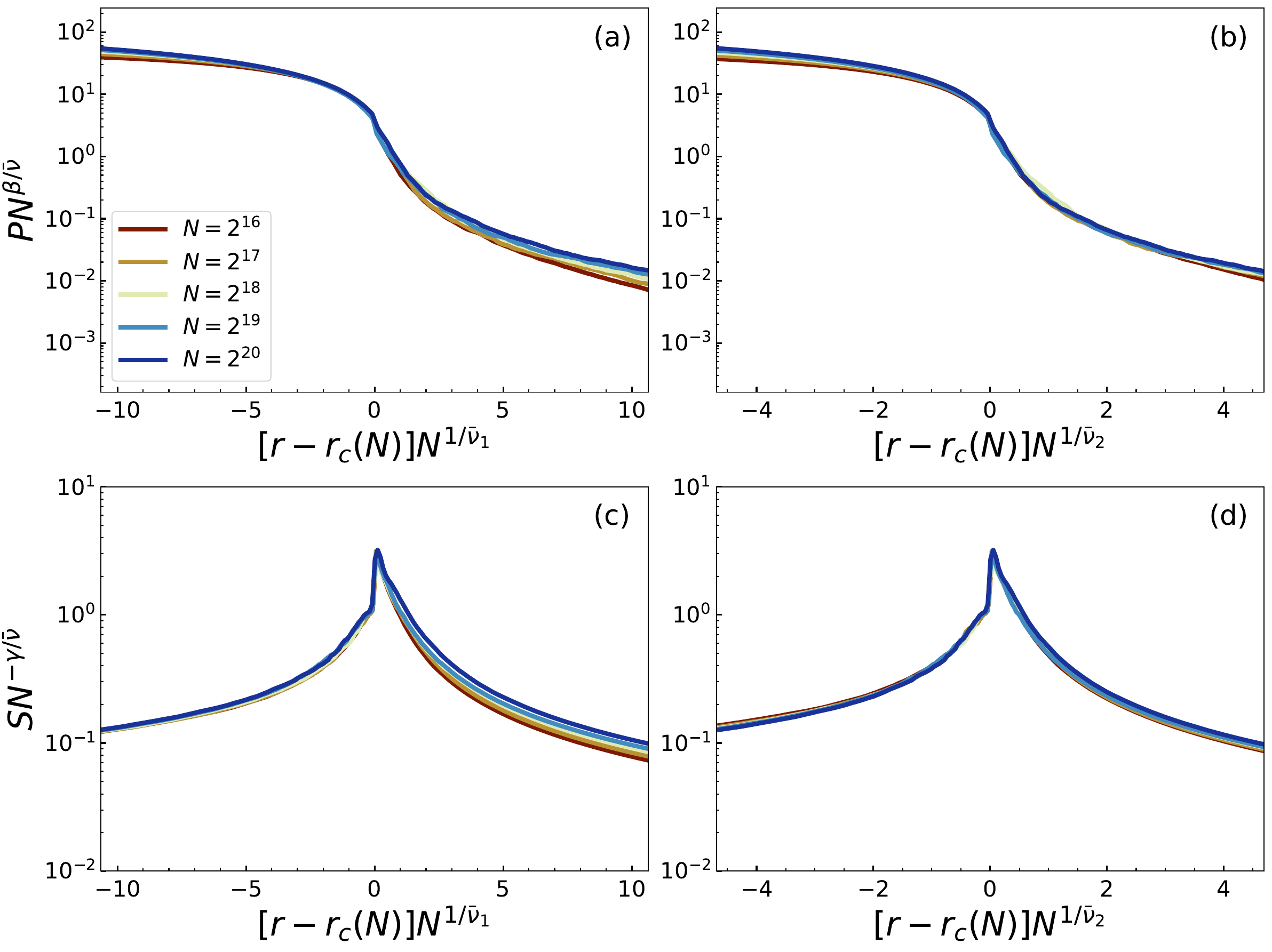}
    \caption{\textbf{Data Collapse of $P$ and $S$.} Similar to Fig.~\ref{sm-fig:collapse_sfn2.1_C2}, but results are valid for SFNs with $\lambda=4.5$ and $k_{min}=4$, and SPP model with $C=1$.}
    \label{sm-fig:collapse_sfn4.5_C1}
\end{figure}

\begin{figure}[h]
    \centering
    \includegraphics[width=0.7\linewidth]{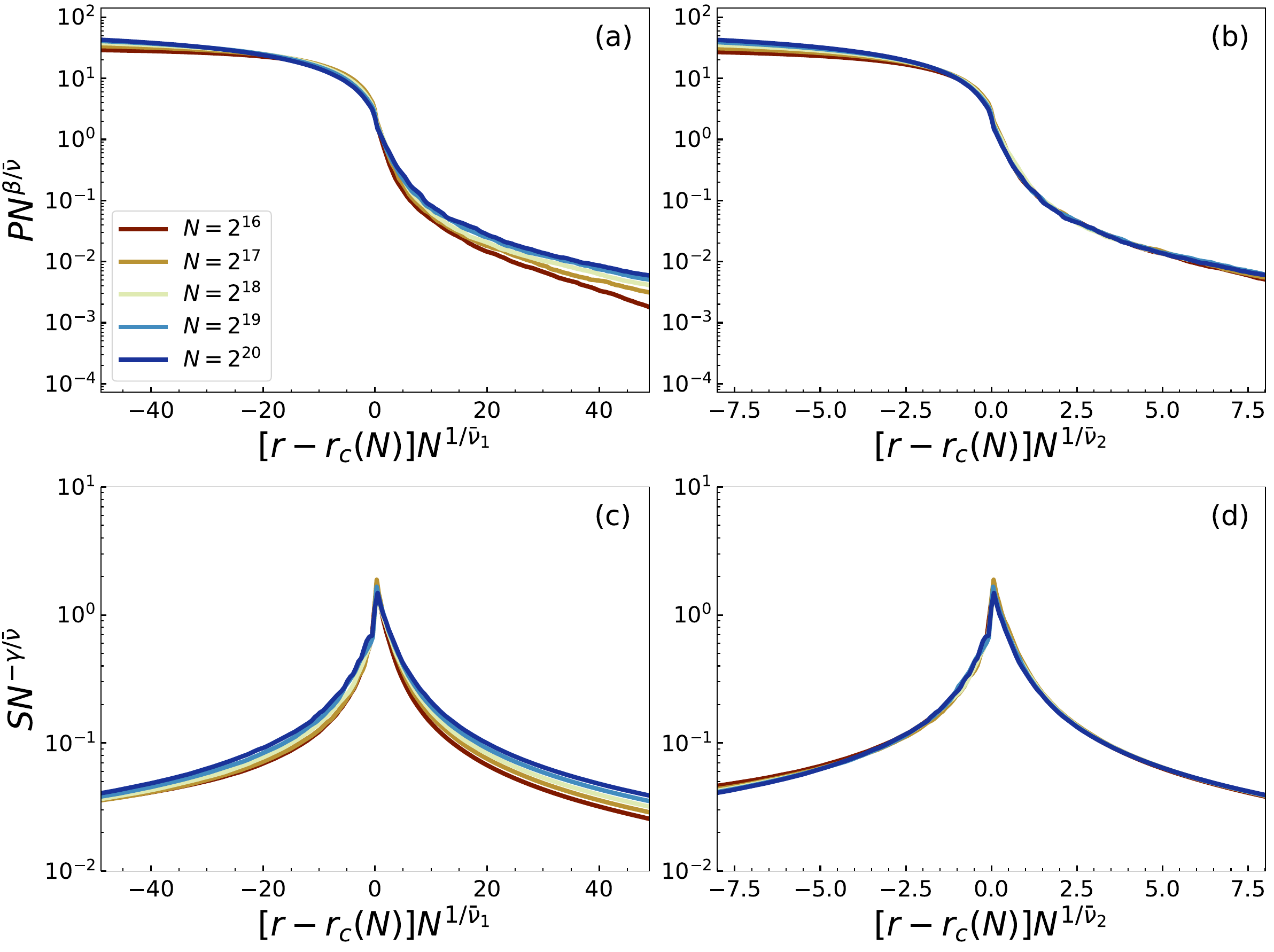}
    \caption{\textbf{Data Collapse of $P$ and $S$.} Similar to Fig.~\ref{sm-fig:collapse_sfn4.5_C1}, but results are valid for SPP model with $C=2$.}
    \label{sm-fig:collapse_sfn4.5_C2}
\end{figure}

\begin{figure}[h]
    \centering
    \includegraphics[width=0.7\linewidth]{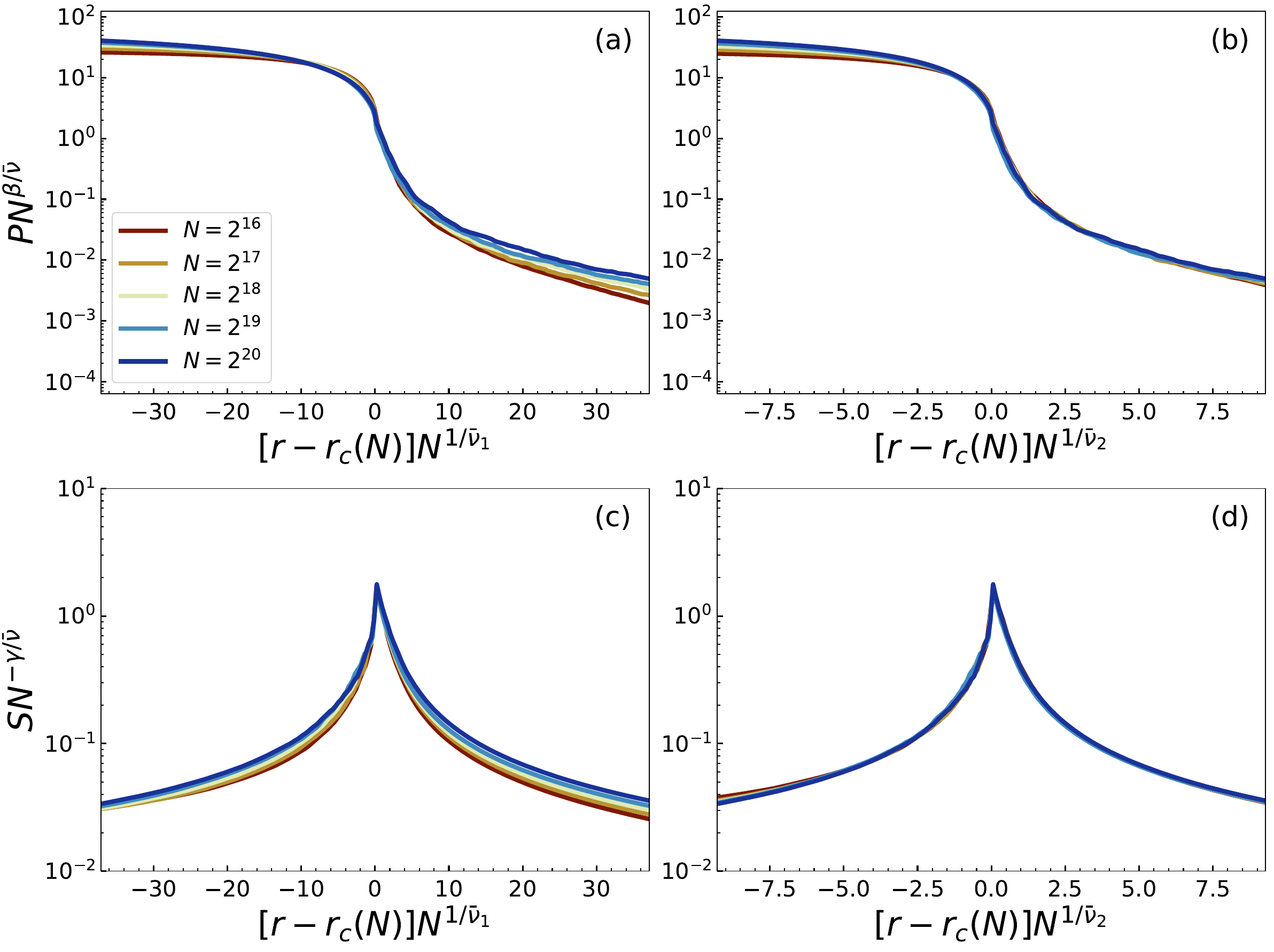}
    \caption{\textbf{Data Collapse of $P$ and $S$.} Similar to Fig.~\ref{sm-fig:collapse_sfn4.5_C1}, but results are valid for SPP model with $C=3$.}
    \label{sm-fig:collapse_sfn4.5_C3}
\end{figure}

\begin{figure}[h]
    \centering
    \includegraphics[width=0.7\linewidth]{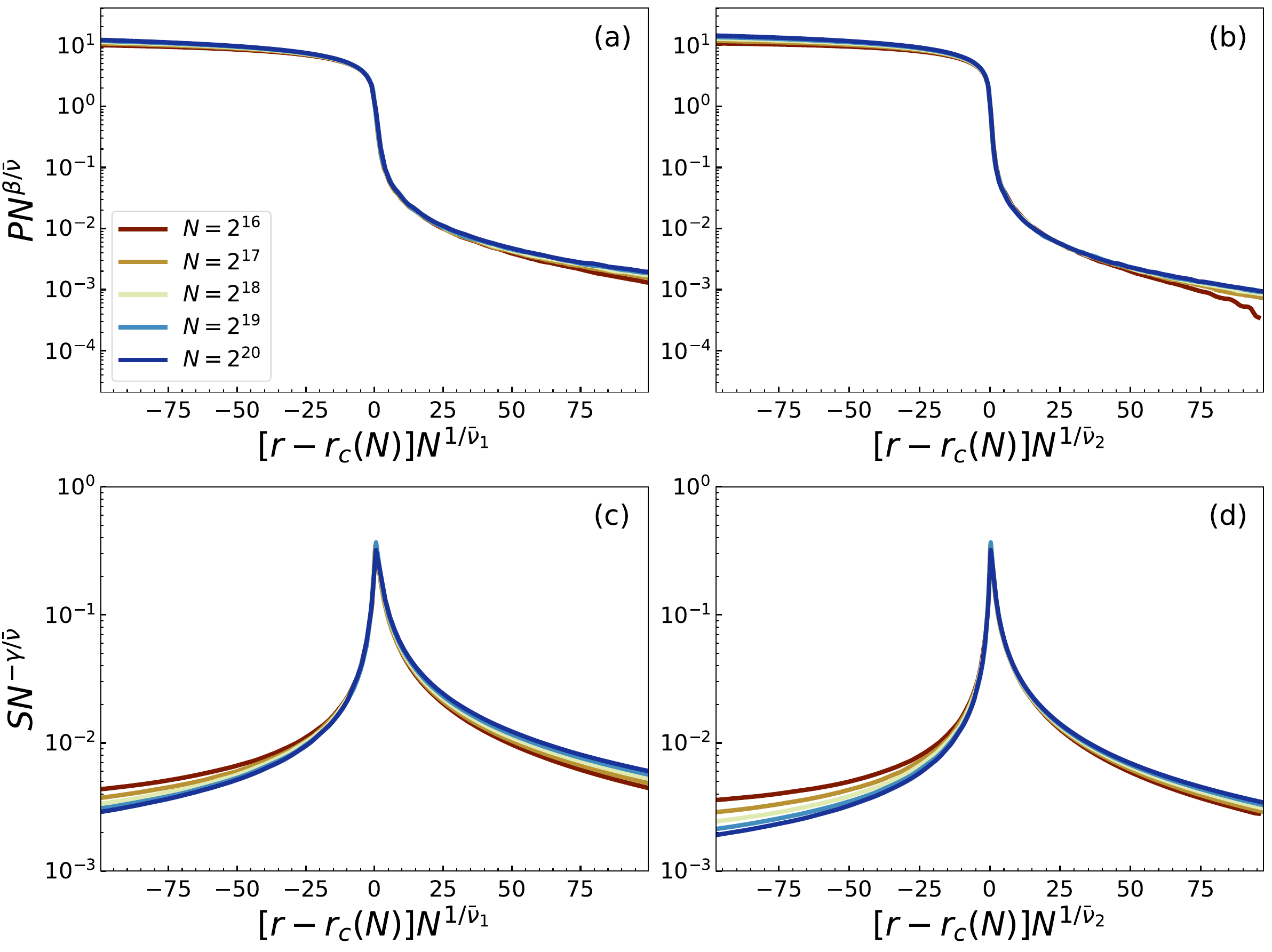}
    \caption{\textbf{Data Collapse of $P$ and $S$.} Similar to Fig.~\ref{sm-fig:collapse_sfn4.5_C1}, but results are valid for SPP model with $C=N$. 
    }
    \label{sm-fig:collapse_sfn4.5_CN}
\end{figure}

\pagebreak
\clearpage

\begin{figure}
    \centering
    \includegraphics[width=0.7\linewidth]{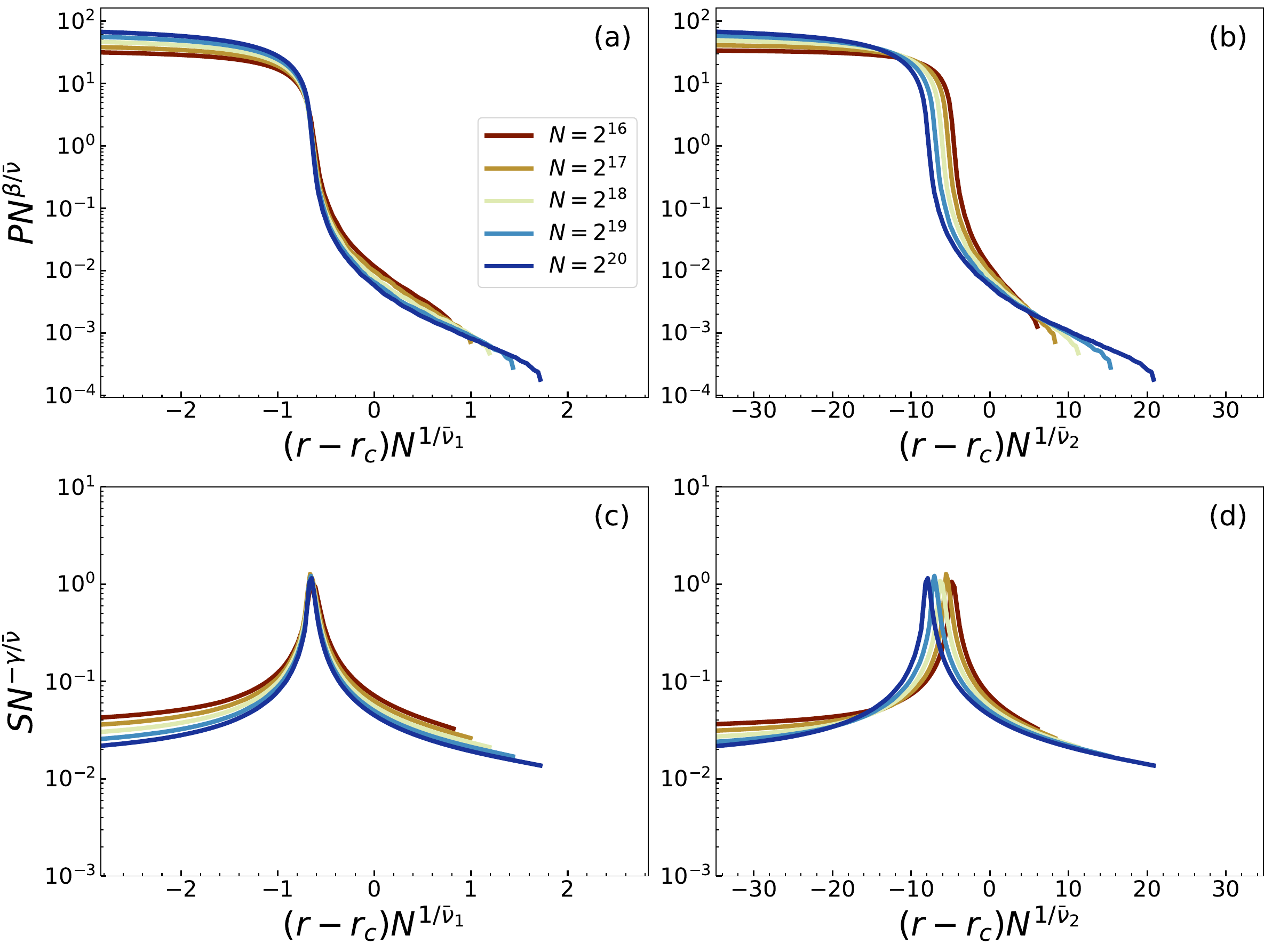}
    \caption{\textbf{Data Collapse of $P$ and $S$.} We plot the rescaled observable $PN^{\beta/\bar\nu}$ as a function of (a) $(r-r_c)N^{1/\bar\nu_{1}}$ and (b) $(r-r_c)N^{1/\bar\nu_{2}}$, respectively. Similarly, we plot the rescaled observable $SN^{-\gamma/\bar\nu}$ as a function of (c) $(r-r_c)N^{1/\bar\nu_{1}}$ and (d) $(r-r_c)N^{1/\bar\nu_{2}}$, respectively. Results are valid for SFNs with $\lambda=2.1$ and $k_{min}=4$, and SPP model with $C=2$. 
    Note that the range of the abscissa is adjusted to display 200 bins of the curves for $N=2^{20}$.
    }
    \label{sm-fig:collapse_conventional_sfn2.1_C2}
\end{figure}

\begin{figure}
    \centering
    \includegraphics[width=0.7\linewidth]{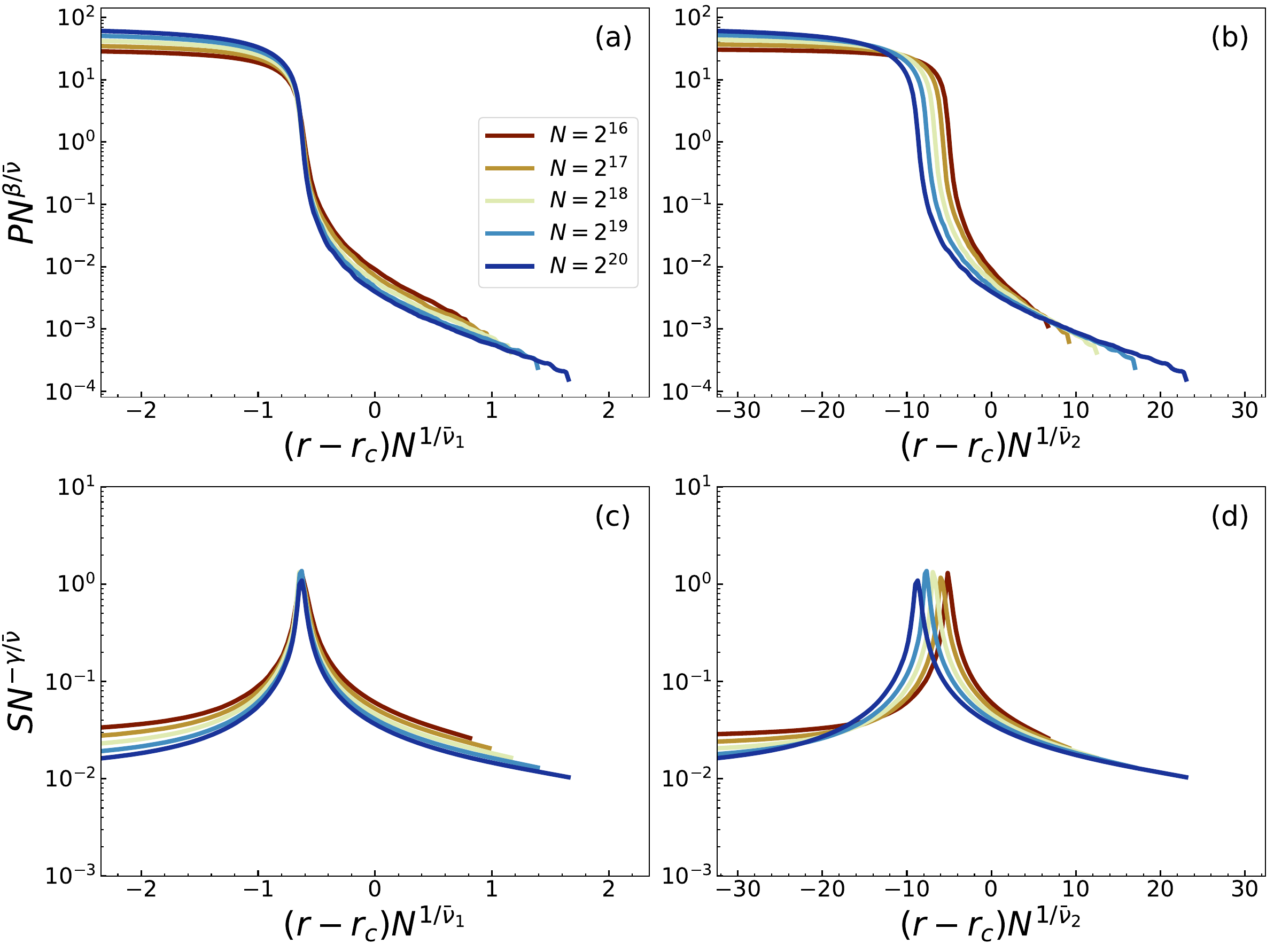}
    \caption{\textbf{Data Collapse of $P$ and $S$.} Similar to Fig.~\ref{sm-fig:collapse_conventional_sfn2.1_C2}, but results are valid for SPP model with $C=3$.}
    \label{sm-fig:collapse_conventional_sfn2.1_C3}
\end{figure}

\begin{figure}
    \centering
    \includegraphics[width=0.7\linewidth]{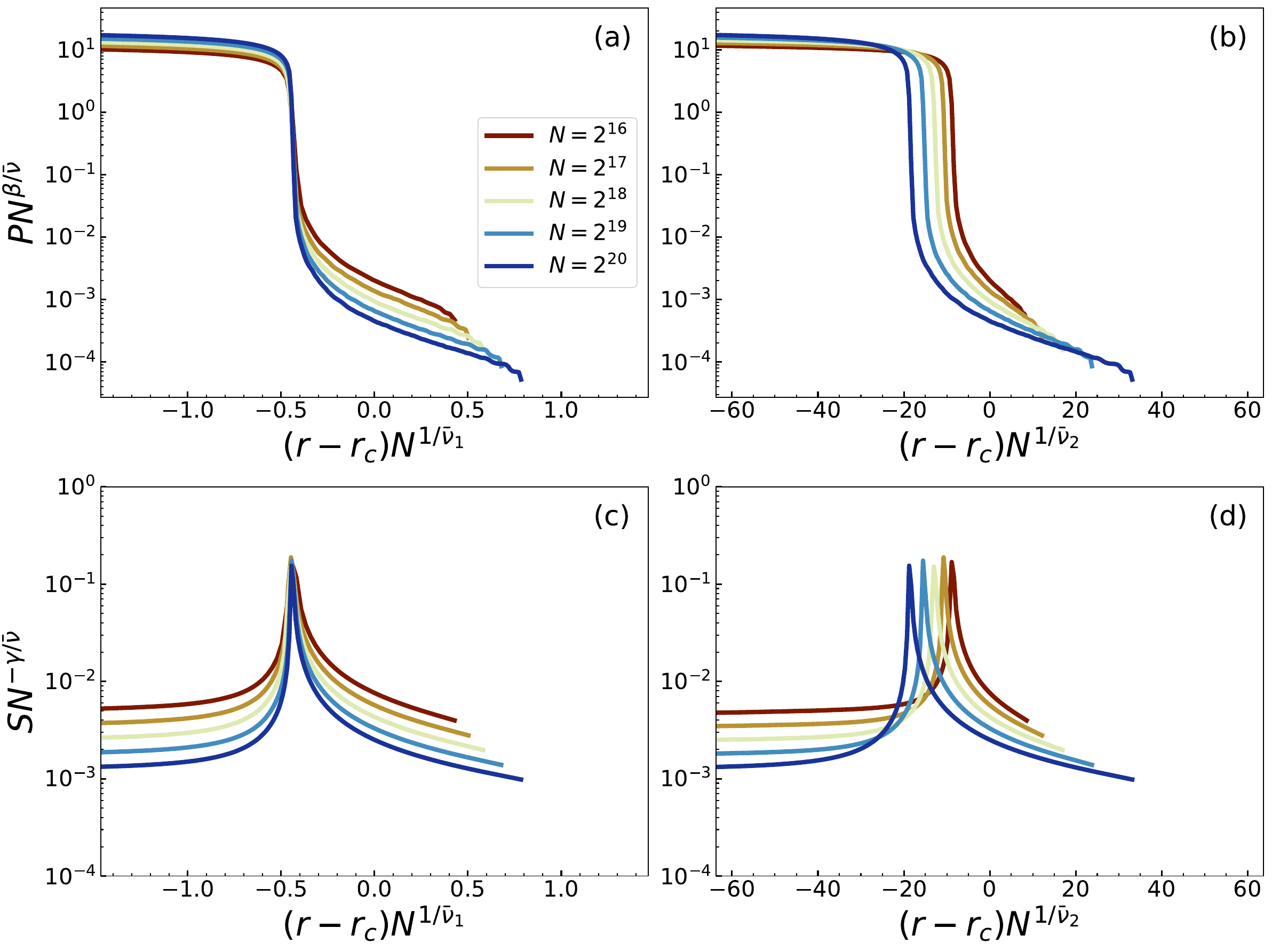}
    \caption{\textbf{Data Collapse of $P$ and $S$.} Similar to Fig.~\ref{sm-fig:collapse_conventional_sfn2.1_C2}, but results are valid for SPP model with $C=N$.}
    \label{sm-fig:collapse_conventional_sfn2.1_CN}
\end{figure}

\pagebreak
\clearpage

\begin{figure}
    \centering
    \includegraphics[width=0.7\linewidth]{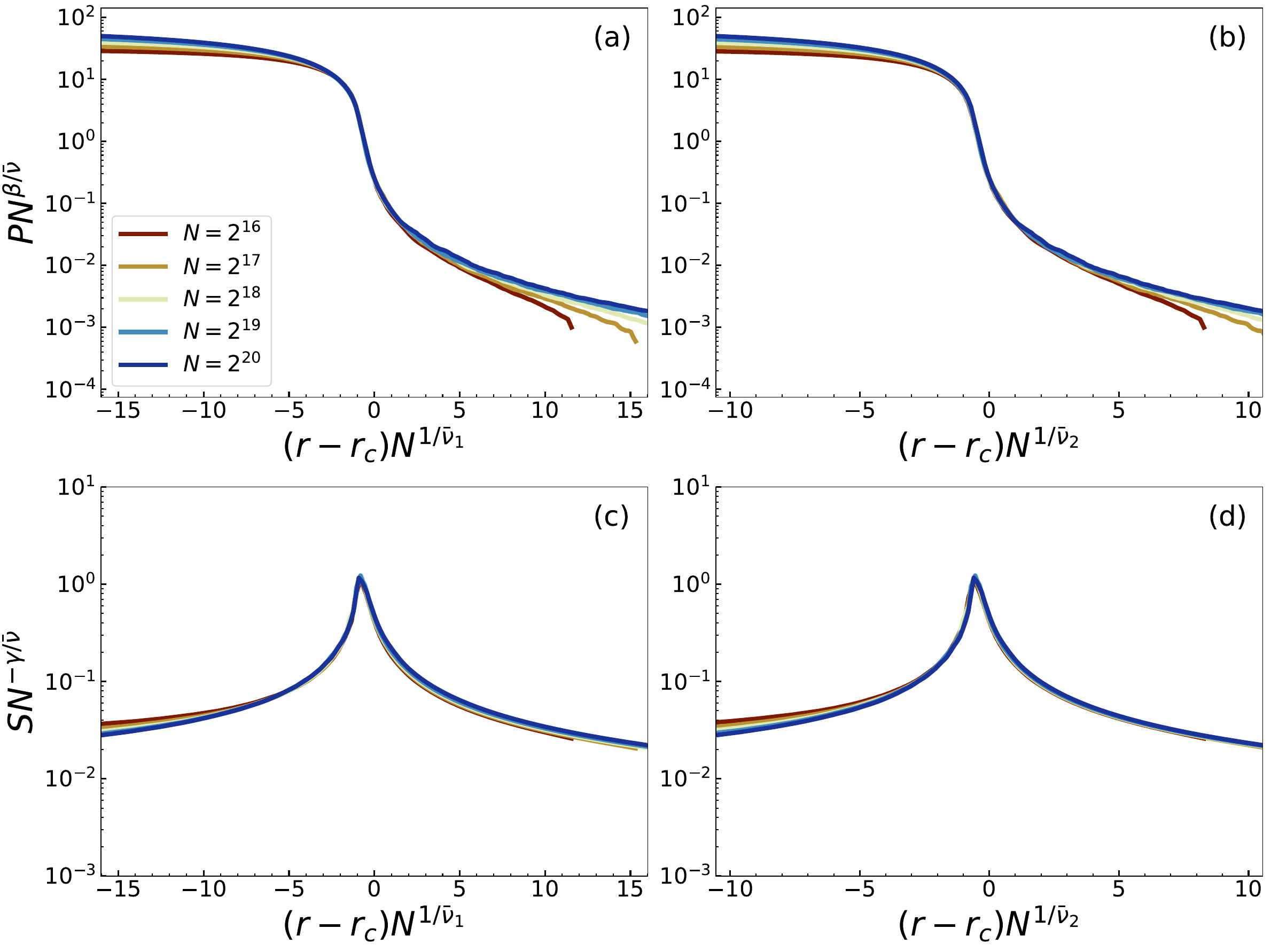}
    \caption{\textbf{Data Collapse of $P$ and $S$.} Similar to Fig.~\ref{sm-fig:collapse_conventional_sfn2.1_C2}, but results are valid for SFNs with $\lambda=2.7$ and $k_{min}=4$, and SPP model with $C=2$.}
    \label{sm-fig:collapse_conventional_sfn2.7_C2}
\end{figure}

\begin{figure}
    \centering
    \includegraphics[width=0.7\linewidth]{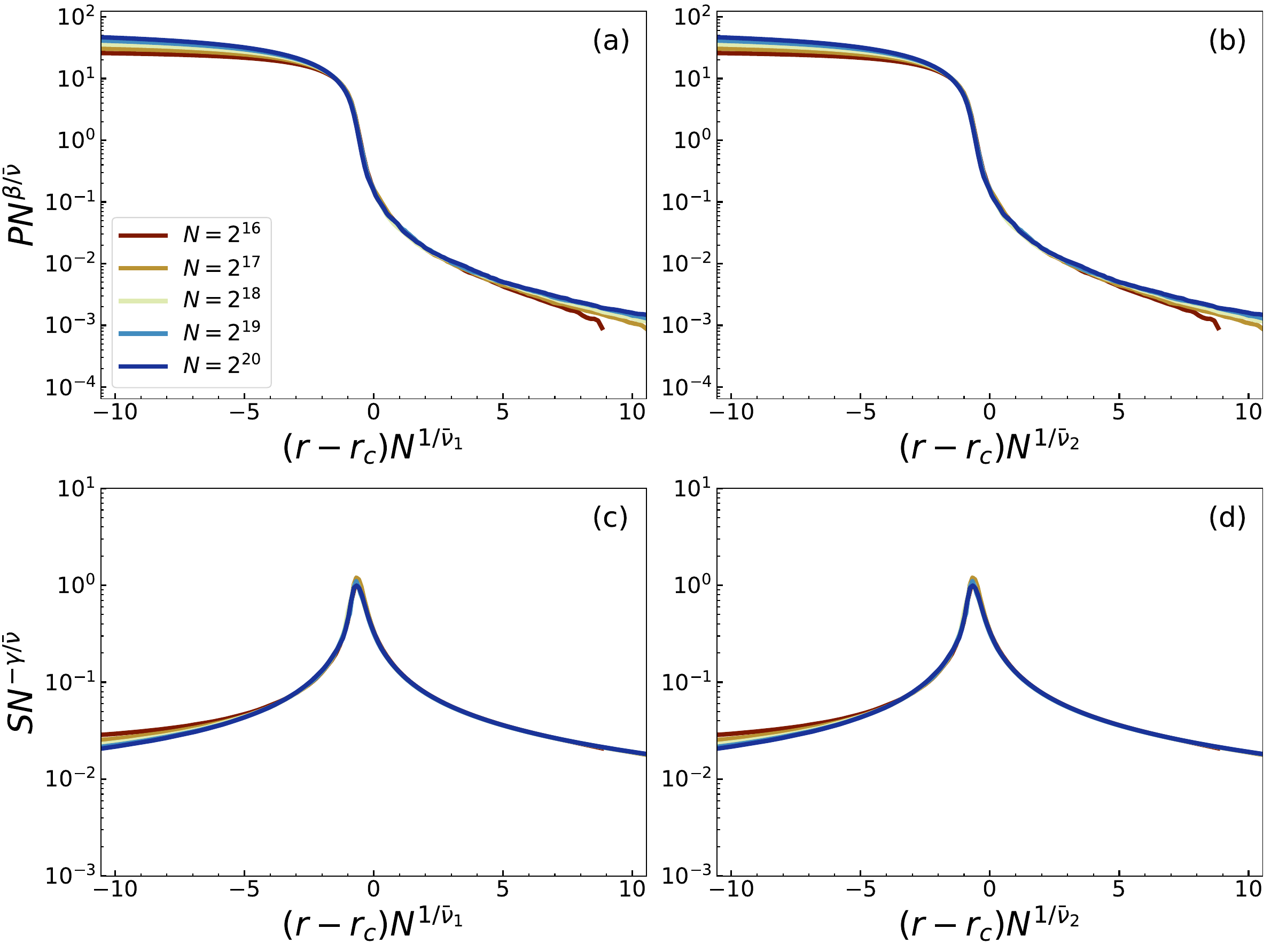}
    \caption{\textbf{Data Collapse of $P$ and $S$.} Similar to Fig.~\ref{sm-fig:collapse_conventional_sfn2.7_C2}, but results are valid for SPP model with $C=3$.}
    \label{sm-fig:collapse_conventional_sfn2.7_C3}
\end{figure}

\begin{figure}
    \centering
    \includegraphics[width=0.7\linewidth]{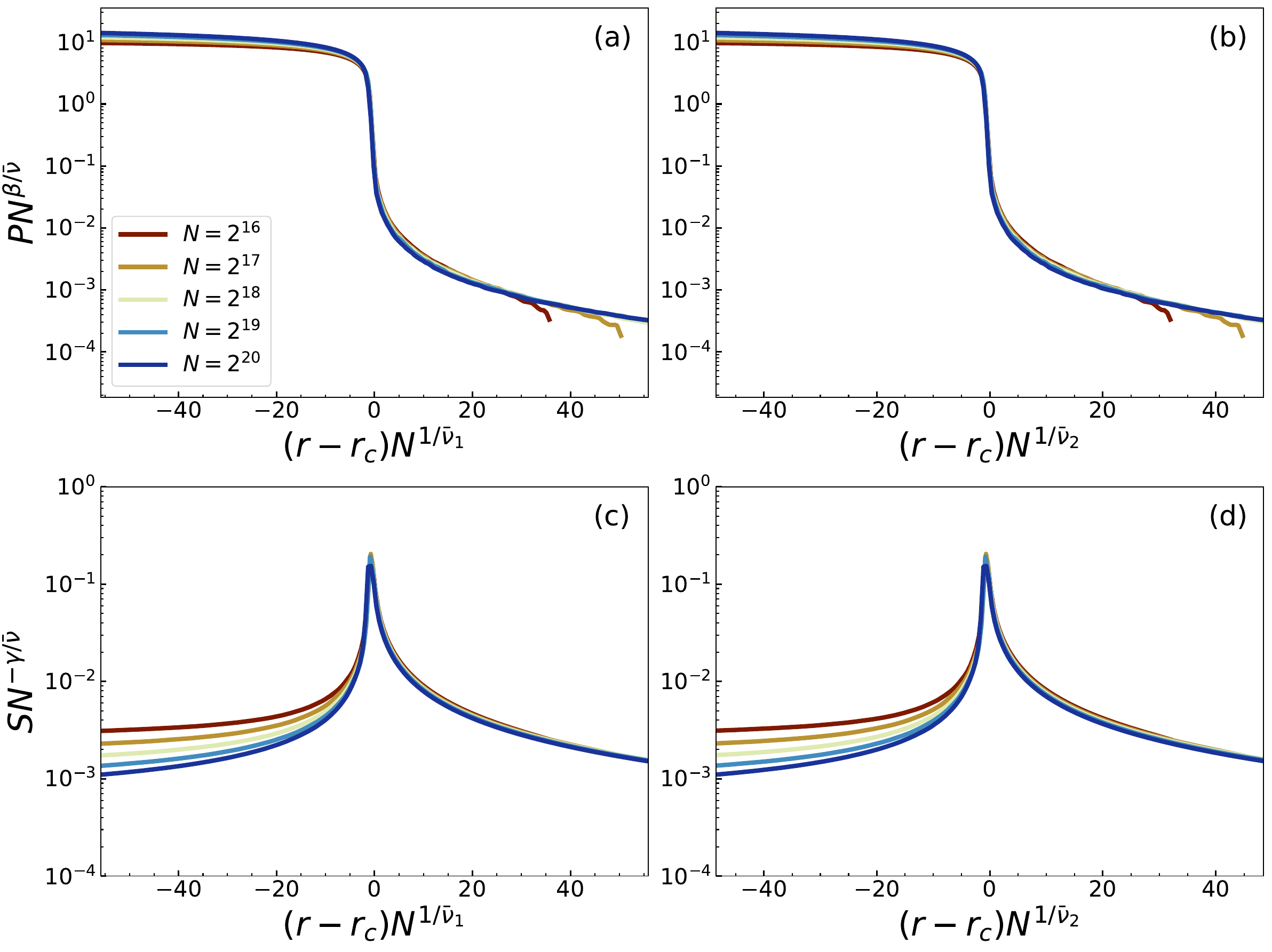}
    \caption{\textbf{Data Collapse of $P$ and $S$.} Similar to Fig.~\ref{sm-fig:collapse_conventional_sfn2.7_C2}, but results are valid for SPP model with $C=N$.}
    \label{sm-fig:collapse_conventional_sfn2.7_CN}
\end{figure}

\pagebreak
\clearpage

\begin{figure}
    \centering
    \includegraphics[width=0.7\linewidth]{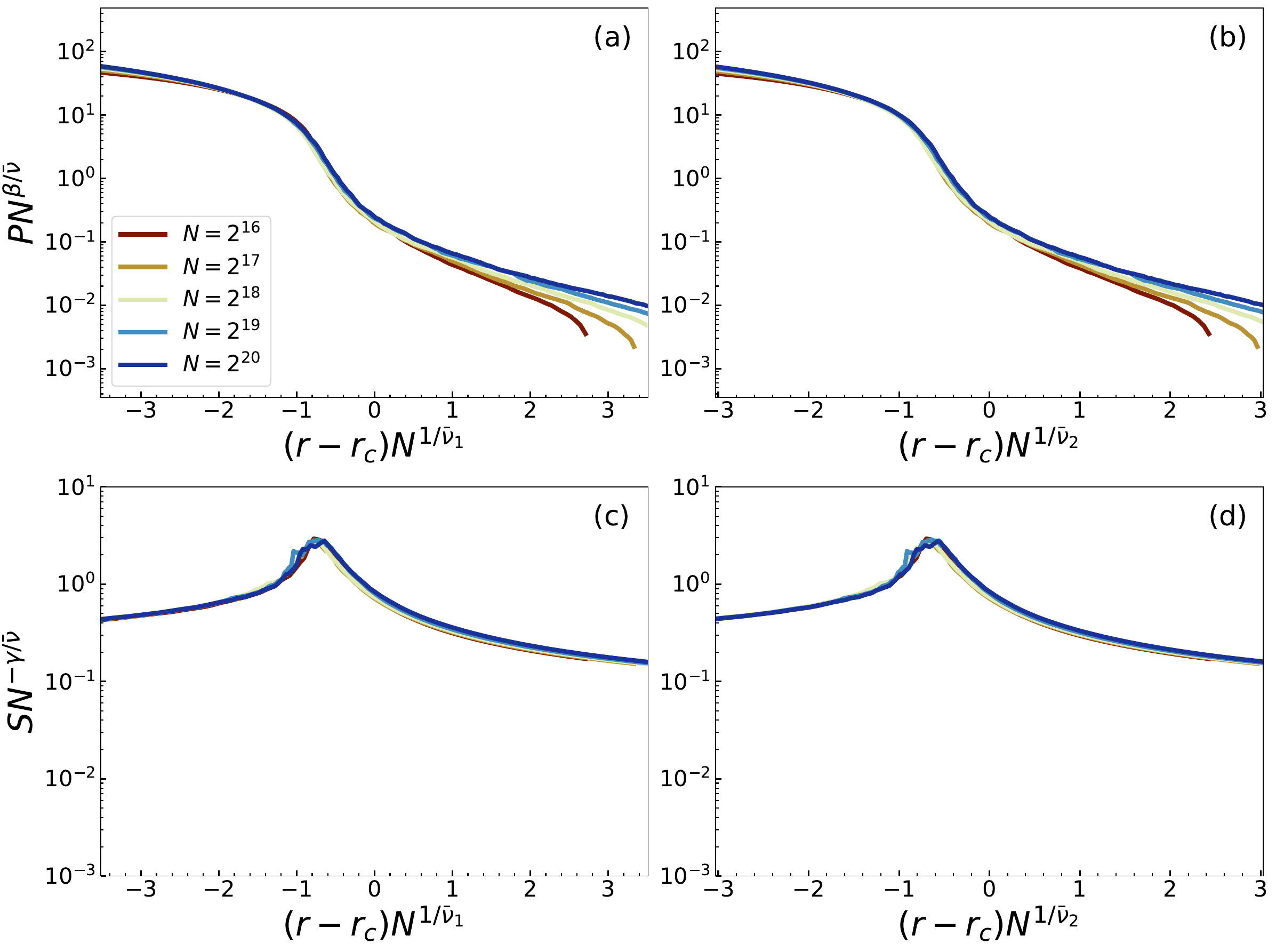}
    \caption{\textbf{Data Collapse of $P$ and $S$.} Similar to Fig.~\ref{sm-fig:collapse_conventional_sfn2.1_C2}, but results are valid for SFNs with $\lambda=3.5$ and $k_{min}=4$, and SPP model with $C=1$.}
    \label{sm-fig:collapse_conventional_sfn3.5_C1}
\end{figure}

\begin{figure}
    \centering
    \includegraphics[width=0.7\linewidth]{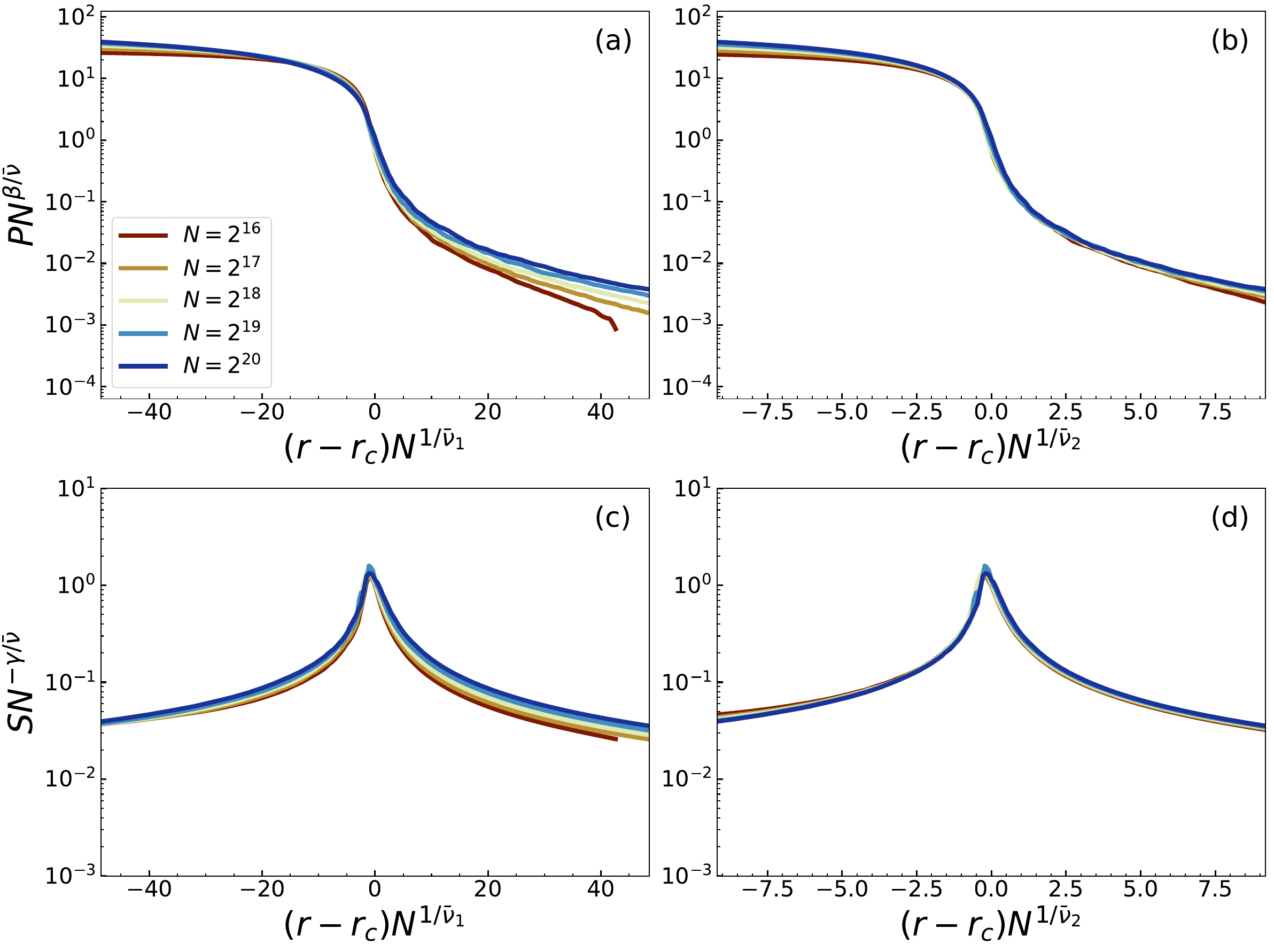}
    \caption{\textbf{Data Collapse of $P$ and $S$.} Similar to Fig.~\ref{sm-fig:collapse_conventional_sfn3.5_C1}, but results are valid for SPP model with $C=2$.}
    \label{sm-fig:collapse_conventional_sfn3.5_C2}
\end{figure}

\begin{figure}
    \centering
    \includegraphics[width=0.7\linewidth]{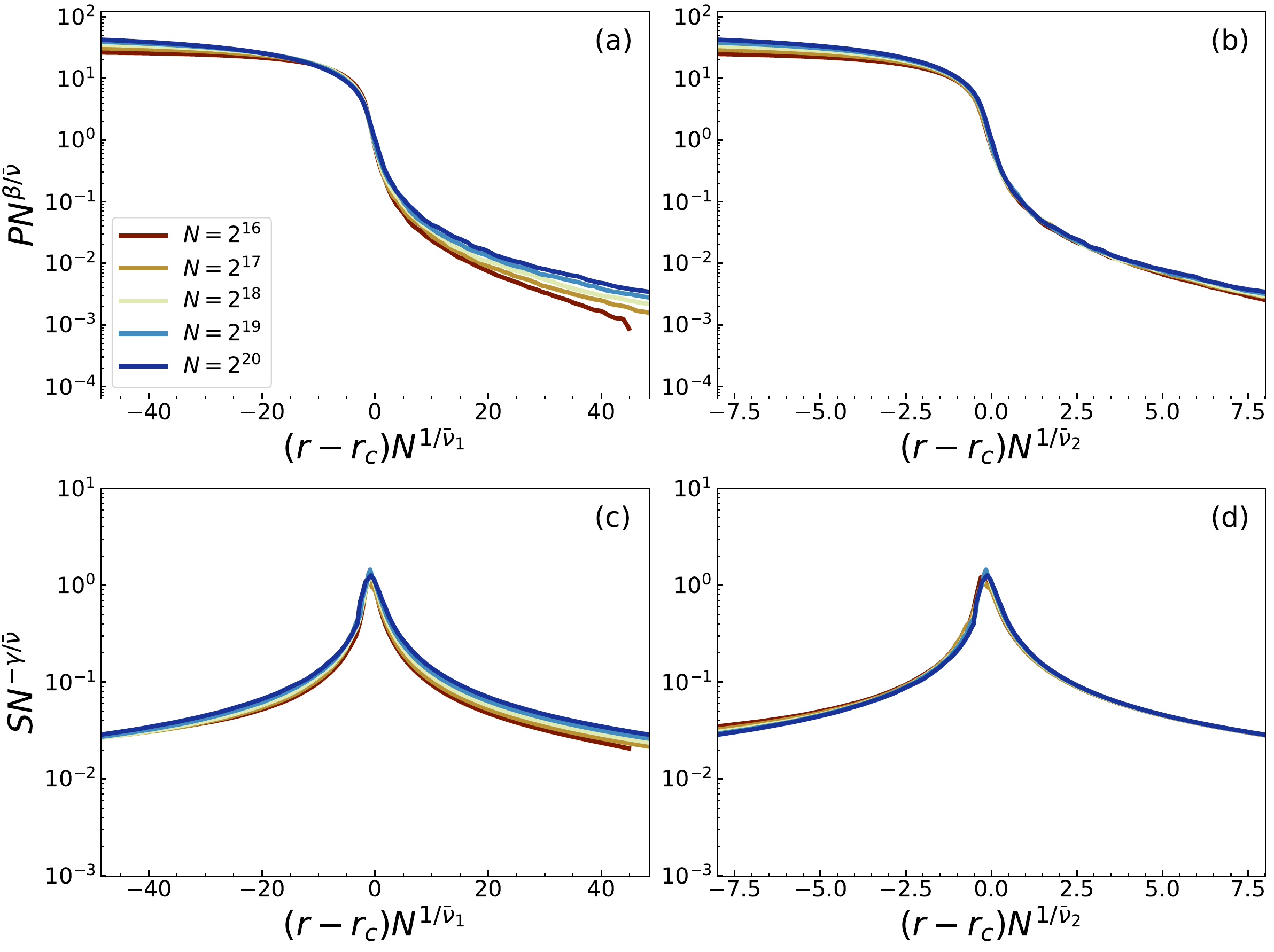}
    \caption{\textbf{Data Collapse of $P$ and $S$.} Similar to Fig.~\ref{sm-fig:collapse_conventional_sfn3.5_C1}, but results are valid for SPP model with $C=3$.}
    \label{sm-fig:collapse_conventional_sfn3.5_C3}
\end{figure}

\begin{figure}
    \centering
    \includegraphics[width=0.7\linewidth]{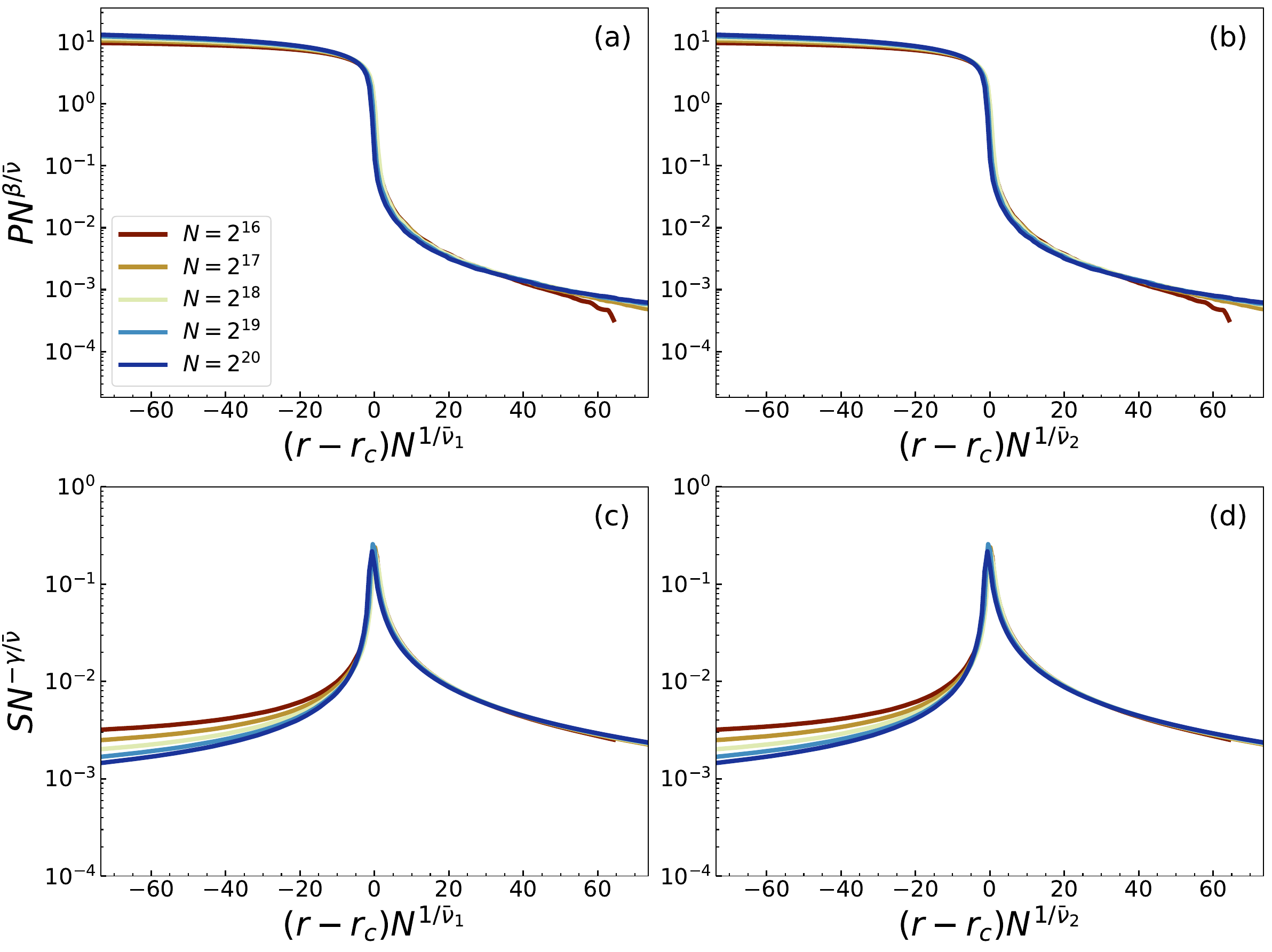}
    \caption{\textbf{Data Collapse of $P$ and $S$.} Similar to Fig.~\ref{sm-fig:collapse_conventional_sfn3.5_C1}, but results are valid for SPP model with $C=N$.}
    \label{sm-fig:collapse_conventional_sfn3.5_C-1}
\end{figure}

\pagebreak
\clearpage

\begin{figure}
    \centering
    \includegraphics[width=0.7\linewidth]{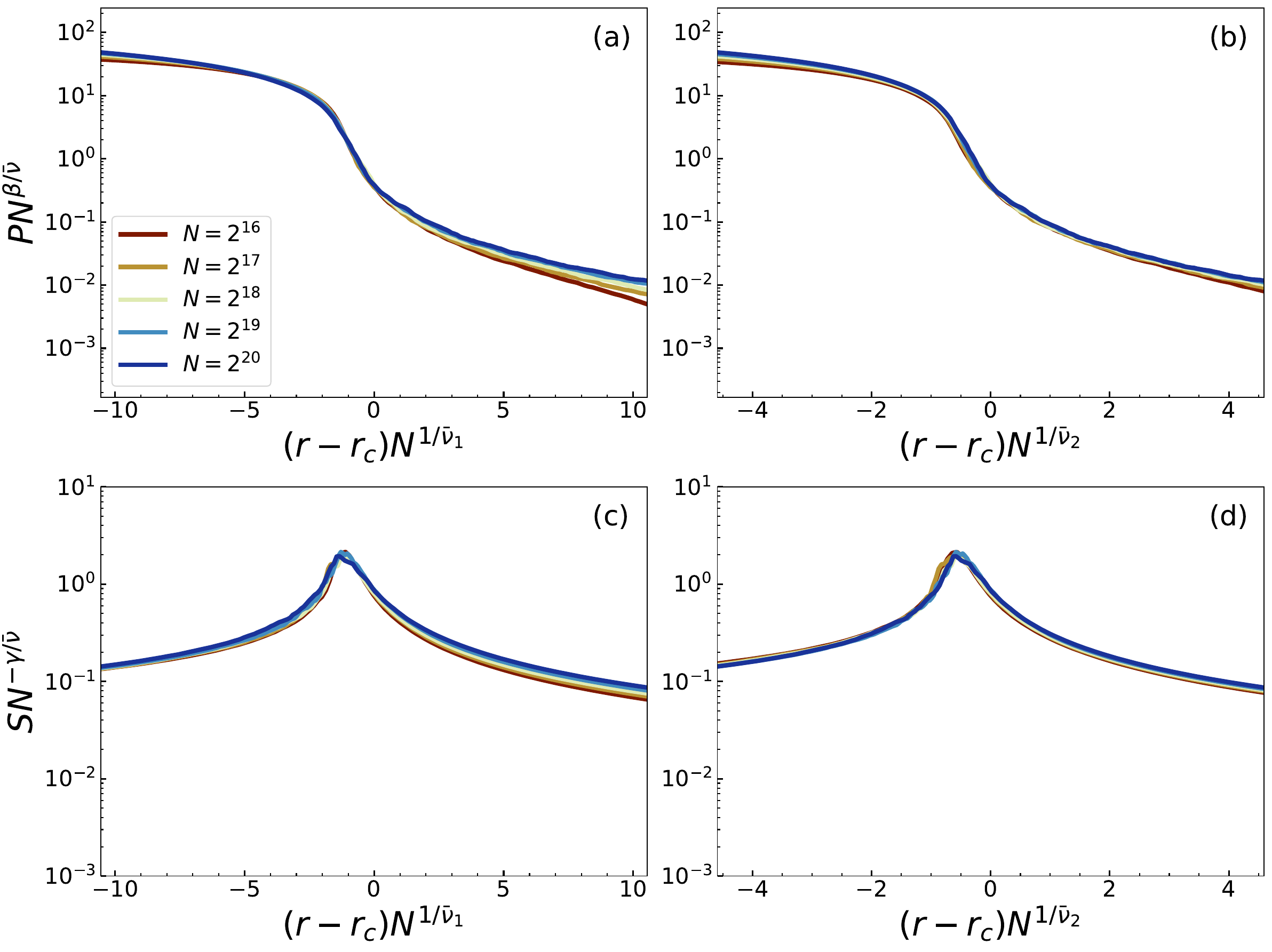}
    \caption{\textbf{Data Collapse of $P$ and $S$.} Similar to Fig.~\ref{sm-fig:collapse_conventional_sfn2.1_C2}, but results are valid for SFNs with $\lambda=4.5$ and $k_{min}=4$, and SPP model with $C=1$.}
    \label{sm-fig:collapse_conventional_sfn4.5_C1}
\end{figure}

\begin{figure}
    \centering
    \includegraphics[width=0.7\linewidth]{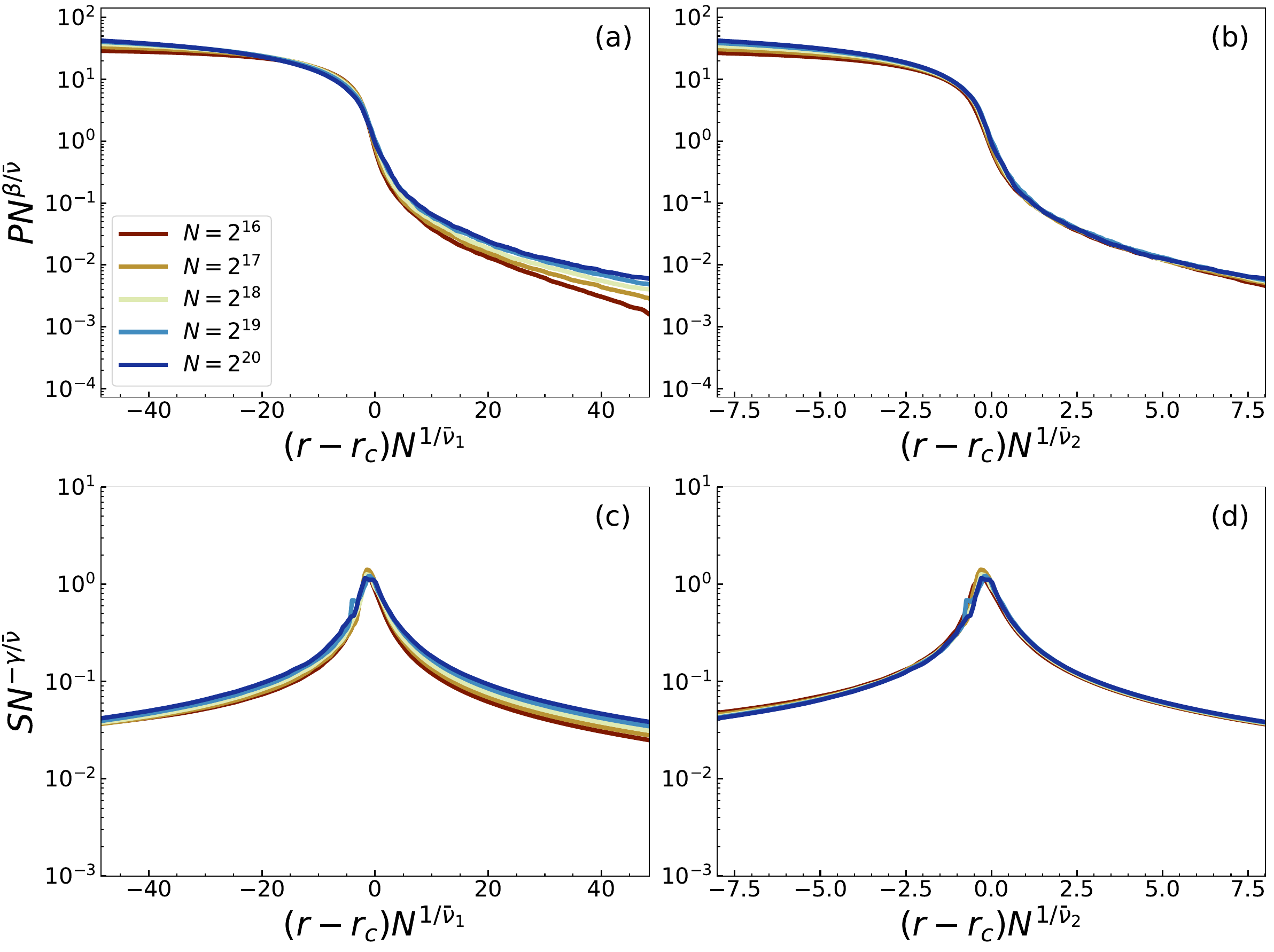}
    \caption{\textbf{Data Collapse of $P$ and $S$.} Similar to Fig.~\ref{sm-fig:collapse_conventional_sfn4.5_C1}, but results are valid for SPP model with $C=2$.}
    \label{sm-fig:collapse_conventional_sfn4.5_C2}
\end{figure}

\begin{figure}
    \centering
    \includegraphics[width=0.7\linewidth]{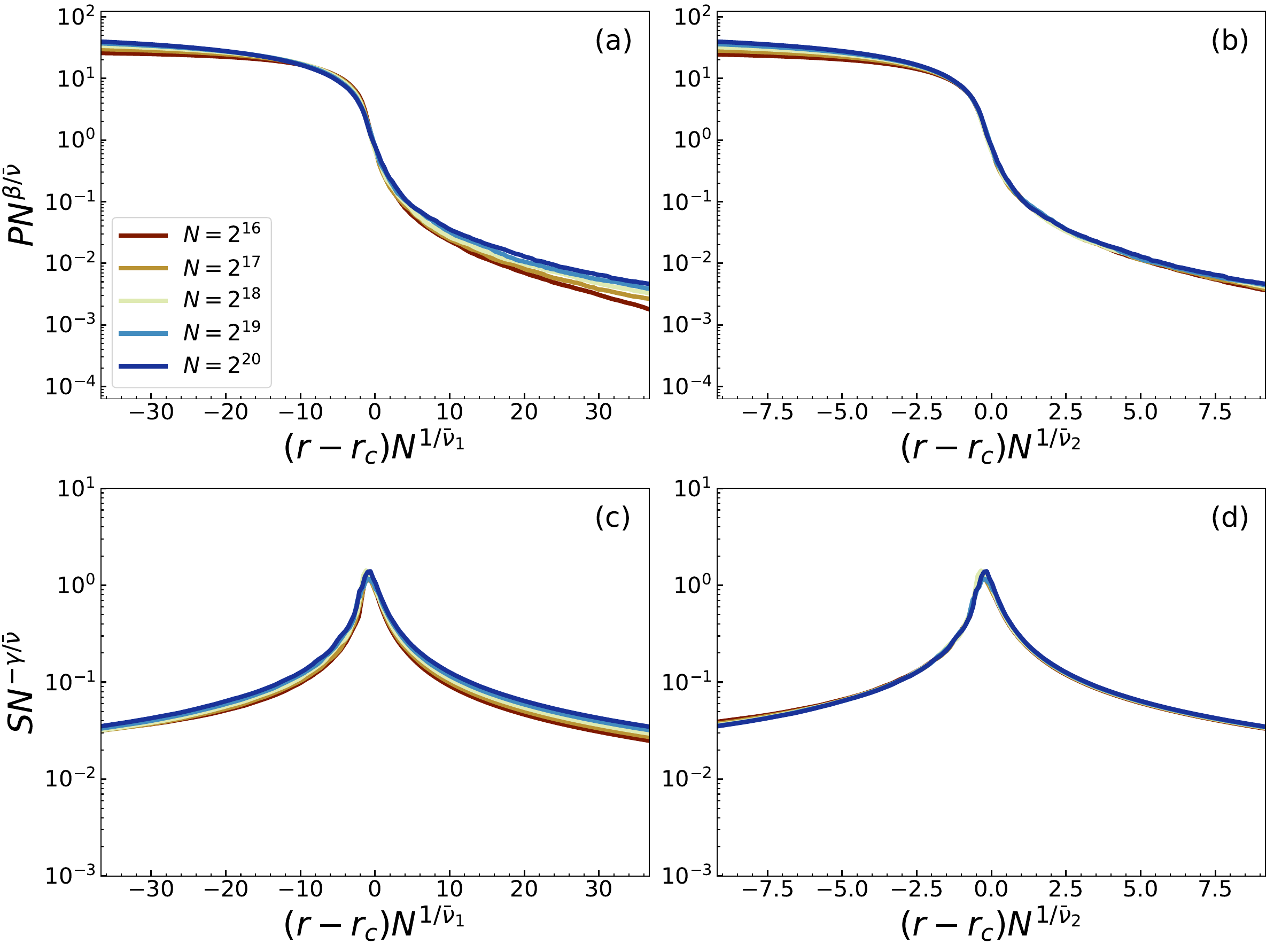}
    \caption{\textbf{Data Collapse of $P$ and $S$.} Similar to Fig.~\ref{sm-fig:collapse_conventional_sfn4.5_C1}, but results are valid for SPP model with $C=3$.}
    \label{sm-fig:collapse_conventional_sfn4.5_C3}
\end{figure}

\begin{figure}
    \centering
    \includegraphics[width=0.7\linewidth]{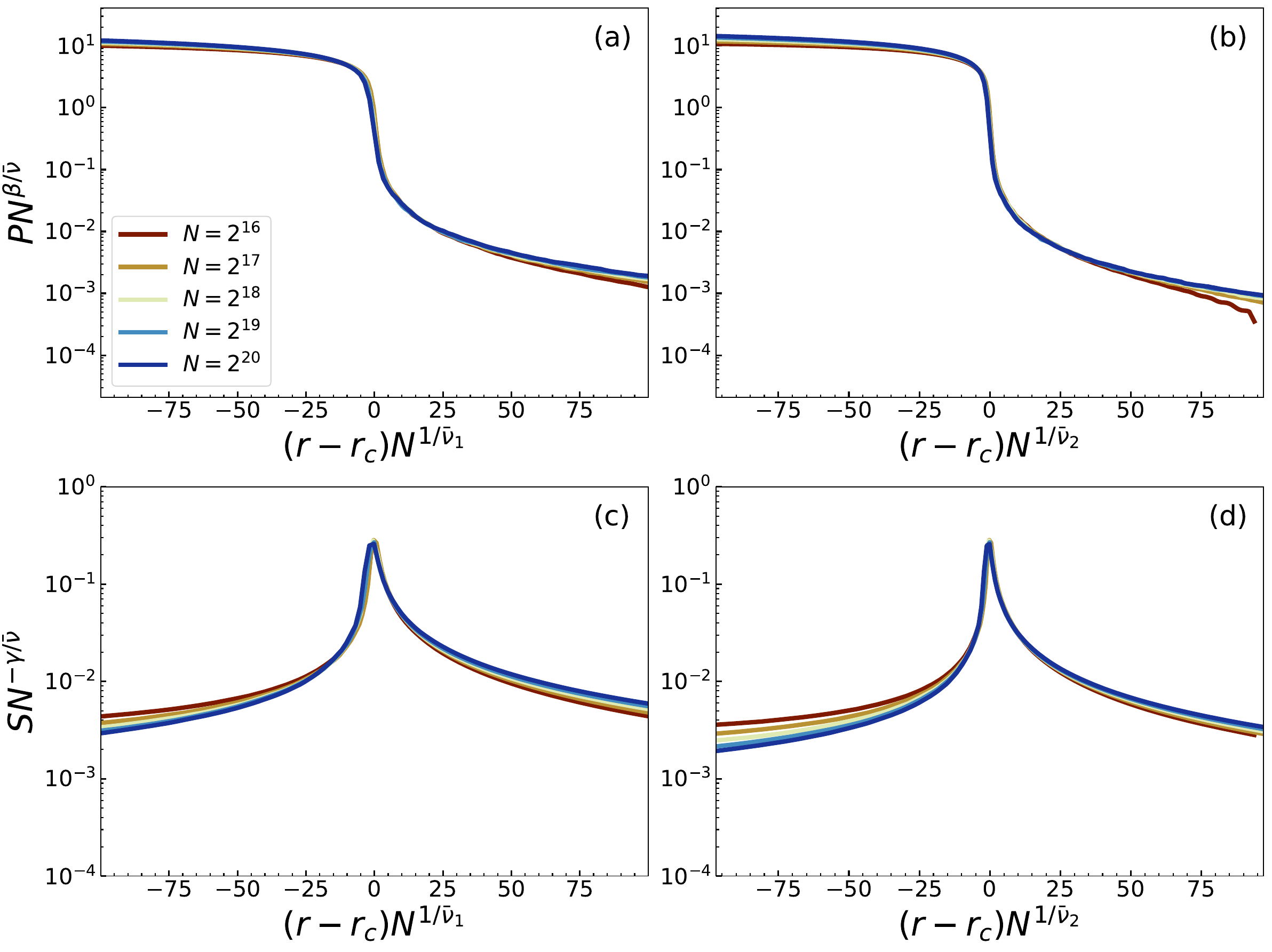}
    \caption{\textbf{Data Collapse of $P$ and $S$.} Similar to Fig.~\ref{sm-fig:collapse_conventional_sfn4.5_C1}, but results are valid for SPP model with $C=N$.}
    \label{sm-fig:collapse_conventional_sfn4.5_C-1}
\end{figure}

\clearpage

\section*{Supplementary tables}
In this section, we report detailed information related to the numerical simulations performed in this study.

\renewcommand{\arraystretch}{1.3}
\begin{table*}[!htb]
\centering
\begin{tabular*}{0.6\textwidth}{@{\extracolsep{\fill}}llll}
\hline\hline
$\lambda$ & $C$ & $N$ & \# of realizations \\
\hline
2.1 & 1 & $2^{10}$--$2^{20}$ & 4000 \\
\hline
    & 2 & $2^{10}$--$2^{17}$ & 4000 \\
    &   & $2^{18}$           & 3000 \\
    &   & $2^{19}$--$2^{20}$ & 1000 \\
\hline    
    & 3 & $2^{10}$--$2^{17}$ & 4000 \\
    &   & $2^{18}$           & 2000 \\
    &   & $2^{19}$           & 800  \\
    &   & $2^{20}$           & 350  \\
\hline
    & $N$ & $2^{10}$--$2^{16}$ & 4000 \\
    &   & $2^{17}$           & 3000 \\
    &   & $2^{18}$           & 1000 \\
    &   & $2^{19}$           & 800  \\
    &   & $2^{20}$           & 300  \\
\hline
\hline
2.7 & 1 & $2^{10}$--$2^{20}$ & 4000 \\
\hline
    & 2 & $2^{10}$--$2^{18}$ & 4000 \\
    &   & $2^{19}$--$2^{20}$ & 2000 \\
\hline
    & 3 & $2^{10}$--$2^{18}$ & 4000 \\
    &   & $2^{19}$--$2^{20}$ & 2000 \\
\hline
    & $N$ & $2^{10}$--$2^{16}$ & 4000 \\
    &   & $2^{17}$--$2^{18}$ & 3000 \\
    &   & $2^{19}$           & 1000 \\
    &   & $2^{20}$           & 700  \\
\hline
\hline
3.5 & 1 & $2^{10}$--$2^{20}$ & 4000 \\
\hline
    & 2 & $2^{10}$--$2^{18}$ & 4000 \\
    &   & $2^{19}$--$2^{20}$ & 2000 \\
\hline
    & 3 & $2^{10}$--$2^{18}$ & 4000 \\
    &   & $2^{19}$--$2^{20}$ & 2000 \\
\hline
    & $N$ & $2^{10}$--$2^{18}$ & 4000 \\
    &   & $2^{19}$--$2^{20}$ & 1000 \\
\hline
\hline
4.5 & 1 & $2^{10}$--$2^{20}$ & 4000 \\
\hline
    & 2 & $2^{10}$--$2^{18}$ & 4000 \\
    &   & $2^{19}$--$2^{20}$ & 2000 \\
\hline
    & 3 & $2^{10}$--$2^{18}$ & 4000 \\
    &   & $2^{19}$--$2^{20}$ & 2000 \\
\hline
    & $N$ & $2^{10}$--$2^{18}$ & 4000 \\
    &   & $2^{19}$--$2^{20}$ & 1000 \\
\hline\hline
\end{tabular*}
\caption{\textbf{Number of realizations.} We report the number of realizations of the shortest-path percolation processes used to estimate the critical point and the ratios of critical exponents. For a given initial network configuration, we repeated at most 10 times of the shortest-path percolation process.}
\label{tab:simulation}
\end{table*}

\renewcommand{\arraystretch}{1.3}
\begin{table*}[!htb]
\centering
\begin{tabular*}{0.5\textwidth}{@{\extracolsep{\fill}}lll}
\hline\hline
$C$ & $N$ & \# of realizations \\
\hline
1 & $2^{10}$--$2^{18}$ & 300 \\
  & $2^{19}$--$2^{20}$ & 100 \\
\hline
2 & $2^{10}$--$2^{18}$ & 300 \\
  & $2^{19}$--$2^{20}$ & 100 \\
\hline
3 & $2^{10}$--$2^{17}$ & 300 \\
  & $2^{18}$            & 250 \\
  & $2^{19}$            & 100 \\
  & $2^{20}$            & 50  \\
\hline
$N$ & $2^{10}$--$2^{17}$ & 300 \\
    & $2^{18}$            & 200 \\
    & $2^{19}$            & 100 \\
    & $2^{20}$            & 50  \\
\hline\hline
\end{tabular*}
\caption{\textbf{Number of realizations.} We report the number of realizations used in Fig.~\ref{fig3} in the main. For a given initial network configuration, we repeated at most 10 times of the shortest-path percolation process.}
\label{tab:k_ratio_realization}
\end{table*}

\renewcommand{\arraystretch}{1.3}
\begin{table*}[!htb]
\centering
\begin{tabular*}{0.5\textwidth}{@{\extracolsep{\fill}}lll}
\hline\hline
$N$ & \# of bins & \# of realizations \\
\hline
$2^{16}$ & 400  & 80 \\
$2^{17}$ & 800  & 40 \\
$2^{18}$ & 800  & 40 \\
$2^{19}$ & 1600 & 40 \\
$2^{20}$ & 1600 & 40 \\
\hline\hline
\end{tabular*}
\caption{\textbf{Binning information.} We report the number of bins and the number of realizations used in Fig.~\ref{fig5} in the main. This also applies to the other similar figures in the Supplementary Material. We clearly note that the binning was performed to reduce the visual noise in the figure, not to obtain the critical point or the ratio of critical exponents.}
\label{tab:binning}
\end{table*}

\renewcommand{\arraystretch}{1.3}
\begin{table*}[!ht]
\centering
\begin{tabular*}{\textwidth}{@{\extracolsep{\fill}}llllll}
\hline\hline
$q$ & $C$ & $\Omega_{o,1}$ & $\xi_{1}$ & $\Omega_{o,2}$ & $\xi_2$ \\
\hline
0.7 & 1 & 0.889(1) & 0.330(3) & 0.884(1) & 0.287(2) \\
    & 2 & 0.000(1) & 0.209(4) & 0.000(1) & 0.198(5) \\
    & 3 & 0.000(1) & 0.210(3) & 0.000(1) & 0.198(3) \\
    & $N$ & 0.000(1) & 0.190(2) & 0.000(1) & 0.178(3) \\
\hline
0.8 & 1 & 0.779(1) & 0.327(2) & 0.768(1) & 0.275(2) \\
    & 2 & 0.000(1) & 0.235(5) & 0.000(1) & 0.224(5) \\
    & 3 & 0.000(1) & 0.236(4) & 0.000(1) & 0.223(4) \\
    & $N$ & 0.000(1) & 0.220(4) & 0.000(1) & 0.206(4) \\
\hline
0.9 & 1 & 0.560(1) & 0.308(1) & 0.535(1) & 0.252(1) \\
    & 2 & 0.000(1) & 0.251(6) & 0.000(1) & 0.247(6) \\
    & 3 & 0.000(1) & 0.251(6) & 0.000(1) & 0.246(5) \\
    & $N$ & 0.000(1) & 0.244(6) & 0.000(1) & 0.236(5) \\
\hline\hline
\end{tabular*}
\caption{
\textbf{Scaling of the $\rho$ ratios.}
We report the estimated variables $\Omega_{o,1}$, $\Omega_{o,2}$, $\xi_1$, and $\xi_2$ using Eq.~(\ref{eq:variance_scaling}).}
\label{tab:k_variance_slope}
\end{table*}

\end{document}